\documentclass[journal]{IEEEtran}

\usepackage{booktabs} % For formal tables
\usepackage{tabularx}
\usepackage[strict]{changepage}
\usepackage{calc}
\usepackage{mathtools,lipsum}
\usepackage{amsmath}
\usepackage{amssymb}
\usepackage{mathrsfs}
\usepackage{upgreek}
\usepackage{graphicx}
\usepackage{algorithmicx}
\usepackage{algorithm}
\usepackage{enumitem}
\usepackage[noend]{algpseudocode}
\usepackage{caption}
\usepackage{subcaption}
\usepackage{amsthm, times, amsfonts, cite}
\usepackage{epstopdf}

\newtheorem{theorem}{Theorem}
\newtheorem{lemma}{Lemma}

\usepackage[T1]{fontenc}
\usepackage[latin9]{inputenc}
\usepackage{array}
\usepackage{float}
\usepackage{amsmath}
\usepackage{amssymb}
\usepackage{graphicx}

\usepackage{amsmath}

\begin{document}

\title{FastTrack: Minimizing Stalls for CDN-based Over-the-top Video Streaming Systems}

\author{Abubakr Alabbasi$^{1}$, Vaneet Aggarwal$^{1}$, Tian Lan$^{2}$, Yu Xiang$^{3}$, Moo-Ryong Ra$^{3}$, and Yih-Farn R. Chen$^{3}$\\
$^{1}$Purdue University, West Lafayette, IN 47907, USA\\
% (email:\{aalabbas,vaneet\}@purdue.edu) \\
$^{2}$George Washington University, Washington DC, 20052, USA\\
% (email:tlan@gwu.edu). \\
$^{3}$AT\&T Research Labs, Bedminster, NJ 07921, USA}
% (e-mail:{yxiang,mra,chen}@research.att.com)}
	\maketitle

\begin{abstract}
Traffic for internet video streaming has been rapidly increasing and 
is further expected to increase with the higher definition videos and 
IoT applications, such as 360 degree videos and augmented virtual reality applications. 
While efficient management of heterogeneous cloud resources to optimize the quality of experience is important, existing work in this problem space often 
left out important factors.
In this paper, we present a model for describing a today's representative system architecture for video streaming applications, typically composed of a centralized origin server and several CDN sites.
Our model comprehensively considers the following factors: 
limited caching spaces at the CDN sites, allocation of CDN for a video request, 
choice of different ports from the CDN, and the central storage and bandwidth allocation. With the model, we focus on minimizing a performance metric, stall duration tail probability (SDTP), 
and present a novel, yet efficient, algorithm to solve the formulated optimization problem.
The theoretical bounds with respect to the SDTP metric are also analyzed and presented. 
Our extensive simulation results demonstrate that the proposed algorithms can significantly improve the SDTP metric, compared to the baseline strategies. Small-scale video streaming system implementation in a real cloud environment further validates our results.
%We take one step further and implement a small-scale video streaming systems 
%in a real cloud environment and validate our results.
\end{abstract}


\section{Introduction}
\label{sec:Intro}

Over-the-top video streaming, e.g., Netflix and YouTube, has been dominating the global IP traffic in recent years. It is shown in \cite{networking2016forecast} that video streaming applications in North America now represents $62\%$ of the Internet traffic, and this figure will continue to grow due to the introduction of even higher resolution video formats such as 4K on the horizon. With the growing popularity of video services, increased congestion and latency related to retrieving content from remote datacenters can lead to degraded end customer experience. Service and content providers often seek to mitigate such performance issues by employing caching at the network edge and by pushing content closer to their customers using content distribution networks (CDNs). More than $50\%$ of over-the-top video traffic are now delivered through CDNs \cite{rep62}.

%[T1] Netflix. Netflix OpenConnect Appliance Deployment Guide, April 2015. Version 3.7. Version 3.7.
%[T2] Cache Content-Selection Policies for Streaming Video Services, INFOCOM 2016

Caching of video content has to address a number of crucial challenges that differ from caching of web objects, see for instance \cite{7524619} and the references therein. First, video streaming services such as Netflix \cite{T1} often adopt a proactive caching strategy, which conciously pushes video files into local caches during off-peak hours, while cache content is updated according to changes in predicted demand. Due to the correlation in user preferences within different regions \cite{7524619}, it calls for new solutions that take into account both regional and global popularity of video files, for jointly optimizing cache content and performance. Second, video files are significantly larger in size than web objects. In order to minimize congestion and latency, caching of video files must be optimized together with network resource allocation and request scheduling, which however, is currently under-explored. Finally, while recent work have considered video-streaming over distributed storage systems \cite{7524619}, they normally focus on network performance metrics similar to those considered by web object caching (e.g., packet delay and cache hit rate), rather than Quality-of-Experience (QoE) metrics that are more relevant to end user experience in over-the-top video streaming.

In this paper, we propose a new QoE metric, the stall duration tail probability (SDTP), which measures the likelihood of end users suffering a worse-than-expected stall duration, and develop a holistic optimization framework for minimizing the overall SDTP over joint caching content placement, network resource optimization and user request scheduling. We consider a Virtualized Content Distribution Network (vCDN) architecture. It consists of a remote datacenter that stores complete original video data and multiple CDN sites (i.e., local cache servers) that only have part of those data and are equipped with solid state drives (SSDs) for higher throughput. A user request for video content not satisfied in the local cache is directed to, and processed by, the remote datacenter 
%where the users are connected to an edge router fetching video content from distributed storage cache servers (as depicted . If the required content is not stored in cache servers, the requested contents are fetched from the origin datacenter. 
(as shown in  Fig. \ref{fig:sysModel}). If the required video content/chunk is not stored in cache servers, multiple parallel connections 
% (or virtual machines (VMs))  % RC:  connections are NOT the same as VMs
are established between a cache server and the edge router, as well as between the cache servers and the origin server, to support multiple video streams simultaneously. Our goal is to develop an optimization framework and QoE metrics that server and content providers could use to answer the following questions: How to quantify the impact of video caching on end user experience? What is the best video caching strategy for CDN? How to optimize QoE metrics over various ``control knobs"? Are there enough benefits to justify the adoption of proposed solutions in practice?

%{\color{blue} Talk about the SDTP metric, why it is important, how to quantify it in our system, key steps and contributions in the analysis, and compare with some related work.}  

It has been shown that in modern cloud applications such as FaceBook, Bing, and Amazon's retail platform, that the long tail of latency is of major concern, with $99.9$th percentile response times that are orders of magnitude worse than the mean \cite{T1,T2}. The key QoE metric considered in this paper is SDTP, denoted by Pr$(\Gamma^{(i)}>\sigma)$, which measures the probability that the stall duration $\Gamma^{(i)}$ of video $i$ is greater than a pre-defined threshold $\sigma$. Despite resource and load-balancing mechanisms, large scale storage systems evaluations shows that there
is a high degree of randomness in delay performance \cite{JK13}. In contrast to web object caching and delivery, the video chunks in the latter part of a video do not have to be downloaded much earlier than their actual play time to maintain the desired QoE, making SDTP highly dependent on the joint optimization with resource management and request scheduling in CDN-based video streaming. %hence multiple parallel streams help achieve better QoE. 
%Hence, the QoE metric of SDTP becomes crucial. 

Quantifying SDTP with ditributed cache/storage is an open problem. Even for single-chunk video files, the problem is equivalent to minimizing the download tail latency, which is still an open problem \cite{MDS-Queue}. The key challenge arises from the difficulty of constructing and analyzing a scheduling policy that (optimally) redirects each request based on {\em dependent system and queuing dynamics} (including cache content, network conditions, request queue status) on the fly. To overcome these challenges, we propose a novel two-stage, probabilistic scheduling approach, where each request of video $i$ is (i) processed by cache server $j$ with probability $\pi_{i,j}$ and (ii) assigned to video stream $v$ with probability $p_{i,j,v}$. The two-stage, probability scheduling allows us to model each cache server and video stream as separate queues, and thus, to characterize the distributions of different video chunks' download time and playback time. 
%The key challenges in quantification of SDTP are the choices of scheduling policy to choose the storage cache servers for each request, as well as the parallel stream from the chosen servers, which we denote in the rest of this paper as the two-stage scheduling probability. 
%For a single video-chunk, the problem is equivalent to minimizing the download tail latency, which is still an open problem \cite{MDS-Queue}. Further, the cache placement optimization makes the problem challenging since the choice of the portion of the video that needs to be cached would depend on many factors including the users' requests behavior and the capacity of the cache servers. 
By optimizing these probabilities, we quantify SDTP through a closed-form, tight upper bound for CDN-based video streaming with arbitrary cache content placement and network resource allocation. We note that the analysis in this paper is fundamentally different from those for distributed file storage, e.g., \cite{Xiang:2014:Sigmetrics:2014,Yu_TON}, because the stall duration of a video relies on the download times of all its chunks, rather than simply the time to download the last chunk of a file. Further, since video chunks are downloaded and played sequentially, the download times and playback times of different video chunks are highly correlated and thus jointly determine the SDTP metric.

We propose FastTrack, a holistic optimization framework for minimizing overall SDTP in CDN-based video streaming. To the best of our knowledge, this is the first framework to jointly consider all key design degrees of freedom, including bandwidth allocation among different parallel streams, cache content placement and update, request scheduling, and the modeling variables associated with the SDTP bound. An efficient algorithm is then proposed to solve this non-convex optimization problem. In particular, the proposed algorithm performs an alternating optimization over the different dimensions, such that each sub-problem is shown to have convex constraints and thus can be efficiently solved using the iNner cOnVex
Approximation (NOVA)  algorithm proposed in \cite{scutNOVA}. 
%The proposed algorithm is shown to converge to a local optimal. 
The proposed algorithm is implemented in a virtualized cloud system manged by Openstack~\cite{openstack}. The experimental results demonstrate significant improvement of QoE metrics as compared to the considered baselines.

%{\color{blue} Talk about our holistic optimization framework, different decision variables involved, our optimization algorithm. Highlight the novelty of this optimization by comparing it with prior work.} To the best of our knowledge, this is the first paper to jointly consider all these degrees of freedom for QoE optimization in video caching and SDN.

%{\color{blue} Discuss implementation details, simulation results, and key insights learned.} 

The main contributions of this paper can be summarized as follows:
%\begin{itemize}
%\item We propose SDTP and quantify it for CDN ...
%\item A holistic optimization framework is developed to ...
%\item We implement the proposed solutions ...
%\end{itemize}

\begin{itemize}[leftmargin=0cm,itemindent=.3cm,labelwidth=\itemindent,labelsep=0cm,align=left]
\item We propose a novel framework for analyzing CDN-based over-the top video streaming systems with the use of parallel streams. A novel two-stage  probabilistic scheduling policy is proposed  to assign each user request to different cache servers and parallel video streams. 

%analyze the bounds on SDTP for the CDN-based over-the top video streaming systems through a new two-stage probabilistic scheduling policy for assigning each user request to different cache servers and parallel video streams. 
%the choice of server and the parallel streams. Further, the cache placement is optimized. 

\item The distribution of (random) download time of different video chunks are analyzed. Then, using ordered statistics, we quantify the playback time of each video segment and derive an analytical upper bound on SDTP for arbitrary cache content placement and the parameters of the two-stage probabilistic scheduling.

%the random variable corresponding to the playback time of each video segment is characterized. This is further used to give a bound on the SDTP. 

\item  A holistic optimization framework is developed to optimize a weighted sum of SDTP of all video files over the request scheduling probabilities, cache content placement, the bandwidth allocation among different streams,  and the modeling parameters in SDTP bound. An efficient algorithm is provided to decouple and solve this non-convex optimization.

\item 
The experimental results validate our theoretical analysis and demonstrate the efficacy of our proposed
algorithm. Further, the proposed algorithm shows to
converge within a few iterations. Moreover, the QoE metric is shown to have significant improvement as compared to competitive strategies. 
\end{itemize}

The rest of this paper is organized as follows. Section \ref{sec:RelWork}
provides related work for this paper. In Section \ref{sec:SysMod}, we describe
the system model used in the paper with a description of
CDN-based Over-the-top video streaming systems. Section \ref{sec:DownLoadPlay} derives expressions for the download and play times of the chunks from storage cache servers as well as origin datacenter, which are used in  Section \ref{sec:StallTailProb}  to find an upper bound on the mean stall duration. Section \ref{sec:SDTP} formulates the  QoE optimization problem as a weighted sum of all SDTP of all files and proposes the iterative algorithmic solution of this problem.   Numerical results are provided in Section \ref{sec:num}. Section \ref{sec:conc} concludes the paper
\section{Related Work}\label{sec:RelWork}
Servicing Video on Demand and Live TV Content from cloud servers have been studied widely \cite{lee2013vod,huang2011cloudstream,he2014cost,chang2016novel,oza2016implementation}. 
% cite Aggarwal:multimedia:2013 here for final version
The placement of content and resource optimization over the cloud servers have been considered. In \cite{jointContent}, authors utilize the social
information propagation pattern to improve the efficiency of
social video distribution. Further, they used replication and user request dispatching mechanism in the cloud content
delivery network architecture to reduce the system operational
cost, while maintaining the averaged service latency. However, this work considers only video download. The benefits of delivering videos at the edge network is shown in \cite{Hu2016}. Authors show that bringing videos at the edge network can significantly improve the content item delivery performance,
in terms of improving quality experienced by
users as well as reducing content item delivery costs.
To the best of our knowledge, reliability of content over the cloud servers have not been considered for video streaming applications. There are novel challenges to characterize and optimize the QoE metrics at the end user. Adaptive streaming algorithms have also been considered for video streaming \cite{chen2012amvsc,wang2013ames,Garcia016} \cite{7939148} which are beyond the scope of this paper and are left for future work. 

Mean latency and tail latency have been characterized in \cite{Xiang:2014:Sigmetrics:2014,Yu_TON} and \cite{Jingxian,Tail_TON}, respectively, for a system with multiple files using probabilistic scheduling. However, these papers consider only file downloading rather than video streaming.  This paper considers CDN-based video streaming. We note that file downloading can follow as a special case of streaming, which makes our model more general. Additionally,  the metrics for video streaming do not only account for the end of the download of the video but also for the download of each segment. Hence, the analysis for the content download cannot be extended to the video streaming directly and the analysis approach in this paper is very different from the prior works in the area of file downloading.

Recently, the authors of \cite{Abubakr_TON} considered video-streaming over distributed storage systems. However, caching placement optimization is not considered. Further, Authors considered only a single stream between each storage server and edge node and hence neither the two-stage probabilistic scheduling nor bandwidth allocation were considered. Thus, the analysis and the problem formulation in this work is different from that in \cite{Abubakr_TON}.

\section{System Model}
\label{sec:SysMod}
\subsection{Target System: Virtualized Content Distribution Network}

Our work is motivated by the architecture of a production system with a vCDN.
The example services include video-on-demand (VoD), live linear streaming services 
(also referred to as over-the-top video streaming services), 
firmware over the air (FOTA) Android updates to mobile devices, etc.  
The main role of this CDN infrastructure is not only to provide users with lower response 
time and higher bandwidth,  but also to distribute the load (especially during peak time) 
across many edge locations. Consequently, the core backbone network will have 
reduced network load and better response time. 
So far, our vCDN system has been deployed to multiple locations in U.S. and will be expanded 
to more locations in the next couple of years. The origin server has original data and 
CDN sites have only part of those data. 
Each CDN site is composed of multiple cache servers. 
Each cache server is typically implemented as a VM backed by 
multiple directly attached solid state drives (SSDs) for higher throughput. 
The cache servers store video segments and a typical duration of each segment 
covers 11 seconds of playback time. 

When a client such as VoD/LiveTV app requests a certain content, 
it goes through multiple steps. First, it contacts \emph{CDN manager}, 
choose the best CDN service\footnote{Our production system utilizes multiple 
clouds, i.e., both 3rd party CDNs and in-house CDN infrastructure.}
to use and retrieve a fully qualified domain name (FQDN). 
Second, with the acquired FQDN, it gets a cache server's IP address from 
a content routing service (called iDNS). Then we use the IP address to connect 
to one of the cache servers\footnote{Our production CDN system architecture 
has two layers of cache servers, i.e., smaller number of parent cache servers 
and more number of child cache servers. In this paper, we abstract our 
infrastructure in a  per-site basis without loss of generality.}
The cache server will directly serve the incoming request if it has data in its local 
storage (cache-hit). If the requested content is not on the cache server (i.e., cache-miss), 
the cache server will fetch the content from the origin server and then serve 
the client\footnote{Some popular contents can be prefetched from the origin 
server a priori but we did not consider this aspect in this paper}.  

In the rest of this section, we present a generic mathematical model 
applicable to not only our vCDN system but also other video streaming systems 
that implement CDN-like caching-tier closer to the users.
%~\cite{
%youtube - https://www.youtube.com/watch?v=w5WVu624fY8, fb-haystack-osdi10}. 

\subsection{System Description}

\begin{figure}[t]
\centering\includegraphics[scale=0.25]{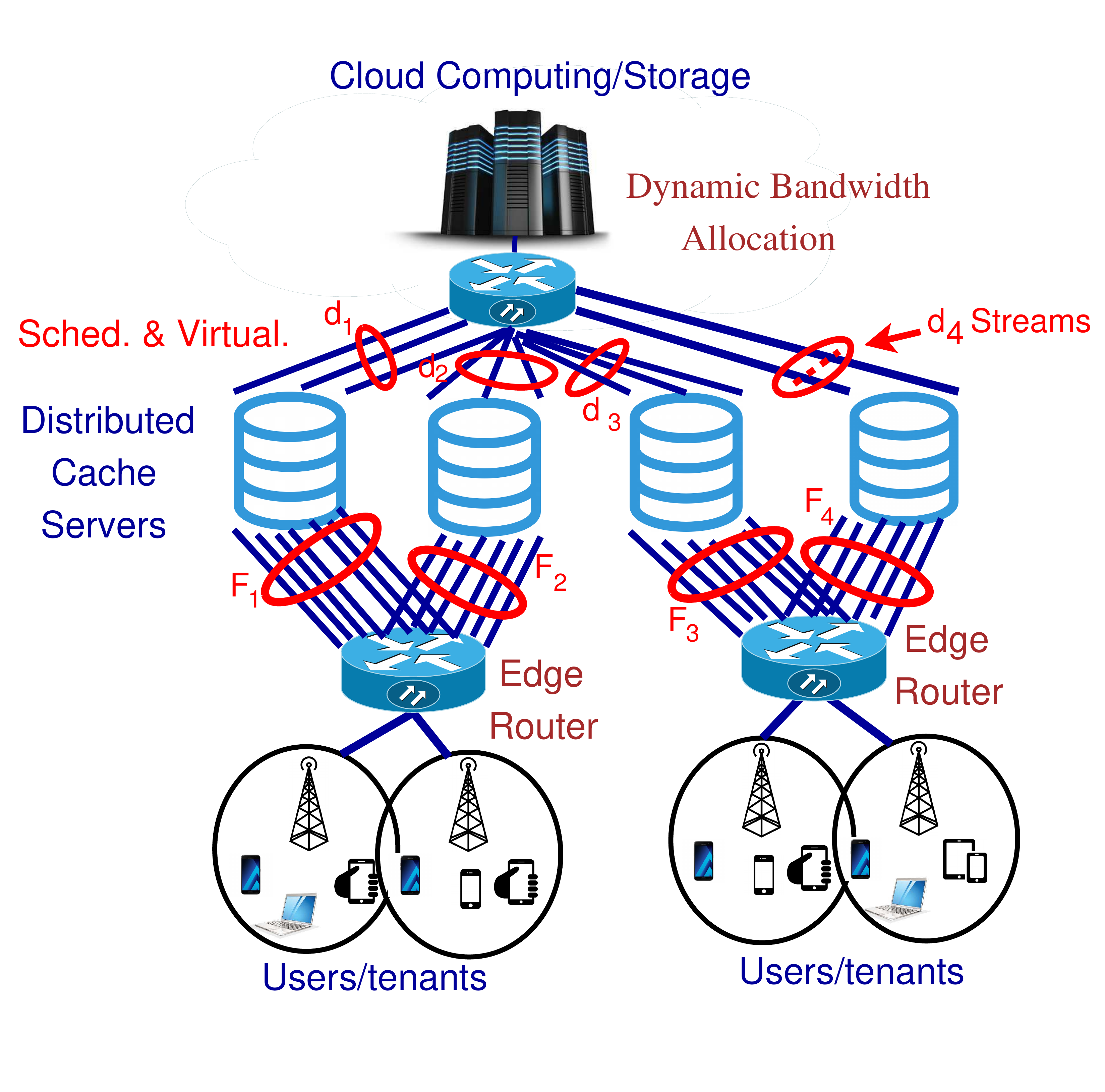}
%\vspace{-.4in}
\caption{An illustration of our system model for video content delivery, consisting
of a datacenter, four cache servers ($m=4$), and $2$ edge routers. $d_j$ and $F_j$ parallel connections are assumed between datacenter and cache server $j$, and datacenter and edge router, respectively. 
\label{fig:sysModel}}
%\vspace{-.20in}
\end{figure}

\begin{figure}[t]
	\centering\includegraphics[trim=0.035in 0.15in 0.25in 0.051in, clip,width=0.481\textwidth]{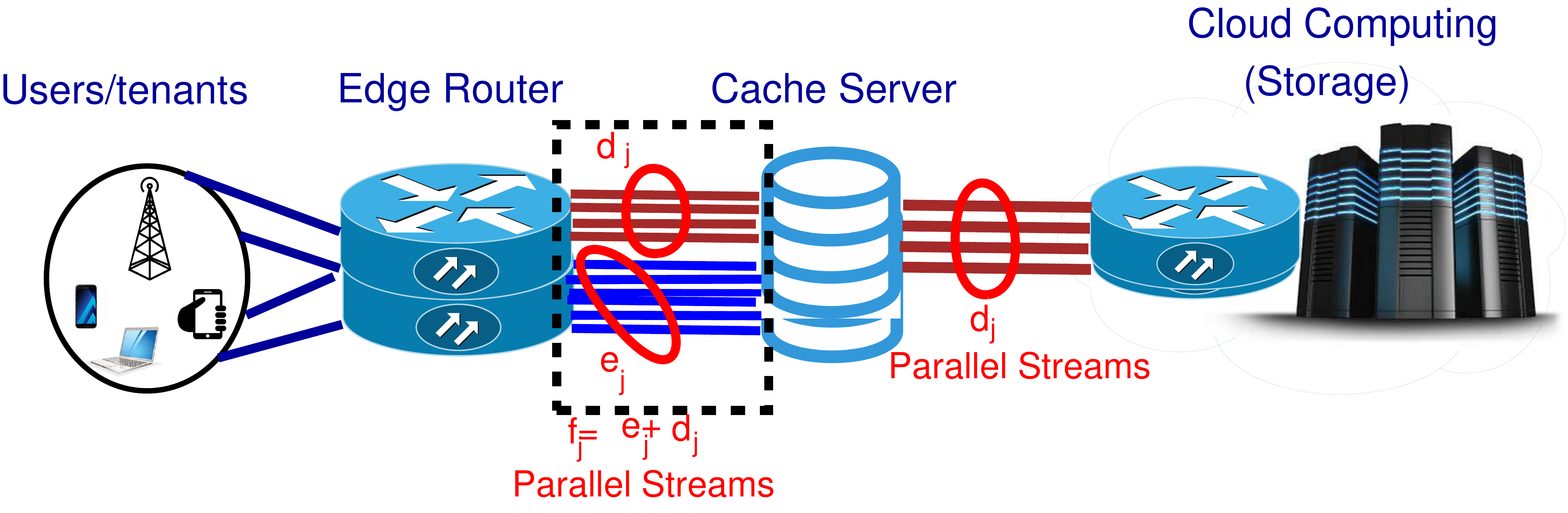}
%	\centering\includegraphics[scale=0.20]{sysModelPllstreams}
	%\vspace{-.4in}
	\caption{ A schematic illustrates the parallel streams setup between the different system model components.
		\label{PSsetup}}
	%\vspace{-.20in}
\end{figure}

We consider a content delivery network as shown in Fig. \ref{fig:sysModel},
consisting of a single datacenter that has an origin server, 
$m$ geographically-distributed cache servers denoted by $j=1,\ldots,m$, and edge routers. The compute
cache servers (also called storage nodes) are located close to the
edge of the network and thus provide lower access latency for end
users. 
%Since there is no sharing of cache servers, we only consider one edge router in this work. 
Further, the connection from the edge router to the users is not considered as a bottleneck. Thus, the edge router is considered as a combination of users and is the last hop for our analysis. We also note that the link from the edge router to the end users is not controlled by the service provider and thus cannot be considered for optimized resource allocation from the network. The service provider wishes to optimize the links it controls for efficient quality of experience to the end user.

A set of $r$ video files (denoted by $i=1,\ldots,r$) are stored
 in the datacenter, where video file $i$ is divided into $L_{i}$ equal-size segments each of length $\tau$ seconds. We assume that the first $L_{j,i}$ chunks of video $i$ are stored on cache server $j$. Even though we consider a fixed cache placement, we note that $L_{j,i}$ are optimization variables and can be updated when sufficient arrival rate change is detected.

We assume that the bandwidth between the data center and the cache server $j$ is split into $d_{j}$ parallel streams, where the streams are  denoted as $PS^{(d,j)}_{\beta_j}$ for $\beta_j = 1, \cdots, d_j$. Further,  the bandwidth between the cache server $j$ and the edge router is divided into $f_j$ parallel streams, denoted as $PS^{(f,j)}_{\zeta_j}$ for $\zeta_j = 1, \cdots, f_j$. Multiple parallel streams are assumed for video streaming since multiple video downloads can happen simultaneously. Since we care about stall duration, obtaining multiple videos simultaneously is helpful as the stall durations of multiple videos can be improved. We further assume that $f_j$ PSs are divided into two set of streams $d_j$ and $e_j$. This setup is captured in Figure \ref{PSsetup}. The first $d_j$ parallel streams are denoted as $PS^{(\overline{d},j)}_{\beta_j}$ for $\beta_j = 1, \cdots, d_j$, while the remaining $\nu_j$ streams are denoted as $PS^{(e,j)}_{\nu_j}$ for $\nu_j = 1, \cdots, e_j$. In order to consider the splits, $PS^{(d,j)}_{\beta_j}$ gets $\{w_{j,\beta_{j}}^{(d)},\,\beta_{j}=1,\ldots,d_{j}\}$ fraction of the bandwidth from the data center to the cache server $j$. Similarly, $PS^{(\overline{d},j)}_{\beta_j}$ gets $\{w_{j,\beta_{j}}^{(\overline{d})},\,\beta_{j}=1,\ldots,d_{j}\}$ fraction of bandwidth from cache server $j$ to the edge router and $PS^{(e,j)}_{\nu_j}$ gets $\{w_{j,\nu_{j}}^{(e)},\,\nu_{j}=1,\ldots,e_{j}\}$ fraction of bandwidth from cache server $j$ to the edge router. Thus, we have 
%\begin{equation}
%\vspace{-.12in}
%\sum_{\beta_j=1}^{d_j}w_{j,\beta_{j}}^{(d)} \le 1
%\end{equation}
%\begin{equation}
%\sum_{\beta_j=1}^{d_j}w_{j,\beta_{j}}^{(c)} +  \sum_{\nu_j=1}^{e_j}w_{j,\nu_{j}}^{(e)} \le 1,
%\end{equation}

\begin{equation}
%\vspace{-.12in}
\sum_{\beta_j=1}^{d_j}w_{j,\beta_{j}}^{(d)} \le 1,\,\,\,\,\,\,\,\,\,\,\,\,\,\,\,\,\,\,\,
\sum_{\beta_j=1}^{d_j}w_{j,\beta_{j}}^{(\overline{d})} +  \sum_{\nu_j=1}^{e_j}w_{j,\nu_{j}}^{(e)} \le 1,
\end{equation}
for all $\beta_j = 1, \cdots, d_j$ and $\nu_j = 1, \cdots, e_j$. We note that the sum of weights may be less than $1$ and some amount of the bandwidth may be wasted. While the optimal solution will satisfy this with equality since for better utilization, we do not need to explicitly enforce the equality constraint. 

We assume that the segment service times between the data center and the cache server $j$ is shifted-exponential with rate $\alpha_j^{(d)}$ and a shift of $\eta_{j}^{(d)}$ while that between the cache server $j$ and the edge router is also shifted-exponential with rate $\alpha_j^{(f_j)}$ and a shift of $\eta_{j}^{(f_j)}$. We validate this assumption in our  analysis via experiments on the testbed which will be explained in Section \ref{testbed}. Our measurement of real
service time distribution shows that the service time can be well approximated by a shifted exponential distribution. We also note that the rate of a parallel stream is proportional to the bandwidth split. Thus, the service time distribution of $PS^{(d,j)}_{\beta_j}$, $PS^{(\overline{d},j)}_{\beta_j}$, and $PS^{(e,j)}_{\nu_j}$, denoted as $\alpha_{j,\beta_{j}}^{(d)}$, $\alpha_{j,\beta_{j}}^{(\overline{d})}$, and $\alpha_{j,\nu_{j}}^{(\overline{d})}$, respectively, and are given as follows.

%\begin{eqnarray}
%\vspace{-.4in}
%\alpha_{j,\beta_{j}}^{(d)}&=& w_{j,\beta_{j}}^{(d)} \alpha_j^{(d)}\label{alphad}\\
%\alpha_{j,\beta_{j}}^{(c)}&=& w_{j,\beta_{j}}^{(c)} \alpha_j^{(c)}\label{alphac}\\
%\alpha_{j,\nu_{j}}^{(e)} &=& w_{j,\nu_{j}}^{(e)} \alpha_j^{(c)}\label{alphae},
%\vspace{-.4in}
%\end{eqnarray}

\begin{eqnarray}
%\vspace{-.4in}
\alpha_{j,\beta_{j}}^{(d)}= w_{j,\beta_{j}}^{(d)} \alpha_j^{(d)}\label{alphad};\,\,\,\,
\alpha_{j,\beta_{j}}^{(\overline{d})}= w_{j,\beta_{j}}^{(\overline{d})} \alpha_j^{(f_j)}\label{alphac};\,\,\,\
\alpha_{j,\nu_{j}}^{(e)} = w_{j,\nu_{j}}^{(e)} \alpha_j^{(f_j)}\label{alphae},
%\vspace{-.4in}
\end{eqnarray}
for all $\beta_j = 1, \cdots, d_j$ and $\nu_j = 1, \cdots, e_j$. We further define the moment generating functions of the service times of $PS^{(d,j)}_{\beta_j}$, $PS^{(\overline{d},j)}_{\beta_j}$, and $PS^{(e,j)}_{\nu_j}$ as $M^{(d)}_{j,\beta_j}$, $M^{(\overline{d})}_{j,\beta_j}$, and $M^{(\overline{d})}_{j,\nu_j}$, which are defined as follows. 
%\begin{eqnarray}
%\vspace{-.4in}
%M_{j,\beta_{j}}^{(d)} & =\frac{\alpha_{j,\beta_{j}}^{(d)}}{\alpha_{j,\beta_{j}}^{(d)}-t}\\
%M_{j,\beta_{j}}^{(c)} & =\frac{\alpha_{j,\beta_{j}}^{(c)}}{\alpha_{j,\beta_{j}}^{(c)}-t}\\
%M_{j,\nu_{j}}^{(e)} & =\frac{\alpha_{j,\nu_{j}}^{(e)}}{\alpha_{j,\nu_{j}}^{(e)}-t}\label{Me_jnuj}
%\vspace{-.4in}
%\end{eqnarray}

\begin{eqnarray}
\vspace{-.4in}
M_{j,\beta_{j}}^{(d)}  =\frac{\alpha_{j,\beta_{j}}^{(d)}e^{\eta_{j,\beta_j}^{(d)}t}}{\alpha_{j,\beta_{j}}^{(d)}-t};\,\,\,\,
\end{eqnarray}

\begin{eqnarray}
M_{j,\beta_{j}}^{(\overline{d})}  =\frac{\alpha_{j,\beta_{j}}^{(\overline{d})}e^{\eta_{j,\beta_j}^{(\overline{d})}t}}{\alpha_{j,\beta_{j}}^{(\overline{d})}-t};\,\,\,\,
\end{eqnarray}

\begin{eqnarray}
M_{j,\nu_{j}}^{(e)}  =\frac{\alpha_{j,\nu_{j}}^{(e)}e^{\eta_{j,\nu_j}^{(e)}t}}{\alpha_{j,\nu_{j}}^{(e)}-t}\label{Me_jnuj}
\vspace{-.4in}
\end{eqnarray}

\subsection{Queuing Model and Two-stage probabilistic scheduling}

If cache server $j$ is chosen for accessing video file $i$, the first $L_{j,i}$ chunks are obtained from one of the $e_j$ parallel streams $PS^{(e,j)}_{\nu_j}$. Further, the remaining $L_i - L_{j,i}$ chunks are obtained from the data canter where a choice of $\beta_j$ is made from $1, \cdots, d_j$ and the chunks are obtained from the stream $PS^{(d,j)}_{\beta_j}$  which after being served from this queue is enqueued in the queue for the stream $PS^{(\overline{d},j)}_{\beta_j}$.

%\begin{align}
%\alpha_{j,\nu_{j}}^{(e)} & =w_{j,\nu_{j}}^{(e)}\alpha_{j}^{(e)}\\
%\alpha_{j,\beta_{j}}^{(c)} & =w_{j,\beta_{j}}^{(c)}\alpha_{j}^{(c)}\\
%\alpha_{j,\zeta_{j}}^{(d)} & =w_{j,\zeta_{j}}^{(d)}\alpha^{(d)}
%\end{align}
%

We assume that the arrival of requests at the edge router for each
video $i$ form an independent Poisson process with a known rate $\lambda_{i}$. In order to serve the request for file $i$, we need to choose three things - (i) Selection of Cache server $j$, (ii) Selection of $\nu_j$ to determine one of $PS^{(e,j)}_{\nu_j}$ streams to deliver cached content, (iii) Selection of $\beta_j$ to determine one of $PS^{(d,j)}_{\beta_j}$ streams from the data-center which automatically selects the stream $PS^{(\overline{d},j)}_{\beta_j}$ from the cache server, to obtain the non-cached content from the datacenter. Thus, we will use a two-stage probabilistic scheduling to select the cache server and the parallel streams. We choose server $j$ with probability $\pi_{i,j}$ for file $i$ randomly. Further, having chosen the cache server, one of the $e_j$ streams is chosen with probability $p_{i,j,\nu_j}$. Similarly, one of the $d_j$ streams is chosen with probability $q_{i,j,\beta_j}$. We note that these probabilities only have to satisfy 

%\begin{eqnarray}
%\vspace{-.3in}
%\sum_{j=1}^m \pi_{i,j} &=& 1 \forall i\\
%\sum_{\nu_j = 1}^{e_j} p_{i,j,\nu_j} &=& 1  \forall i,j
%\vspace{-.3in}
%\end{eqnarray}
%
%\begin{eqnarray}
%\vspace{-.3in}
%\sum_{\beta_j = 1}^{d_j} q_{i,j,\beta_j} &=& 1  \forall i,j\\
%\pi_{i,j}, p_{i,j,\nu_j}, q_{i,j,\beta_j} &\ge& 0  \forall i,j, \beta_j, \nu_j
%\vspace{-.3in}
%\end{eqnarray}

%\begin{eqnarray}
%\vspace{-.3in}
%\sum_{j=1}^m \pi_{i,j} &=& 1 \forall i\\
%\sum_{\nu_j = 1}^{e_j} p_{i,j,\nu_j} &=& 1  \forall i,j\\
%\sum_{\beta_j = 1}^{d_j} q_{i,j,\beta_j} &=& 1  \forall i,j\\
%\pi_{i,j}, p_{i,j,\nu_j}, q_{i,j,\beta_j} &\ge& 0  \forall i,j, \beta_j, \nu_j
%\vspace{-.3in}
%\end{eqnarray}
\begin{eqnarray}
\sum_{j=1}^m \pi_{i,j} = 1 \forall i;\,\,\,\,
\sum_{\nu_j = 1}^{e_j} p_{i,j,\nu_j} = 1  \forall i,j;\,\,\,\,
\sum_{\beta_j = 1}^{d_j} q_{i,j,\beta_j} = 1  \forall i,j,\\
\pi_{i,j}, p_{i,j,\nu_j}, q_{i,j,\beta_j} \ge 0  \forall i,j, \beta_j, \nu_j
\end{eqnarray}

%In order to serve the incoming request at edge-router whose connected
%to server $j$, we need to assign one of the $(e_{j})$ PSs
%to bring the cached segments from the cache and we need also to select
%one of the $d_{j}$ PSs to bring the remaining segments from the DC,
%if any. Let $p_{j,\nu_{j}}$ be the probability of choosing the stream
%$\nu_{j}$-out-of-$(e_{j})$ PSs associated with server $j$
%and $q_{j,\beta_{j}}$ be the probability of choosing the stream $\beta_{j}$-out-of-$d_{j}$
%PSs that are dedicated to serve the non-cached segements from the DC, and let $f_{j,\zeta_{j}}$ be the probability of choosing the stream $\zeta_j$-out-of-$d_j$ PSs that are connected to the datacenter (as shown in Figure \ref{fig:sysPS}).
%These choices are feasible if and only if 
%
%\begin{align}
%\sum_{\nu_{j}=1}^{e_{j}}p_{j,\nu_{j}} & =1\,,\forall j\\
%\sum_{\beta_{j}=1}^{d_{j}}q_{j,\beta_{j}} & =1\,,\forall j\\
%\sum_{\zeta_{j}=1}^{d_{j}}f_{j,\zeta_{j}} & =1\,,\forall j
%\end{align}
%

%\subsection{Queuing Model }

Since sampling of Poisson process is Poisson, and superposition of independent Poisson processes is Poisson, we get the aggregate arrival rate at $PS^{(d,j)}_{\beta_j}$, $PS^{(\overline{d},j)}_{\beta_j}$, and $PS^{(e,j)}_{\nu_j}$, denoted as $\Lambda^{(d)}_{j,\beta_{j}}$,  $\Lambda^{(\overline{d})}_{j,\beta_{j}}$, and  $\Lambda^{(e)}_{j,\nu_{j}}$, respectively are given as follows. 

%\begin{eqnarray}
%\vspace{-.12in}
%\Lambda^{(d)}_{j,\beta_{j}} &=& \sum_{i=1}^r \lambda_{i} \pi_{i,j}q_{i,j,\beta_{j}}\label{lambdad_betaj}\\
%\Lambda^{(c)}_{j,\beta_{j}} &=& \Lambda^{d}_{j,\beta_{j}} \label{lambdac_betaj}\\
%\Lambda^{(e)}_{j,\nu_{j}} &=& \sum_{i=1}^r \lambda_{i} \pi_{i,j}p_{i,j,\nu_{j}}\label{lambdae_nuj}
%\vspace{-.12in}
%\end{eqnarray}

\begin{eqnarray}
\Lambda^{(d)}_{j,\beta_{j}} = \sum_{i=1}^r \lambda_{i} \pi_{i,j}q_{i,j,\beta_{j}}\label{lambdad_betaj};\,\,\,\,
\Lambda^{(\overline{d})}_{j,\beta_{j}} = \Lambda^{(d)}_{j,\beta_{j}} \label{lambdac_betaj},\\
\Lambda^{(e)}_{j,\nu_{j}} = \sum_{i=1}^r \lambda_{i} \pi_{i,j}p_{i,j,\nu_{j}}\label{lambdae_nuj}
\end{eqnarray}
%Then, the arrival of file requests at PS $\nu_{j}$ at node $j$ forms
%a Poisson Process with rate $\Lambda_{j,\nu_{j}}=\sum_{i}\lambda_{i} \pi_{i,j}p_{j,\nu_{j}}$
%which is the superposition of $r(e_{j})$ Poisson processes
%each with rate $\lambda_{i} \pi_{i,j} p_{j,\nu_{j}}.$ 
%

%If video $i$ segments are not all stored in the cache, we
%request the non-cached segments from the DC. Let $L_{c,i}$ denotes
%the number of cached segments of video file $i$, i.e., $L_{c,i}\leq L_{i}$.
%Hence, the remaining $(L_{i}-L_{c,i})$ segments need to be brought
%(served) from the DC. In order to schedule the request for the $(L_{i}-L_{c,i})$
%segments of video file $i$ from a PS $\beta_{j}$, we need first
%to characterize the arrival rate at the PS $\zeta_{j}$. Since the
%arrival process and the service time of the cache server are memoryless
%(i.e., M/M/1), it is easy to show that the arrival of file requests
%at PS $\zeta_{j}$ at node $j$ forms a Poisson Process with rate $\Lambda_{j,\zeta_{j}}=\sum_{i}\lambda_{i}\pi_{i,j}f_{j,\zeta_{j}}$
%which is the superposition of $rd_{j}$ Poisson processes each with
%rate $\lambda_{i}\pi_{i,j}f_{j,\beta_{j}}$. Here, we also assume that the service time of a segment
%at node $j$ at PS $\zeta_{j}$, i.e., $X_{j,\zeta_{j}}^{(d)}$, follows
%also exponential distribution with rate $\alpha_{j,\zeta_{j}}^{(d)}$
%and $M_{j,\zeta_{j}}^{(d)}(t)$ can be expressed as 
%\begin{equation}
%M_{j,\zeta_{j}}^{(d)}(t)=\frac{\alpha_{j,\beta_{j}}^{(d)}}{\alpha_{j,\zeta_{j}}^{(d)}-t}
%\end{equation}

\begin{lemma}
	\label{PoisArr2ndQ}
 When the service time distribution of datacenter server (first queue) is given by shifted exponential distribution, the arrivals at the cache servers (second queue) are Poisson.
\end{lemma}

\begin{proof}
	The proof is provided in  Appendix \ref{2ndQArr}.
\end{proof}

We also assume that there is a start-up delay of $d_{s}$ (in seconds)
for the video which is the duration in which the content can be buffered
but not played. This work will characterize the stall duration tail
probability using the two-stage probabilistic scheduling and allocation of bandwidth
weights. We note that the arrival rates are given in terms of the video files,
and the service rate above is provided in terms of segment at each
server. The analysis would require detailed consideration of the different segments in a video.

\section{Download and Play Times of the Segment}

\label{sec:DownLoadPlay}

In order to characterize the stall duration tail probability, we need
to find the download time and the play time of different video segments, for any server $j$ and streams with the choice of $\beta_j$ and $\nu_j$. The optimization over these decision variables will be considered in the next Section. 

%In the following, we will first estimate the download time of segments from the cache servers and from datacenter, respectively. Then, based on the download time of each segment, we derive the play time of different segments and then give the stall duration expression for each segment request.

\subsection{Download Times of first $L_{j,i}$ Segments }

We consider a queuing model, where $W_{j,\nu_{j}}^{(e)}$ denotes the random  waiting time of all video files in the queue of $PS^{(e,j)}_{\nu_j}$ before file $i$ request is served, 
and $Y_{i,j,\nu_{j}}^{(e,g)}$ be the (random service time of a coded chunk
$g$ for file $i$ from server $j$ and queue $\nu_{j}$. Then, for $g\le L_{j,i}$, the random download time of the first $L_{j,i}$ segments $g\in\{1,\ldots,L_{j,i}\}$ if file $i$ from stream $PS^{(e,j)}_{\nu_j}$ is given as 
 \begin{equation}
D_{i,j,\beta_j,\nu_{j}}^{(g)}=W_{j,\nu_{j}}^{(e)}+\sum_{v=1}^{g}Y_{i,j,\nu_{j}}^{(e,v)}
\end{equation}
Since video file $i$ consists of $L_{j,i}$ segments stored at cache
server $j$, the total service time for video file $i$ request at queue
$PS^{(e,j)}_{\nu_j}$, denoted by $ST_{i,j,\nu_{j}}$, is given as 
\begin{equation}
ST_{i,j,\nu_{j}}=\sum_{v=1}^{L_{j,i}}Y_{i,j,\nu_j}^{(e,v)}
\end{equation}
Hence, the service time of the video files at the parallel stream $PS^{(e,j)}_{\nu_j}$ is given as  
\begin{equation}
R_{j,\nu_{j}}^{(e)}=\begin{cases}
ST_{i,j,\nu_{j}}^{(e,L_{j,i})} \ \   \text { with prob. } \frac{\lambda_{i}\pi_{i,j}p_{i,j,\nu_j}}{\Lambda_{j,\nu_{j}}^{(e)}}& \forall i\end{cases}
\end{equation}

We can show that the moment generating function of the service time for all video files from parallel stream $PS^{(e,j)}_{\nu_j}$   is given by 
\begin{equation}
B_{j,\nu_{j}}^{(e)}(t)={\mathbb E}[e^{tR_{j,\nu_{j}}^{(e)}}] =\sum_{i=1}^{r}\frac{\lambda_{i}\pi_{i,j}p_{i,j,\nu_j}}{\Lambda_{j,\nu_{j}}^{(e)}}\left(\frac{\alpha_{j,\nu_{j}}^{(e)}e^{\eta_{j,\nu_j}^{(e)}t}}{\alpha_{j,\nu_{j}}^{(e)}-t}\right)^{L_{j,i}}\label{Be}
\end{equation}
Further, based on our 2-stage scheduling policy, the load intensity at $PS^{(e,j)}_{\nu_j}$  is as follows 
\begin{equation}
\rho_{j,\nu_{j}}^{(e)}  = \Lambda_{j,\nu_{j}}^{(e)} B_{j,\nu_{j}}^{(e)'}(0)=  \sum_{i=1}^{r}\lambda_{i}\pi_{i,j}p_{i,j,\nu_j}L_{j,i}\left( \eta_{j,\nu_j}^{(e)}+\frac{1}{\alpha_{j,\nu_{j}}^{(e)}}\right) \label{rhoe}
\end{equation}

Since the arrival is Poisson and the service time is shifted-exponentially distributed, the moment generating function (MGF) of the waiting time at queue $PS^{(e,j)}_{\nu_j}$ can be calculated usingthe  Pollaczek-Khinchine formula, i.e.,
%(since M/M/1 is a special case of M/G/1) 
%and is given by
\begin{align}
\mathbb{E}\left[e^{tW_{j,\nu_{j}}^{(e)}}\right] & =\frac{(1-\rho_{j,\nu_{j}}^{(e)})tB_{j,\nu_{j}}^{(e)}(t)}{t-\Lambda_{j,\nu_{j}}^{(e)}(B_{j,\nu_{j}}^{(e)}(t)-1)}
\end{align}
From the MGF of $W_{j,\nu_{j}}^{(e)}$ and the service
time, the MGF of the download time of segment $g$ from the queue $PS^{(e,j)}_{\nu_j}$ for file $i$ is then
\begin{equation}
\mathbb{E}\left[e^{tD_{i,j,\beta_j,\nu_{j}}^{(g)}}\right]=\frac{(1-\rho_{j,\nu_{j}}^{(e)})tB_{j,\nu_{j}}^{(e)}(t)}{t-\Lambda_{j,\nu_{j}}^{(e)}(B_{j,\nu_{j}}^{(e)}(t)-1)}\left(\frac{\alpha_{j,\nu_{j}}^{(e)}e^{\eta_{j,\nu_j}^{(e)}t}}{\alpha_{j,\nu_{j}}^{(e)}-t}\right)^{g}
\label{D_nu_j}.
\end{equation}
We note that the above is defined only when MGFs exist, i.e., 
\begin{eqnarray}
t&<& \alpha_{j,\nu_{j}}^{(e)}\label{alpha_j_nuj}\\
0 &<& t-\Lambda_{j,\nu_{j}}^{(e)}(B_{j,\nu_{j}}^{(e)}(t)-1)
\label{mgf_gz_e_queue}
\end{eqnarray}

%Finally, the overall download time of all the $L_{j,i}$ segments of video $i$ at the client,
%$D_{i}^{(g)}$ , is given by
%
%
%\begin{equation}
%D_{i}^{(g)}=\sum_{j=1}^{m}\pi_{i,j}\sum_{\nu_{j}=1}^{e_{j}}p_{j,\nu_{j}}D_{i,j,\nu_{j}}^{(g)},\,\,\,\,\,g\leq L_{j,i}
%\end{equation}

\subsection{Download Times of last $(L_i - L_{j,i})$ Segments }

Since the later video segments $(L_i - L_{j,i})$ are downloaded from the data center, we need to schedule them to the $\beta_j$ streams using the proposed probabilistic scheduling policy.
%and thus the choice is among the 
We first determine the time it takes for chunk $g$ to depart the first queue (i.e., $\beta_j$ queue at datacenter). For that, we define the time of chunk $g$ to depart the first queue as
\begin{equation}
E_{i,j,\beta_{j}}^{(g)}=W_{j,\beta_{j}}^{(d)}+\sum_{v=L_{j,i}+1}^{L_i}Y^{(d,v)}_{j,\beta_{j}},
\end{equation}
where $W_{j,\beta_{j}}^{(d)}$ is the waiting time from $PS^{(d,j)}_{\beta_j}$ for the earlier video segments, and $Y^{(d,v)}_{j,\beta_{j}}$ is the service time for obtaining segment $v$ from the queue of $PS^{(d,j)}_{\beta_j}$. Using similar analysis for that of deriving the MGF of download time of chunk $g$  as in the last section, we obtain
 \begin{equation}
 \mathbb{E}\left[e^{tE_{i,j,\beta_{j}}^{(g)}}\right]=\frac{(1-\rho_{j,\beta_{j}}^{(d)})tB_{j,\beta_{j}}^{(d)}(t)}{t-\Lambda^{(d)}_{j,\beta_{j}}(B_{j,\beta_{j}}^{(d)}(t)-1)}\left(\frac{\alpha_{j,\beta_{j}}^{(d)}e^{\eta_{j,\beta_j}^{(d)}t}}{\alpha_{j,\beta_{j}}^{(d)}-t}\right)^{g-L_{j,i}-1},
 \end{equation}
where the load intensity at queue $\beta_j$ at datacenter, $\rho_{j,\beta_{j}}^{(d)}$
 \begin{align}
\rho_{j,\beta_{j}}^{(d)} & =\sum_{i=1}^{r}\lambda_{i}\pi_{i,j}q_{i,j,\beta_j}\left( L_i-L_{j,i}\right)\left( \eta_{j,\beta_j}^{(d)}+ \frac{1}{\alpha_{j,\beta_{j}}^{(d)}}\right) \label{rhod}\\
B_{j,\beta_{j}}^{(d)}(t) & =\sum_{i=1}^{r}\frac{\lambda_{i}\pi_{i,j}q_{i,j,\beta_j}}{\Lambda_{j,\beta_{j}}^{(d)}}\left(\frac{\alpha_{j,\beta_{j}}^{(d)}e^{\eta_{j,\beta_j}^{(d)}t}}{\alpha_{j,\beta_{j}}^{(d)}-t}\right)^{L_{i}-L_{j,i}}\label{Bd}
\end{align}
% $\rho_{j,\beta_{j}}^{(d)} = ??$ and $B_{j,\beta_{j}}^{(d)}(t) = ??$.
 
%{\color{blue} Add a couple of sentences here to explain why it can be derived recursively.} 

To find the download time of video segments from the second queue (at cache server $j$), we notice that the download time for segment $g$ includes the waiting to download all previous segments and the idle time if the segment $g$ is not yet downloaded from the first queue ($PS^{(d,j)}_{\beta_j}$), as well as the service time of the segment from $PS^{(\overline{d},j)}_{\beta_j}$. Then, the download time of the video segments from the second queue (i.e.,$PS^{(\overline{d},j)}_{\beta_j}$) can be derived by a set of recursive equations, with download time of the first (initial) segment $(L_{j,i}+1)$ as 
 \begin{equation}
D_{i,j,\beta_{j},\nu_j}^{(L_{j,i}+1)} = \max(W_{j,\beta_{j}}^{(\overline{d})},  E_{j,\beta_{j}}^{(L_{j,i}+1)}) + Y^{(\overline{d},L_{j,i}+1)}_{j,\beta_{j}},
 \end{equation}
where $W_{j,\beta_{j}}^{(\overline{d})}$ is the waiting time from queue $PS^{(\overline{d},j)}_{\beta_j}$ for the previous video segments, and $Y^{(\overline{d},v)}_{j,\beta_{j}}$ is the required service time for obtaining segment $v$ from the queue of $PS^{(\overline{d},j)}_{\beta_j}$. The download time of the following segments ($g>L_{j,i}+1$) is given by the following recursive equation 
\begin{equation}
D_{i,j,\beta_{j},\nu_j}^{(g)} = \max(D_{i,j,\beta_{j},\nu_j}^{(g-1)},  E_{i,j,\beta_{j}}^{(g)}) + Y^{(\overline{d},g)}_{j,\beta_{j}}.
\end{equation}
%This is because the download time for segment $g$ has to 

With the above recursive equations from $y=L_{j,i}$ to $y=g$, we can obtain that 
\begin{equation}
D_{i,j,\beta_{j},\nu_j}^{(g)} = \max_{y=L_{j,i}}^g U_{i,j,\beta_j,g,y},
\end{equation}
where
\begin{equation}
U_{i,j,\beta_j,g,L_{j,i}} = W_{j,\beta_{j}}^{(\overline{d})} + \sum_{h= L_{j,i}+1}^gY^{(\overline{d},h)}_{j,\beta_{j}}
\end{equation}
Similarly, for $y>L_{j,i}$, we have
\begin{equation}
U_{i,j,\beta_j,g,y} = E_{i,j,\beta_{j}}^{(y)} + \sum_{h= y}^gY^{(\overline{d},h)}_{j,\beta_{j}}.
\end{equation}

It is easy to see that the moment generating function of $U_{i,j,\beta_j,g,y}$ for $y=L_{j,i}$ is given by
\begin{equation}
\mathbb{E}[e^{tU_{i,j,\beta_{j},g,L_{j,i}}}]=  \!\!\frac{(1-\rho_{j,\beta_{j}}^{(\overline{d})})tB_{j,\beta_{j}}^{(\overline{d})}(t)}{t-\Lambda_{j,\beta_{j}}^{(c)}(B_{j,\beta_{j}}^{(\overline{d})}(t)-1)}\!\!\left(\!\!\!\frac{\alpha_{j,\beta_{j}}^{(\overline{d})}e^{\eta_{j,\beta_{j}}^{(\overline{d})}t}}{\alpha_{j,\beta_{j}}^{(\overline{d})}-t}\!\!\right)^{g-L_{j,i}},
\label{U_MGF_Lji}
\end{equation}
where the load intensity at queue $\beta_j$ at cache server $j$, $\rho_{j,\beta_{j}}^{(\overline{d})}$ is given by 
\begin{align}
\rho_{j,\beta_{j}}^{(\overline{d})} & =\sum_{i=1}^{r}\lambda_{i}\pi_{i,j}q_{i,j,\beta_j}\left( L_i-L_{j,i}\right)\left( \eta_{j,\beta_j}^{(\overline{d})}+ \frac{1}{\alpha_{j,\beta_{j}}^{(\overline{d})}}\right)\label{rhoc}\\
B_{j,\beta_{j}}^{(\overline{d})}(t) & =\sum_{i=1}^{r}\frac{\lambda_{i}\pi_{i,j}q_{i,j,\beta_j}}{\Lambda_{j,\beta_{j}}^{(c)}}\left(\frac{\alpha_{j,\beta_{j}}^{(\overline{d})}e^{\eta_{j,\beta_j}^{(\overline{d})}t}}{\alpha_{j,\beta_{j}}^{(\overline{d})}-t}\right)^{L_{i}-L_{j,i}}\label{Bc}
\end{align}
Similarly, the moment generating function of $U_{i,j,\beta_j,g,y}$ for $y>L_{j,i}$ is given as 
\begin{align}
\mathbb{E}[e^{tU_{i,j,\beta_{j},g,y}}] & =\overline{W}_{j,\beta_{j}}^{(d)}\times \nonumber\\
& \left(\frac{\alpha_{j,\beta_{j}}^{(d)}e^{\eta_{j,\beta_{j}}^{(d)}t}}{\alpha_{j,\beta_{j}}^{(d)}-t}\right)^{y-L_{j,i}-1}\left(\frac{\alpha_{j,\beta_{j}}^{(\overline{d})}e^{\eta_{j,\beta_{j}}^{(\overline{d})}t}}{\alpha_{j,\beta_{j}}^{(\overline{d})}-t}\right)^{g-y+1}
\label{U_MGF_all}
\end{align}
%{\mathbb E}[e^{tU_{i,j,\beta_j,g,y}}] = \frac{(1-\rho_{j,\beta_{j}}^{(d)})tB_{j,\beta_{j}}^{(d)}(t)}{t-\Lambda^{(d)}_{j,\beta_{j}}(B_{j,\beta_{j}}^{(d)}(t)-1)}\left(\frac{\alpha_{j,\beta_{j}}^{(d)}}{\alpha_{j,\beta_{j}}^{(d)}-t}\right)^{y-L_{j,i}-1}\left(\frac{\alpha_{j,\beta_{j}}^{(c)}}{\alpha_{j,\beta_{j}}^{(c)}-t}\right)^{g-y+1}.
where $\overline{W}_{j,\beta_{j}}^{(d)}$
$\overline{W}_{j,\beta_{j}}^{(d)}=\frac{(1-\rho_{j,\beta_{j}}^{(d)})tB_{j,\beta_{j}}^{(d)}(t)}{t-\Lambda_{j,\beta_{j}}^{(d)}(B_{j,\beta_{j}}^{(d)}(t)-1)}$.
We further note that these moment generating functions are only defined when the MGF functions exist, i.e., 
%\begin{eqnarray}
%t&<& \alpha_{j,\beta_{j}}^{(d)}\label{alpha_betaj_d}\\
%0 &<& t-\Lambda_{j,\beta_{j}}^{(d)}(B_{j,\beta_{j}}^{(d)}(t)-1)\\
%t&<& \alpha_{j,\beta_{j}}^{(\overline{d})}\\
%0 &<& t-\Lambda_{j,\beta_{j}}^{(\overline{d})}(B_{j,\beta_{j}}^{(\overline{d})}(t)-1)
%\label{mgf_beta_j_c}
%\end{eqnarray}
\begin{eqnarray}
t<\alpha_{j,\beta_{j}}^{(d)}\label{alpha_betaj_d};\,\,\,\,
0 < t-\Lambda_{j,\beta_{j}}^{(d)}(B_{j,\beta_{j}}^{(d)}(t)-1)\\
t< \alpha_{j,\beta_{j}}^{(\overline{d})};\,\,\,\,
0 < t-\Lambda_{j,\beta_{j}}^{(\overline{d})}(B_{j,\beta_{j}}^{(\overline{d})}(t)-1)
\label{mgf_beta_j_c}
\end{eqnarray}

\subsection{Play Times of different Segments}

Next, we find the play time of different video segments. Recall that $D_{i,j,\beta_{j},\nu_{j}}^{(g)}$ is the download time of segment
$g$ from $\nu_{j}$ and $\beta_{j}$ queues at client $i$. We further define $T_{i,j,\beta_j,\nu_j}^{\left(g\right)}$ as the time that segment $g$ begins to play at the client $i$, given that it is downloaded from $\beta_j$ and $\nu_j$ queues. This start-up delay of the video is denoted by $d_{s}$. Then, the first segment is ready for play at the maximum
of the startup delay and the time that the first segment can be downloaded. This means
\begin{equation}
T_{i,j,\beta_j,\nu_j}^{(1)}=\mbox{max }\left(d_{s},\,D_{i,j,\beta_j,\nu_j}^{(1)}\right).
\end{equation}

For $1<g\le L_{i}$, the play time of segment $g$ of video file $i$ is
given by the maximum of (i) the time to download the segment
and (ii) the time to play all previous segment plus the time
to play segment $g$ (i.e., $\tau$ seconds). Thus, the play time of segment
$g$ of video file $i$, when requested from server $j$ and from $\nu_j$ and $\beta_j$ queues, can be expressed as 
%$T_{i,j,\beta_j,\nu_j}^{(q)}$
\begin{equation}
T_{i,j,\beta_j,\nu_j}^{(q)}=\mbox{max }\left(T_{i,j,\beta_j,\nu_j}^{(q-1)}+\tau,\,D_{i,j,\beta_j,\nu_j}^{(q)}\right).\label{eq:T_ij_general}
\end{equation}
%Equation (\ref{eq:T_ij_general}) 
This results in a set of recursive equations, which further yield by
\begin{align}
T_{i,j,\beta_{j},\nu_{j}}^{(L_{i})} & =\mbox{max }\left(T_{i,j,\beta_{j},\nu_{j}}^{(L_{i}-1)}+\tau,\,D_{i,j,\beta_{j},\nu_{j}}^{(L_{i})}\right)\nonumber \\
 & =\mbox{max }\left(T_{i,j,\beta_{j},\nu_{j}}^{(L_{i}-2)}+2\tau,\,D_{i,j,\beta_{j},\nu_{j}}^{(L_{i}-1)}+\tau,\,D_{i,j,\beta_{j},\nu_{j}}^{(L_{i})}\right)\nonumber \\
 & =\!\mbox{max}\left(d_{s}+(L_{i}-1)\tau,\,\max_{z=2}^{L_{i}+1}D_{i,j,\beta_{j},\nu_{j}}^{(z-1)}+(L_{i}-z+1)\tau\right)\nonumber \\
 & =\overset{L_{i}+1}{\underset{z=1}{\text{max}}}\,\mathcal{F}_{i,j,\beta_{j},\nu_{j},z}
 \label{T_ijbetanuF}
\end{align}
where $\mathcal{F}_{i,j,\beta_{j},\nu_{j},z}$ is expressed as
\begin{equation}
\mathcal{F}_{i,j,\beta_{j},\nu_{j},z}=\begin{cases}
d_{s}+(L_{i}-1)\tau & \,,z=1\\
D_{i,j,\beta_{j},\nu_{j}}^{(z-1)}+(L_{i}-z+1)\tau & 1<z\leq L_{i}
\end{cases} \label{F_1}
\end{equation}

We now get the MGFs of the 
$\mathcal{F}_{i,j,\nu_{j},\beta_{j},z}$ to use in characterizing the play time of the different segments. Towards this goal, we plug Equation (\ref{F_1}) into $\mathbb{E}\left[e^{t\mathcal{F}_{i,j,\nu_{j},\beta_{j},z}}\right]$ and obtain
%can be calculated as follows
\begin{equation}
\mathbb{E}\left[e^{t\mathcal{F}_{i,j,\nu_{j},\beta_{j},z}}\right]=\begin{cases}
e^{(d_{s}+(L_{i}-1)\tau)t} & \,,z=1\\
e^{(L_{i}-z+1)\tau t}\mathbb{E}\left[e^{tD_{i,j,\beta_{j},\nu_{j}}^{(z-1)}}\right] & 1<z\leq (L_{i}+1)
\end{cases}
\label{MGF_F}
\end{equation}
where $\mathbb{E}\left[e^{tD_{i,j,\beta_{j},\nu_{j}}^{(z-1)}}\right]$ can be calculated using equation (\ref{D_nu_j}) when $1<z\leq (L_{j,i}+1)$ and using equation (\ref{U_MGF_all}) when $z>L_{j,i}+1$ .
%
%where $Z_{D_{i,j,\nu_{j}}^{(z-1)}}$ is given by (\ref{D_nu_j}) and $Z_{D_{i,j,\beta_{j},\nu_j}^{(z-1)}}$ is given as follows.
%
%\begin{align}
%Z_{D_{i,j,\beta_{j},\nu_{j}}^{(z-1)}} & =\mathbb{E}\left[e^{tD_{i,j,\beta_{j},\nu_{j}}^{(z-1)}}\right]=\mathbb{E}\left[e^{t(\underset{y}{\text{\ensuremath{\max}}}U_{i,j,\beta_{j},g,y})}\right]\\
% & =\mathbb{E}\left[\underset{y}{\text{\ensuremath{\max}}}\,\,e^{tU_{i,j,\beta_{j},g,y}}\right]\\
% & \leq\sum_{y}e^{tU_{i,j,\beta_{j},g,y}}\\
% & =\left[e^{tU_{i,j,\beta_{j},g,L_{j,i}}}\right]+\sum_{w=L_{j,i}+1}^{g}\frac{(1-\rho_{j,\beta_{j}}^{(d)})tB_{j,\beta_{j}}^{(d)}(t)}{t-\Lambda_{j,\beta_{j}}^{(d)}(B_{j,\beta_{j}}^{(d)}(t)-1)}\left(\frac{\alpha_{j,\beta_{j}}^{(d)}}{\alpha_{j,\beta_{j}}^{(d)}-t}\right)^{w-L_{j,i}-1}\left(\frac{\alpha_{j,\beta_{j}}^{(c)}}{\alpha_{j,\beta_{j}}^{(c)}-t}\right)^{g-w+1}.
% \label{MGF_Z_D_last}
%\end{align}
%where $\left[e^{tU_{i,j,\beta_{j},g,L_{j,i}}}\right]$ is given in (\ref{U_MGF_Lji}).

The last segment should be completed by time $d_{s}+L_{i}\tau$ (which is the time at which the playing of the $L_i-1$ segment finishes). Thus, 
the difference between the play time of the last segment $T_{i,j,\beta_{j},\nu_{j}}^{(L_{i})}$ and
$d_{s}+\left(L_{i}-1\right)\tau$ gives the stall duration. We note that
the stalls may occur before any segment and hence this difference will give
the sum of durations of all the stall periods before any segment. Thus, the stall duration for the request of file $i$ from $\beta_{j}$
queue, $\nu_{j}$ queue and server $j$, i.e., $\Gamma^{(i,j,\beta_{j},\nu_{j})}$ is given as
\begin{equation}
\Gamma^{(i,j,\beta_{j},\nu_{j})}=T_{i,j,\beta_{j},\nu_{j}}^{(L_{i})}-d_{s}-\left(L_{i}-1\right)\tau
\label{stallTime}
\end{equation}

In the next section, we will use this stall time to determine a bound on the SDTP, i.e., the stall duration tail probability.
\section{Stall Duration Tail Probability} \label{sec:StallTailProb}

%{\bf To be revised. }

The stall duration tail probability of a video file $i$ is defined as the
probability that the stall duration tail $\Gamma^{(i,j,\beta_j,\nu_j)}$ is greater
than a pre-defined threshold $\sigma$. Since exact evaluation of stall duration  
is hard
\cite{Joshi:13, Abubakr_TON}
, we cannot evaluate $\text{\text{Pr}}\left(\Gamma^{(i,j,\beta_j,\nu_j)}\geq\sigma\right)$
in closed-form. In this section, we derive a tight upper bound on the SDTP through the two-stage Probabilistic Scheduling as follows. First, the SDTP for the request of file $i$ from $\beta_{j}$
queue, $\nu_{j}$ queue and server $j$, can be written as
\begin{equation}
\text{\text{Pr}}\left(\Gamma^{(i,j,\beta_{j},\nu_{j})}\geq \sigma\right)\overset{(a)}{=}\text{\text{Pr}}\left(T_{i,j,\beta_{j},\nu_{j}}^{(L_{i})}\geq\sigma+d_{s}+\left(L_{i}-1\right)\tau\right)\label{eq:Gamma_ij1}
\end{equation}
where (a) follows from (\ref{stallTime}). We next define an auxiliary variable $\overline{\sigma}=\text{\ensuremath{\sigma}}+d_{s}+\left(L_{i}-1\right)\tau$.
Then, we have 
\begin{align}
\text{\text{Pr}}\left(T_{i,j,\beta_{j},\nu_{j}}^{(L_{i})}\geq\overline{\sigma}\right) & \overset{(b)}{=}\text{\text{Pr}}\left(\overset{L_i+1}{\underset{z=1}{\max}\,\,}\text{\ensuremath{\mathcal{F}_{i,j,\beta_{j},\nu_{j},z}}}\geq\overline{\sigma}\right)\nonumber \\
 & =\mathbb{E}\left[\text{\ensuremath{\boldsymbol{1}_{\left(\overset{L_i+1}{\underset{z=1}{\max}}\,\text{\ensuremath{\mathcal{F}_{i,j,\beta_{j},\nu_{j},z}}}\geq\overline{\sigma}\right)}}}\right]\nonumber \\
 & \overset{(c)}{\leq}\mathbb{E}\sum_{z=1}^{L_i+1}\,\boldsymbol{1}_{\left(\text{\ensuremath{\mathcal{F}_{i,j,\beta_{j},\nu_{j},z}}}\geq\overline{\sigma}\right)}\nonumber \\
 & \overset{}{=}\sum_{z=1}^{L_{i}+1}\mathbb{P}\left(\text{\ensuremath{\mathcal{F}_{i,j,\beta_{j},\nu_{j},z}}}\geq\overline{\sigma}\right)
\label{eq:P_T_ij}
\end{align}
where (b) follows from (\ref{T_ijbetanuF}) and (c) follows by replacing
the $\underset{z}{\text{max}}$ by $\sum_{z}$. Moreover, by using Markov Lemma, we get 
\begin{equation}
\mathbb{P}\left(\text{\ensuremath{\mathcal{F}_{i,j,\beta_{j},\nu_{j},z}}}\geq\overline{\sigma}\right)\leq\frac{\mathbb{E}\left[e^{t_{i}\text{\ensuremath{\mathcal{F}_{i,j,\beta_{j},\nu_{j},z}}}}\right]}{e^{t_{i}\overline{\sigma}}}
\label{eq:markovLamma}
\end{equation}

%Then,

%\begin{align}
 %& %\mathbb{P}\left(\text{\ensuremath{\mathcal{F}_{i,j,\beta_{j},\nu_{j},z}}}\geq\overline{\sigma}\right)\leq\frac{\mathbb{E}\left[e^{t_{i}\text{\ensuremath{\mathcal{F}_{i,j,\beta_{j},\nu_{j},z}}}}\right]}{e^{t_{i}\overline{\sigma}}}\\

% & =e^{-t_{i}\sigma}+e^{(\widetilde{\sigma}-\overline{\sigma})t_{i}}\mathbb{E}\left[e^{t_iD_{i,j,\beta_{j},\nu_{j}}^{(z-1)}}\right]
%\label{eq:P_ijvbz}
%\end{align}
%where (e) follows from (\ref{MGF_F}) and  $\widetilde{\sigma}=(L_{i}-z+1)\tau
%$. Finally, substituting
%(\ref{eq:P_ijvbz}) in (\ref{eq:P_T_ij}), we get the stall duration
%tail probability as follows
Further, 
\begin{align}
&\text{\text{Pr}}\left(T_{i,j,\beta_{j},\nu_{j},z}^{(L_{i})}\geq\overline{\sigma}\right) \\
& \leq\sum_{z=1}^{L_{i}+1}\mathbb{P}\left(\text{\ensuremath{\mathcal{F}_{i,j,\beta_{j},\nu_{j},z}}}\geq\overline{\sigma}\right)\\
 & \overset{(e)}{=}e^{t_{i}(d_{s}+(L_{i}-1)\tau-\overline{\sigma})}\nonumber\\
 & +\sum_{z=2}^{L_{i}+1}e^{(L_{i}-z+1)\tau t_{i}-\overline{\sigma}t_{i}}\mathbb{E}\left[e^{t_{i}D_{i,j,\beta_{j},\nu_{j}}^{(z-1)}}\right]\\
 & =e^{-t_{i}\sigma}+\sum_{z=2}^{L_{i}+1}e^{(\widetilde{\sigma}-\overline{\sigma})t_{i}}\mathbb{E}\left[e^{t_{i}D_{i,j,\beta_{j},\nu_{j}}^{(z-1)}}\right]\\
 & =\left[e^{-t_{i}\sigma}+\sum_{v=1}^{L_{i}}e^{(\widetilde{\sigma}-\overline{\sigma})t_{i}}\mathbb{E}\left[e^{t_{i}D_{i,j,\beta_{j},\nu_{j}}^{(v)}}\right]\right],
\label{eq:stall_final}
\end{align}
where (e) follows from (\ref{MGF_F}) and  $\widetilde{\sigma}=(L_{i}-z+1)\tau
$. Using the two-stage probabilistic scheduling, the stall distribution tail probability for video file $i$ is bounded by 
%\text{\text{Pr}}\left(\Gamma^{(i)}\geq\sigma\right) & \le \sum_{j=1}^{m}\pi_{i,j}\left[e^{-t_{i}\sigma}+\sum_{\nu_{j}=1}^{e_{j}}p_{i,j,\nu_{j}}\sum_{\beta_{j}=1}^{d_{j}}q_{i,j,\beta_{j}}\sum_{v=1}^{L_{i}}e^{(\widetilde{\sigma}-\overline{\sigma})t_{i}}\mathbb{E}\left[e^{t_iD_{i,j,\beta_{j},\nu_{j}}^{(v)}}\right]\right].
\begin{align}
\text{\text{Pr}}\left(\Gamma^{(i)}\geq\sigma\right) & \le\sum_{j=1}^{m}\pi_{i,j}\times \nonumber\\
& \left[e^{-t_{i}\sigma}+\sum_{\nu_{j}=1}^{e_{j}}p_{i,j,\nu_{j}}\sum_{\beta_{j}=1}^{d_{j}}q_{i,j,\beta_{j}}\sum_{v=1}^{L_{i}}\overline{D}_{i,j,\beta_{j},\nu_{j}}^{(v,t)}\right].
 \label{Pr_Tijbetanuj1}
\end{align}
where $\overline{D}_{i,j,\beta_{j},\nu_{j}}^{(v,t)}=e^{(\widetilde{\sigma}-\overline{\sigma})t_{i}}\mathbb{E}\left[e^{t_{i}D_{i,j,\beta_{j},\nu_{j}}^{(v)}}\right]$. 
%where $(f)$ follows from the two-stage Probabilistic Scheduling, $\widetilde{\sigma}=(L_{i}-z+1)\tau
%$, $\overline{\sigma}=\text{\ensuremath{\sigma}}+d_{s}+\left(L_{i}-1\right)\tau$,  is given by (\ref{D_nu_j}). 
Further, we derive $\mathbb{E}\left[e^{t_iD_{i,j,\beta_{j},\nu_{j}}^{(v)}}\right]$ using the following two lemmas. The key idea is that we characterize the download and play times of each segments and use them in determining the SDTP of each file request.  
%can be calculated as shown in the following two lemmas. 

\begin{lemma}
For $v\leq L_{j,i}$,   $\mathbb{E}\left[e^{t_iD_{i,j,\beta_{j},\nu_{j}}^{(v)}}\right]$ is given by 
\begin{equation}
\mathbb{E}\left[e^{t_iD_{i,j,\beta_{j},\nu_{j}}^{(v)}}\right]=
\frac{(1-\rho_{j,\nu_{j}}^{(e)})t_iB_{j,\nu_{j}}^{(e)}(t_i)}{t_i-\Lambda_{j,\nu_{j}}^{(e)}(B_{j,\nu_{j}}^{(e)}(t_i)-1)}\left(\frac{\alpha_{j,\nu_{j}}^{(e)}e^{\eta_{j,\nu_j}^{(e)}t}}{\alpha_{j,\nu_{j}}^{(e)}-t_i}\right)^{v}
\end{equation}
\end{lemma}
\begin{proof}
The proof follows from \eqref{D_nu_j} by replacing $g$ by $v$ and rearranging the terms in the result.
\end{proof}

\begin{lemma}
	\label{tailLemma}
For $v > L_{j,i}$,  $\mathbb{E}\left[e^{t_iD_{i,j,\beta_{j},\nu_{j}}^{(v)}}\right]$ is given by 
\begin{align}
\text{\text{Pr}}\left(\Gamma^{(i)}\geq\sigma\right) & \overset{}{\leq}\sum_{j=1}^{m}\pi_{i,j}\left[e^{-t_{i}\sigma}+\sum_{\nu_{j}=1}^{e_{j}}p_{i,j,\nu_{j}}\sum_{\beta_{j}=1}^{d_{j}}q_{i,j,\beta_{j}}e^{(\widetilde{\sigma}-\overline{\sigma})t_{i}}\times\right. \nonumber \\
 & \left.\left(\delta_{1}^{(e)}+\delta_{2}^{(\overline{d})}+\delta_{3}^{(d,\overline{d})}+\delta_{4}^{(d,\overline{d})}\right)\right]
\end{align}
where we define auxiliary variables

$\delta_{1}^{(e)}=\frac{M_{j,\nu_{j}}^{(e)}(t_{i})(1-\rho_{j,\nu_{j}}^{(e)})t_{i}B_{j,\nu_{j}}^{(e)}(t_{i})}{t_{i}-\Lambda_{j,\nu_{j}}^{(e)}(B_{j,\nu_{j}}^{(e)}(t_{i})-1)}\frac{\left((M_{j,\nu_{j}}^{(e)}(t_{i}))^{L_{j,i}}-1\right)}{M_{j,\nu_{j}}^{(e)}(t_{i})-1}{\bf 1}_{L_{j,i}>0}$, 
$\delta_{2}^{(\overline{d})}=\frac{(1-\rho_{j,\beta_{j}}^{(\overline{d})})t_{i}B_{j,\beta_{j}}^{(\overline{d})}(t_{i})(v-L_{j,i})}{t_{i}-\Lambda_{j,\beta_{j}}^{(\overline{d})}(B_{j,\beta_{j}}^{(\overline{d})}(t_{i})-1)}\left(\frac{\alpha_{j,\beta_{j}}^{(\overline{d})}e^{\eta_{j,\beta_j}^{(\overline{d})}t}}{\alpha_{j,\beta_{j}}^{(\overline{d})}-t_{i}}\right)^{v-L_{j,i}}{\bf 1}_{v>L_{j,i}}\,,$

$\delta_{3}^{(d,\overline{d})}=\frac{(1-\rho_{j,\beta_{j}}^{(d)})t_{i}B_{j,\beta_{j}}^{(d)}(t_{i})\left(\widetilde{M}_{j,\beta_{j}}^{(d,\overline{d})}(t_{i})\right)^{L_{j,i}+1}}{t_{i}-\Lambda_{j,\beta_{j}}^{(d)}(B_{j,\beta_{j}}^{(d)}(t_{i})-1)}\times$

$\qquad\left(\frac{\alpha_{j,\beta_{j}}^{(d)}-t_{i}}{\alpha_{j,\beta_{j}}^{(d)}e^{\eta_{j,\beta_j}^{(\overline{d})}t}}\right)^{L_{j,i}+1}\left(\frac{\alpha_{j,\beta_{j}}^{(\overline{d})}e^{\eta_{j,\beta_j}^{(\overline{d})}t}}{\alpha_{j,\beta_{j}}^{(\overline{d})}-t_{i}}\right)^{v+1}{\bf 1}_{v>L_{j,i}}$

$\delta_{4}^{(d,\overline{d})}=\frac{\left(\widetilde{M}_{j,\beta_{j}}^{(d,\overline{d})}(t_{i})\right)^{v-L_{j,i}}-(v-L_{j,i})}{\widetilde{M}_{j,\beta_{j}}^{(d,\overline{d})}(t_{i})-1}\\\qquad +\frac{\widetilde{M}_{j,\beta_{j}}^{(d,\overline{d})}(t_{i})\left(\left(\widetilde{M}_{j,\beta_{j}}^{(d,\overline{d})}(t_{i})\right)^{v-L_{j,i}-1}-1\right)}{\left(\widetilde{M}_{j,\beta_{j}}^{(d,\overline{d})}(t_{i})-1\right)^{2}}{\bf 1}_{v>L_{j,i}}$

and
\begin{equation}
\widetilde{M}_{j,\beta_{j}}^{(d,\overline{d})}(t)=\frac{\alpha_{j,\beta_{j}}^{(d)}(\alpha_{j,\beta_{j}}^{(\overline{d})}-t)e^{\eta_{j,\beta_j}^{(d)}t}}{\alpha_{j,\beta_{j}}^{(\overline{d})}(\alpha_{j,\beta_{j}}^{(d)}-t)e^{\eta_{j,\beta_j}^{(\overline{d})}t}},\,\,\,\forall j,\beta_{j}\label{tailEqn}
\end{equation}
\label{lemmatailEqn}
\end{lemma}
\begin{proof}
The proof is provided in Appendix \refeq{lemmatailEqn}.
\end{proof}

Now, we are ready to state the main theorem of this paper as follows.

\begin{theorem}

The stall distribution tail probability for video
file $i$ is bounded by

\begin{align*}
& \!\text{\text{Pr}}\left(\Gamma^{(i)}\geq\sigma\right)\leq\!\sum_{j=1}^{m}\pi_{i,j}\!\!\left[e^{-t_{i}\sigma}\!+\!\!\sum_{\nu_{j}=1}^{e_{j}}p_{i,j,\nu_{j}}\!\!\sum_{\beta_{j}=1}^{d_{j}}\!\!q_{i,j,\beta_{j}}e^{(\widetilde{\sigma}-\overline{\sigma})t_{i}}\right.\\
& \times\left.\left(\delta_{1}^{(e)}+\delta_{2}^{(\overline{d})}+\delta_{3}^{(d,\overline{d})}+\delta_{4}^{(d,\overline{d})}\right)\right]
\end{align*}
where $\delta_{1}^{(e)}$, $\delta_{2}^{(\overline{d})}$, $\delta_{3}^{(d,\overline{d})}$ and $\delta_{4}^{(d,\overline{d})}$ (after setting $v=L_i$) are given by \eqref{tailEqn}
and $\rho_{j,\beta_{j}}^{(d)}<1$, $\rho_{j,\beta_{j}}^{(\overline{d})}<1$, $\rho_{j,\nu_{j}}^{(e)}<1$. Further, the moment generating function terms exist by satisfying of (\ref{alpha_j_nuj}--\ref{mgf_gz_e_queue}), (\ref{alpha_betaj_d})--(\ref{mgf_beta_j_c}) for $t=t_i$.
\end{theorem}
We note that the indicators can be omitted since the terms should always be zero if the indicators are not true. For instance, $\delta_{1}^{(e)}=\delta_{2}^{(\overline{d})}=0$, if the cache is empty and $\delta_{3}^{(d)}$, $\delta_{4}^{(d)}$ have nonzero values only if some segments need to be served from the origin datacenter. Hence, we omit the indicators in the rest of this paper. 

%the indicators were kept to see that the terms should be zero if it is not the case, while that can be easily seen to be the case and thus the indicators can be omitted. 

%We also note that the last two terms in Theorem above are corresponding to the stall duration probability that is incurred due to streaming video segments from the datacenter. 

%\begin{equation}
%\mathbb{E}\left[e^{t_iD_{i,j,\beta_{j},\nu_{j}}^{(v)}}\right]=\begin{cases}
%\frac{(1-\rho_{j,\nu_{j}}^{(e)})t_iB_{j,\nu_{j}}^{(e)}(t_i)}{t_i-\Lambda_{j,\nu_{j}}^{(e)}(B_{j,\nu_{j}}^{(e)}(t_i)-1)}\left(\frac{\alpha_{j,\nu_{j}}^{(e)}}{\alpha_{j,\nu_{j}}^{(e)}-t_i}\right)^{g} & ,\,g\leq L_{j,i}\\
%\left[e^{t_iU_{i,j,\beta_{j},g,L_{j,i}}}\right]+\sum_{w=L_{j,i}+1}^{g}\frac{(1-\rho_{j,\beta_{j}}^{(d)})t_iB_{j,\beta_{j}}^{(d)}(t_i)}{t_i-\Lambda_{j,\beta_{j}}^{(d)}(B_{j,\beta_{j}}^{(d)}(t_i)-1)}\left(\frac{\alpha_{j,\beta_{j}}^{(d)}}{\alpha_{j,\beta_{j}}^{(d)}-t_i}\right)^{w-L_{j,i}-1}\left(\frac{\alpha_{j,\beta_{j}}^{(c)}}{\alpha_{j,\beta_{j}}^{(c)}-t_i}\right)^{g-w+1} & g>L_{j,i}
%\end{cases}
%\end{equation}
%
%We note that for $g>L_{j,i}$, the given expression in () can be  

%either equation (\ref{D_nu_j}) or equation (\ref{U_MGF_all}), based on the value of $v$.
\section{Optimization Problem Formulation and Proposed Algorithm }
\label{sec:SDTP}
\subsection{Problem Formulation}
We define 
%$\boldsymbol{\pi}=(\pi_{i,j}\forall i=1,\cdots,r\text{ and }j=1,\cdots,m)$,
%$\boldsymbol{p}=(p_{i,,j,\nu_{j}}\forall i=1,\cdots,r\,,j=1,\cdots,m\,,\,\nu_{j}=1,\ldots e_{j}),$
%$\boldsymbol{q}=(q_{i,,j,\beta_{j}}\forall i=1,\cdots,r\,,j=1,\cdots,m\,,\,\beta_{j}=1,\ldots d_{j}),$ $\boldsymbol{t}=\left(t_{1},t_{2},\ldots,t_{r}\right)$, 
%
%$\boldsymbol{w}_{j,\beta_{j}}^{(c)}=(w_{j,1}^{(c)},w_{j,2}^{(c)},\ldots,w_{j,d_{j}}^{(c)},\text{ and }j=1,\cdots,m)$,
%$\boldsymbol{w}_{j,\beta_{j}}^{(e)}=(w_{j,1}^{(e)},w_{j,2}^{(e)},\ldots,w_{j,d_{j}}^{(e)},\text{ and }j=1,\cdots,m)$,
%$\boldsymbol{w}_{j,\beta_{j}}^{(d)}=(w_{j,1}^{(d)},w_{j,2}^{(d)},\ldots,w_{j,d_{j}}^{(d)},\text{ and }j=1,\cdots,m)$,
%$\boldsymbol{L}_{j,i}=(L_{j,i},\,\forall i,\,\forall j)$

$\ensuremath{\boldsymbol{\pi}=(\pi_{i,j}\forall i=1,\cdots,r\text{ and }j=1,\cdots,m)},$, $\boldsymbol{p}=(p_{i,,j,\nu_{j}}\forall i=1,\cdots,r\,,j=1,\cdots,m\,,\,\nu_{j}=1,\ldots e_{j})$, $\ensuremath{\boldsymbol{q}=(q_{i,,j,\beta_{j}}\forall i=1,\cdots,r\,,j=1,\cdots,m\,,\,\beta_{j}=1,\ldots d_{j}),}$, $\boldsymbol{t}=\left(t_{1},t_{2},\ldots,t_{r}\right)\,$, 
$\ensuremath{\boldsymbol{w}^{(\overline{d})}=(w_{j,1}^{(\overline{d})},\ldots,w_{j,d_{j}}^{(c)}$,$\text{ and }j=1,\cdots,m)}\,,$, $\boldsymbol{w}^{(e)}=(w_{j,1}^{(e)},w_{j,2}^{(e)},\ldots,w_{j,d_{j}}^{(e)},\text{ and }j=1,\cdots,m)\,$, $\ensuremath{\boldsymbol{w}^{(d)}=(w_{j,1}^{(d)},\ldots,w_{j,d_{j}}^{(d)},\text{ and }j=1,\cdots,m)},$, $\boldsymbol{L}=(L_{j,i},\,\forall i=1,\,\text{ and }j=1,\cdots,m)$.
Our goal is to minimize the stall duration tail probability over the
choice of cache and datacenter access decisions, bandwidth allocation
weights, portion (number) of cached segments, and auxiliary bound parameters.

To incorporate for weighted fairness and differentiated services,
we assign a positive weight $\omega_{i}$ for each file $i$. Without
loss of generality, each file $i$ is weighted by the arrival rate
$\lambda_{i}$ in the objective (so larger arrival rates are weighted
higher). However, any other weights can be incorporated to accommodate
for weighted fairness or differentiated services. Let $\overline{\lambda}=\sum_{i}\lambda_{i}$
be the total arrival rate. Hence, $\omega_{i}=\lambda_{i}/\overline{\lambda}$
is the ratio of file $i$ requests. Hence, the objective is the minimization
of stall duration tail probability, averaged over all the file requests,
and is given as $\sum_{i}\frac{\lambda_{i}}{\overline{\lambda}}\,\text{\text{Pr}}\left(\Gamma^{(i)}\geq\sigma\right)$.
By using the expression for stall duration tail probability in Section \ref{sec:StallTailProb},
the optimization problem can be formulated as follows.

 \vspace{-.1in}
 
 \begin{align}
 & \text{\textbf{min}\,\,\,\,\,}\sum_{i=1}^{r}\frac{\lambda_{i}}{\overline{\lambda}}\sum_{j=1}^{m}\pi_{i,j}\left[e^{-t_{i}\sigma}+\sum_{\nu_{j}=1}^{e_{j}}p_{i,j,\nu_{j}}\times\right.\nonumber\\
 & \qquad\qquad\left.\sum_{\beta_{j}=1}^{d_{j}}q_{i,j,\beta_{j}}e^{(\widetilde{\sigma}-\overline{\sigma})t_{i}}\left(\delta_{1}^{(\overline{d})}+\delta_{2}^{(\overline{d})}+\delta_{3}^{(d,\overline{d})}+\delta_{4}^{(d,\overline{d})}\right)\right]\label{eq:optfun}\\
 & \boldsymbol{\text{s.t.}}\nonumber\\
 &
\eqref{alphad}-
\eqref{Me_jnuj},\,
\eqref{lambdac_betaj},\,
\eqref{lambdae_nuj},\,
\eqref{Be},
%\eqref{rhoe},
%\eqref{rhod},
\eqref{Bd},
%\eqref{rhoc},
\eqref{Bc},\label{formulas}\\ 
&
\widetilde{M}_{j,\beta_{j}}^{(d,\overline{d})}(t)=\frac{\alpha_{j,\beta_{j}}^{(d)}(\alpha_{j,\beta_{j}}^{(\overline{d})}-t)e^{\eta_{j,\beta_j}^{(d)}t}}{\alpha_{j,\beta_{j}}^{(\overline{d})}(\alpha_{j,\beta_{j}}^{(d)}-t)e^{\eta_{j,\beta_j}^{(\overline{d})}t}},\,\,\,\forall j,\beta_{j}\\
& \rho_{j,\beta_{j}}^{(d)}=\sum_{i=1}^{r}\lambda_{i}\pi_{i,j}q_{i,j,\beta_{j}}\frac{L_{i}-L_{j,i}}{\alpha_{j,\beta_{j}}^{(d)}}<1,\,\,\,\forall j,\beta_{j}\label{eq:rho_beta_d}\\
& \rho_{j,\beta_{j}}^{(\overline{d})}=\sum_{i=1}^{r}\lambda_{i}\pi_{i,j}q_{i,j,\beta_{j}}\frac{L_{i}-L_{j,i}}{\alpha_{j,\beta_{j}}^{(\overline{d})}}<1\,\forall j,\beta_{j}\label{eq:rho_beta_c}\\
& \rho_{j,\nu_{j}}^{(e)}=\sum_{i=1}^{r}\lambda_{i}\pi_{i,j}p_{i,j,\nu_{j}}\frac{L_{j,i}}{\alpha_{j,\nu_{j}}^{(e)}}<1,\,\,\,\forall j,\nu_{j}\label{eq:rho_nu_e}\\
& \sum_{j=1}^{m}\pi_{i,j}=1,\,\,\,\forall i,\label{eq:sum_pi_ij}\\
& \sum_{\nu_{j}=1}^{e_{j}}p_{i,,j,\nu_{j}}=1,\,\,\,\forall i,j\label{eq:sum_piij_nu}\\
& \sum_{\beta_{j}=1}^{d_{j}}q_{i,,j,\beta_{j}}=1,\,\,\,\forall i,j\label{eq:sum_q_ij_beta}\\
 & \pi_{i,j},\,p_{i,,j,\nu_{j}},\,q_{i,,j,\beta_{j}}\geq0,\,\,\,\forall i,j,\beta_{j},\nu_{j}\label{eq:pi_q_p_gz}\\
& w_{j,\beta_{j}}^{(\overline{d})},\,w_{j,\beta_{j}}^{(d)},\,w_{j,\nu_{j}}^{(e)}\geq0,\,\,\,\forall j,\beta_{j},\nu_{j}\label{eq:w_cde_gz}\\
& \sum_{\beta_{j}=1}^{d_{j}}w_{j,\beta_{j}}^{(d)}\leq1,\,\,\,\forall j
\label{eq:sum_w_d}\\
& \sum_{\beta_{j}=1}^{d_{j}}w_{j,\beta_{j}}^{(\overline{d})}+\sum_{\nu_{j}=1}^{e_{j}}w_{j,\nu_{j}}^{(e)}\leq1,\,\,\,\forall j\label{eq:sum_w_de}\\
 & \sum_{i}L_{j,i}\leq C_{j},\,L_{j,i}\geq0,\,\,\,\forall i,j\label{eq:capConst}\\
& t_{i}<\alpha_{j,\beta_{j}}^{(d)},\,\,\,\forall i,j,\beta_{j}\label{eq:t_leq_alpha_beta_d}\\
& t_{i}<\alpha_{j,\beta_{j}}^{(\overline{d})},\,\,\,\forall i,j,\beta_{j}\label{eq:t_leq_alpha_beta_c}
  \end{align}
\begin{align}
& t_{i}<\alpha_{j,\nu_{j}}^{(e)},\,\,\,\forall i,j,\nu_{j}\label{eq:t_leq_alpha_nu_e}\\
 & 0<t_{i}-\Lambda_{j,\beta_{j}}^{(d)}(B_{j,\beta_{j}}^{(d)}(t_{i})-1),\,\,\,\forall i,j,\beta_{j}\label{eq:mgf_beta_d}\\
 & 0<t_{i}-\Lambda_{j,\beta_{j}}^{(\overline{d})}(B_{j,\beta_{j}}^{(\overline{d})}(t_{i})-1),\,\,\,\forall i,j,\beta_{j}\label{eq:mgf_beta_c}\\
 & 0<t_{i}-\Lambda_{j,\nu_{j}}^{(e)}(B_{j,\nu_{j}}^{(e)}(t_{i})-1),\,\,\,\forall i,j,\nu_{j}\label{eq:mgf_beta_e}\\
 &L_{j,i}\in {\mathbb Z}\label{eq:Lij}\\
\boldsymbol{\text{var}} & \text{\,\,\,\,\,}\boldsymbol{\pi,}\,\boldsymbol{q},\,\boldsymbol{p},\,\boldsymbol{t},\,\boldsymbol{w}^{(c)},\,\boldsymbol{w}^{(d)},\,\boldsymbol{w}^{(e)},\,\boldsymbol{L}
\end{align}

Here, the formulas mentioned in equation (\ref{formulas}) give the functions in the defined objective function. Constraints (\ref{eq:rho_beta_d})--(\ref{eq:rho_nu_e}) ensure the stablity of the systems queue (do not blow up to infinity) while Constraints (\ref{eq:sum_pi_ij})--(\ref{eq:pi_q_p_gz}) guarantee that the two-stage probabilistic scheduling exists. Further, constriants (\ref{eq:w_cde_gz})--(\ref{eq:sum_w_de}) define the bandwidth split and ensure the feasibility of bandiwdth allocation weights. Constraint (\ref{eq:capConst}) limits the cached segments to the storage server cache capacity. Constraints (\ref{eq:t_leq_alpha_beta_d})--(\ref{eq:mgf_beta_e}) ensure that the moment generating functions exist. We note that some optimization variables can be combined to form a single optimization variable which results in having only four independent and separable variables as shown below. In the next subsection, we will describe the proposed algorithm for this optimization problem.

\subsection{Proposed Algorithm}
\label{sec:algo}

We first note that the two-stage probabilistic scheduling variables are independent and separable, thus we can combine them and define a single variable $\widetilde{\boldsymbol{\pi}}$ such that  
$\widetilde{\boldsymbol{\pi}}=\left(\boldsymbol{\pi}\,,\boldsymbol{\,p}\,\,,\boldsymbol{q}\right)$. Similarly, since the bandwidth allocation weights are independent and separable, we concatenate them in a single optimization variable $\boldsymbol{w}$, where  
$\boldsymbol{w}=\left(\boldsymbol{w}^{(e)}\,,\boldsymbol{\,w}^{(\overline{d})}\,\,,\boldsymbol{w}^{(d)}\right)$. Hence, the weighted stall duration tail probability optimization problem given in \eqref{eq:optfun}-\eqref{eq:mgf_beta_e}  is optimized over four set of variables:  server and PSs scheduling probabilities $\widetilde{\boldsymbol{\pi}}$ (two-stage scheduling probabilities), auxiliary parameters $\boldsymbol{t}$, bandwidth allocation weights $\boldsymbol{w}$, and cache placement $\boldsymbol{L}$.

 Clearly, the problem is non-convex in all the parameters jointly, which can be easily seen in the terms which are product of the different variables. Since the problem is non-convex, we propose an iterative algorithm to solve the problem. The proposed algorithm divides the problem into four sub-problems that optimize one variable while fixing the remaining four. These sub-problems are labeled as   (i) Server and PSs Access Optimization: optimizes  $\widetilde{\boldsymbol{\pi}}$, for given  $\boldsymbol{t}$, $\boldsymbol{w}$ and $\boldsymbol{L}$,  
(ii) Auxiliary Variables Optimization: optimizes  $\boldsymbol{t}$ for given $\widetilde{\boldsymbol{\pi}}$, $\boldsymbol{w}$, and $\boldsymbol{L}$,  and (iii) Bandwidth Allocation Optimization: optimizes  $\boldsymbol{w}$ for given $\widetilde{\boldsymbol{\pi}}$, $\boldsymbol{t}$, and $\boldsymbol{L}$. 
(iv) Cache Placement Optimization: optimizes  $\boldsymbol{L}$ for given $\widetilde{\boldsymbol{\pi}}$, $\boldsymbol{t}$, and $\boldsymbol{w}$. The algorithm is summarized as follows. 

%, where $\boldsymbol{\mathcal{S}}=\left(\mathcal{S}_{1},\mathcal{S}_{2},\ldots,\mathcal{S}_{r}\right)$,
% (ii) access-scheduling probabilities  ($\boldsymbol{\pi}$- Optimization ), and finally (iii) auxiliary variables optimization ($\boldsymbol{t}$- Optimization). These sub-problems have an easier-to-handle structure and can be solved efficiently. Next, we develop an efficient algorithm solution for our optimization problem.
%
%
%The joint mean-tail stall duration optimization problem given in (\ref{eq:joint_otp_prob}) is carried out over three set of variables: chunk placement $\mathcal{S}_{i}$, $\forall$ $i$, scheduling probabilities $\pi_{i,j}$, $\forall$ $i,j$, and auxiliary parameter $t_{i}$, $\forall$ $i$. In order to solve the non-convex problem, we develop an alternating minimization algorithm which is shown to converge to a stationary solution. The proposed algorithm can be written as follows.

\begin{enumerate}[leftmargin=0cm,itemindent=.5cm,labelwidth=\itemindent,labelsep=0cm,align=left]
\item \textbf{Initialization}: Initialize $\boldsymbol{t}$,\,$\widetilde{\boldsymbol{\pi}}$, $\boldsymbol{w}$, 
and $\boldsymbol{L}$ in the feasible set. 
\item \textbf{While Objective Converges}
\begin{enumerate}

\item  Run Server Access Optimization using current values of $\boldsymbol{t}$, $\boldsymbol{w}$,
and $\boldsymbol{L}$ to get new values of $\widetilde{\boldsymbol{\pi}}$

\item  Run Auxiliary Variables Optimization using current values of $\widetilde{\boldsymbol{\pi}}$, $\boldsymbol{w}$,
and $\boldsymbol{L}$ to get new values of $\boldsymbol{t}$

\item  Run Bandwidth Allocation Optimization using current values
of $\widetilde{\boldsymbol{\pi}}$, $\boldsymbol{t}$, and $\boldsymbol{L}$ to get new values of $\boldsymbol{w}$.

\item  Run Cache Placement Optimization using current values
of $\widetilde{\boldsymbol{\pi}}$, $\boldsymbol{t}$, and  $\boldsymbol{w}$ to get new values of $\boldsymbol{L}$.

\end{enumerate}
\end{enumerate}

We next describe the four sub-problems along with the proposed solutions for the sub-problems.

\subsubsection{Server-PSs Access Optimization}

Given the bandwidth allocation weights, the cache placement, and the auxiliary variables, this sub-problem can be written as follows. 

\textbf{Input: $\boldsymbol{t}$, $\boldsymbol{w}$, and  $\boldsymbol{L}$}

\textbf{Objective:} $\qquad\quad\;\;$min $\left(\ref{eq:optfun}\right)$

$\hphantom{\boldsymbol{\text{Objective:}\,}}\qquad\qquad$s.t. \eqref{eq:rho_beta_d}--\eqref{eq:pi_q_p_gz}, \eqref{eq:mgf_beta_d}--
\eqref{eq:mgf_beta_e}\\
$\hphantom{\boldsymbol{\text{Objective:}\,}}\qquad\qquad$var. $\widetilde{\boldsymbol{\pi}}$

%Now, we explain our approach in solving $\boldsymbol{\pi}$- Optimization sub-problem. We optimize over $\pi_{ij}$, $\forall i,j$ for given placement $\boldsymbol{\mathcal{S}}$ and fixed $\boldsymbol{t}$. This problem is non-convex as shown in Section \ref{sec:probForm}. Hence, 

In order to solve this problem, we use iNner cOnVex
Approximation (NOVA)  algorithm proposed in \cite{scutNOVA}. The key idea for this algorithm is that the non-convex objective function is replaced by suitable convex approximations at which convergence to a stationary solution of the original non-convex optimization is established. NOVA solves the approximated function efficiently and maintains feasibility in each iteration. The objective function can be approximated by a convex one (e.g., proximal gradient-like approximation) such that the first order properties are preserved \cite{scutNOVA},  and this convex approximation can be used in NOVA algorithm.

Let $\widetilde{U_{q}}\left(\widetilde{\boldsymbol{\pi}},\boldsymbol{\widetilde{\pi}^{\nu}}\right)$ be the
convex approximation at iterate $\boldsymbol{{\pi}^\nu}$ to the original non-convex problem $U\left(\widetilde{\boldsymbol{\pi}}\right)$, where $U\left(\widetilde{\boldsymbol{\pi}}\right)$ is given by (\ref{eq:optfun}). Then, a valid choice of $U\left(\widetilde{\boldsymbol{\pi}};\boldsymbol{{\pi}^\nu}\right)$ is the first order approximation of  $U\left(\widetilde{\boldsymbol{\pi}}\right)$, e.g., (proximal) gradient-like approximation, i.e.,

\begin{equation}
\widetilde{U_{q}}\left(\widetilde{\boldsymbol{\pi}},\boldsymbol{\widetilde{\pi}^{\nu}}\right)=\nabla_{\widetilde{\boldsymbol{\pi}}}U\left(\widetilde{\boldsymbol{\pi}}^{\nu}\right)^{T}\left(\widetilde{\boldsymbol{\pi}}-\widetilde{\boldsymbol{\pi}}^{\nu}\right)+\frac{\tau_{u}}{2}\left\Vert \widetilde{\boldsymbol{\pi}}-\widetilde{\boldsymbol{\pi}}^{\nu}\right\Vert ^{2}
,\label{eq:U_x_u_bar}
\end{equation}
where $\tau_u$ is a regularization parameter. Note that all the constraints \eqref{eq:rho_beta_d}--\eqref{eq:pi_q_p_gz} are separable and
linear in $\widetilde{\pi}_{i,j,k}$.  
 The NOVA Algorithm for optimizing $\widetilde{\boldsymbol{\pi}}$ is described in  Algorithm \ref{alg:NOVA_Alg1Pi} (given in Appendix \refeq{apdx_alg}). Using the convex approximation $\widetilde{U_{\pi}}\left(\boldsymbol{\pi};\boldsymbol{{\pi}^\nu}\right)$, the minimization steps in Algorithm \ref{alg:NOVA_Alg1Pi} are convex, with linear constraints and thus can be solved using a projected gradient descent algorithm. A step-size ($\gamma
$)  is also used in the update of the iterate $\widetilde{\boldsymbol{\pi}}^{\nu}$. Note that the iterates $\left\{ \boldsymbol{\pi}^{\nu}\right\} $ generated by the algorithm are all feasible for the original problem and, further, convergence is guaranteed, as shown in \cite{scutNOVA} and described in   lemma \ref{lem_pi}. 

In order to use NOVA, there are some assumptions (given in \cite{scutNOVA}) that have to be satisfied in both original function and its approximation. These assumptions can be classified into two categories. The first category is the set of conditions that ensure that the original problem and its constraints are continuously differentiable on the domain of the function, which are satisfied in our problem. The second category is the set of conditions that ensures that the approximation of the original problem is uniformly strongly convex on the domain of the function. The latter set of conditions are also satisfied as the chosen function is strongly convex and its domain is also convex. To see this, we need to show that the constraints \eqref{eq:rho_beta_d}--\eqref{eq:pi_q_p_gz} form a convex domain in $\widetilde{\boldsymbol{\pi}}$
 which is easy to see from the linearity of the constraints. Further details on the assumptions and function approximation can be found in \cite{scutNOVA}. Thus, the following result holds.

\begin{lemma}  \label{lem_pi}
For fixed $\boldsymbol{t}$, $\boldsymbol{w}$, and  $\boldsymbol{L}$, the optimization of our problem over 
$\widetilde{\boldsymbol{\pi}}$
 generates a sequence of  decreasing
objective values and therefore is guaranteed to converge to a stationary point.
\end{lemma}

\subsubsection{Auxiliary Variables Optimization}
Given the probability distribution of the server-PSs scheduling probabilities, the bandwidth allocation weights, and the cache placement, this subproblem can be written as follows. 

\textbf{Input: $\widetilde{\boldsymbol{\pi}}$, $\boldsymbol{w}$, and $\boldsymbol{L}$ }

\textbf{Objective:} $\qquad\quad\;\;$min $\left(\ref{eq:optfun}\right)$

$\hphantom{\boldsymbol{\text{Objective:}\,}}\qquad\qquad$s.t.  \eqref{eq:t_leq_alpha_beta_d}--\eqref{eq:mgf_beta_e}, \\
$\hphantom{\boldsymbol{\text{Objective:}\,}}\qquad\qquad$var. $\boldsymbol{t}$

Similar to Server-PSs Access Optimization, this optimization can be solved using NOVA algorithm. The constraints  \eqref{eq:t_leq_alpha_beta_d}--\eqref{eq:t_leq_alpha_nu_e} are linear in $\boldsymbol{t}$. Further, the next  Lemma show that the constraints  \eqref{eq:mgf_beta_d}-- \eqref{eq:mgf_beta_e} are convex in $\boldsymbol{t}$, respectively.

\begin{lemma}
%	\label{const68_70}
The  constraints \eqref{eq:mgf_beta_d}--\eqref{eq:mgf_beta_e} are convex with respect to  $\boldsymbol{{t}}$.  \label{cnv_t}
\end{lemma}
\begin{proof}
	The proof is given in Appendix \ref{apd_ww}. 
\end{proof}

 Algorithm \ref{alg:NOVA_Alg1} (given in Appendix \refeq{apdx_alg}) shows the used procedure to solve for $\boldsymbol{t}$. Let $\overline{U}\left(\boldsymbol{t};\boldsymbol{t^\nu}\right)$ be the
convex approximation at iterate $\boldsymbol{t^\nu}$ to the original non-convex problem $U\left(\boldsymbol{t}\right)$, where $U\left(\boldsymbol{t}\right)$ is given by (\ref{eq:optfun}), assuming other parameters constant. Then, a valid choice of  $\overline{U}\left(\boldsymbol{t};\boldsymbol{t^\nu}\right)$ is the first order approximation of  $U\left(\boldsymbol{t}\right)$, i.e.,  
\begin{equation}
\overline{U}\left(\boldsymbol{t},\boldsymbol{t^\nu}\right)=\nabla_{\boldsymbol{t}}U\left(\boldsymbol{t^\nu}\right)^{T}\left(\boldsymbol{t}-\boldsymbol{t^\nu}\right)+\frac{\tau_{t}}{2}\left\Vert \boldsymbol{t}-\boldsymbol{t^\nu}\right\Vert ^{2}.\label{eq:U_t_u_bar}
\end{equation}
where $\tau_t$ is a regularization parameter. The detailed steps can be seen in Algorithm \ref{alg:NOVA_Alg1}. Since all the constraints  \eqref{eq:t_leq_alpha_beta_d}-- \eqref{eq:mgf_beta_e} have been shown to be convex in $\boldsymbol{t}$, the optimization problem in Step 1 of  Algorithm \ref{alg:NOVA_Alg1} can be solved by the standard projected gradient descent algorithm. 

%Similar to Access  Optimization problem, we get the minimizer for (\ref{eq:U_t_u_bar}), step 1 in Algorithm \ref{alg:NOVA_Alg1}), and then continue until convergence. 

%In this sub-problem, we solve for all $t_i$, for fixed placement and given accessing probabilities. Similar to $\boldsymbol{\pi}$- Optimization problem, we used NOVA algorithm to solve for $t$.  Algorithm \ref{alg:NOVA_Alg1}, shows the used procedure to solve for $\boldsymbol{t}$. After initialization with feasible choice, NOVA keeps generating  a sequence of monotonically decreasing $\left\{ \boldsymbol{t}^{\nu+1}\right\} $ until  convergence.

\begin{lemma} \label{lem_t}

For fixed $\widetilde{\boldsymbol{\pi}}$, $\boldsymbol{w}$, and  $\boldsymbol{L}$, the optimization of our problem over $\boldsymbol{t}$ generates a sequence of monotonically decreasing objective values and therefore is guaranteed to converge to a stationary point.
\end{lemma}

\subsubsection{ Bandwidth Allocation Weights Optimization}
Given the auxiliary variables,  the server access and PSs selection probabilities, and cache placement, this subproblem can be written as follows. 

\textbf{Input: $\widetilde{\boldsymbol{\pi}}$, $\boldsymbol{L}$, and $\boldsymbol{t}$}

\textbf{Objective:} $\qquad\quad\;\;$min $\left(\ref{eq:optfun}\right)$

$\hphantom{\boldsymbol{\text{Objective:}\,}}\qquad\qquad$s.t. 
\eqref{eq:w_cde_gz}--
\eqref{eq:sum_w_de},\,\eqref{eq:t_leq_alpha_beta_d}--\eqref{eq:mgf_beta_e}, \\
$\hphantom{\boldsymbol{\text{Objective:}\,}}\qquad\qquad$var. $\boldsymbol{w}$

This optimization problem can be solved using NOVA algorithm. It is easy to notice that the constraints    \eqref{eq:w_cde_gz}-- \eqref{eq:sum_w_de} and \eqref{eq:t_leq_alpha_beta_d}-- \eqref{eq:t_leq_alpha_nu_e} are linear and thus convex with respect to $\boldsymbol{w}$. Further, the next two Lemmas show that the constraints  \eqref{eq:t_leq_alpha_beta_d}--\eqref{eq:mgf_beta_e}, are convex in $\boldsymbol{w}$, respectively.

\begin{lemma}
The  constraints \eqref{eq:t_leq_alpha_beta_d}--\eqref{eq:mgf_beta_e} are convex with respect to $\boldsymbol{{w}}$.  \label{cnv_ww}
\end{lemma}

\begin{proof}
The proof is given in Appendix \ref{apd_ww}. 
\end{proof}

Algorithm \ref{alg:NOVA_Alg3} (given in Appendix \refeq{apdx_alg}) shows the used procedure to solve for $\boldsymbol{w}$. Let ${U_w}\left(\boldsymbol{w};\boldsymbol{w^\nu}\right)$ be the
convex approximation at iterate $\boldsymbol{w^\nu}$ to the original non-convex problem $U\left(\boldsymbol{w}\right)$, where $U\left(\boldsymbol{w}\right)$ is given by (\ref{eq:optfun}), assuming other parameters constant. Then, a valid choice of  ${U_w}\left(\boldsymbol{w};\boldsymbol{w^\nu}\right)$ is the first order approximation of  $U\left(\boldsymbol{w}\right)$, i.e.,  
\begin{equation}
{U_w}\left(\boldsymbol{w},\boldsymbol{w^\nu}\right)=\nabla_{\boldsymbol{w}}U\left(\boldsymbol{w^\nu}\right)^{T}\left(\boldsymbol{w}-\boldsymbol{w^\nu}\right)+\frac{\tau_{w}}{2}\left\Vert \boldsymbol{w}-\boldsymbol{w^\nu}\right\Vert ^{2}.\label{eq:U_b}
\end{equation}
where $\tau_t$ is a regularization parameter. The detailed steps can be seen in Algorithm \ref{alg:NOVA_Alg3} (given in Appendix \refeq{apdx_alg}). Since all the constraints  have been shown to be convex, the optimization problem in Step 1 of  Algorithm \ref{alg:NOVA_Alg3} can be solved by the standard projected gradient descent algorithm. 

\begin{lemma} \label{lem_t}

For fixed $\widetilde{\boldsymbol{\pi}}$
, $\boldsymbol{t}$, and  $\boldsymbol{L}$, the optimization of our problem over 
$\boldsymbol{w}$ generates a sequence of  decreasing objective values and therefore is guaranteed to converge to a stationary point.
\end{lemma}

\subsubsection{Cache Placement Optimization}
Given the auxiliary variables, the server access and PS  selection probabilities, and the bandwidth allocation weights, this subproblem can be written as follows.  

\textbf{Input: $\ensuremath{\widetilde{\boldsymbol{\pi}}}$, $\boldsymbol{t}$, and $\boldsymbol{w}$}

\textbf{Objective:} $\qquad\quad\;\;$min $\left(\ref{eq:optfun}\right)$

$\hphantom{\boldsymbol{\text{Objective:}\,}}\qquad\qquad$s.t. 
\eqref{eq:rho_beta_d}--
\eqref{eq:rho_nu_e},
\eqref{eq:capConst},
\eqref{eq:mgf_beta_d}-- \eqref{eq:mgf_beta_e}\\
$\hphantom{\boldsymbol{\text{Objective:}\,}}\qquad\qquad$var. $\boldsymbol{L}$

Similar to the aforementioned  Optimization sub-problems, this optimization can be solved using NOVA algorithm. Constraints  \eqref{eq:rho_beta_d}--
\eqref{eq:rho_nu_e}, are linear in $\boldsymbol{L}$, and hence, form a convex domain. Also, Constraint \eqref{eq:capConst} is relaxed to have it convex. Furthermore, the constraints  \eqref{eq:mgf_beta_d}-- \eqref{eq:mgf_beta_e} are convex as shown in the following Lemmas in this subsection.

Algorithm \ref{alg:NOVA_Alg5} (given in Appendix \refeq{apdx_alg}) shows the used procedure to solve for $\boldsymbol{L}$. Let ${U_L}\left(\boldsymbol{L};\boldsymbol{L^\nu}\right)$ be the
convex approximation at iterate $\boldsymbol{L^\nu}$ to the original non-convex problem $U\left(\boldsymbol{L}\right)$, where $U\left(\boldsymbol{L}\right)$ is given by (\ref{eq:optfun}), assuming other parameters constant. Then, a valid choice of  ${U_L}\left(\boldsymbol{L};\boldsymbol{L^\nu}\right)$ is the first order approximation of  $U\left(\boldsymbol{L}\right)$, i.e.,  
\begin{equation}
{U_L}\left(\boldsymbol{L},\boldsymbol{L^\nu}\right)=\nabla_{\boldsymbol{L}}U\left(\boldsymbol{L^\nu}\right)^{T}\left(\boldsymbol{L}-\boldsymbol{L^\nu}\right)+\frac{\tau_{L}}{2}\left\Vert \boldsymbol{L}-\boldsymbol{L^\nu}\right\Vert ^{2}.\label{eq:U_L}
\end{equation}
where $\tau_L$ is a regularization parameter. The detailed steps can be seen in Algorithm \ref{alg:NOVA_Alg3}. Since all the constraints  have been shown to be convex in $\boldsymbol{L}$, the optimization problem in Step 1 of  Algorithm \ref{alg:NOVA_Alg3} can be solved by the standard projected gradient descent algorithm. 

\begin{lemma} \label{lem_b}

For fixed $\boldsymbol{t}$, $\ensuremath{\widetilde{\boldsymbol{\pi}}}$, and  $\boldsymbol{w}$, the optimization of our problem over 
$\boldsymbol{L}$ generates a sequence of monotonically decreasing
objective values and therefore is guaranteed to converge to a stationary point.
\end{lemma}

\subsubsection{Convergence of the Proposed Algorithm}

We first initialize 
$\widetilde{\boldsymbol{\pi}}$, $\boldsymbol{w}$, $\boldsymbol{t}$, and $\boldsymbol{L}$ $\forall i, j, \nu_j, \beta_j$ such that the choice is feasible for the problem. Then, we do alternating minimization over the four sub-problems defined above. Since each sub-problem can only decrease the objective and the overall problem is bounded from below, we have the following result.

\begin{theorem} 
 The proposed algorithm  converges to a  local optimal solution.
\end{theorem}

\section{Implementation and Evaluation}\label{sec:num}

In this section, we evaluate our proposed  algorithm for weighted stall duration tail probability. 

{\scriptsize{}}
\begin{table}[b]
{\scriptsize{}\caption{{\small{}
The value of $\alpha_j$  used in the evaluation results with units of 1/ms. We set $\eta_{j,\beta_j}^{(d)}=\eta_{j,\beta_j}^{(\overline{d})}=\eta_{j,\nu_j}^{(e)}=14$ ms.
%\vspace{-.2in}
%For $1<\ell\leq10$, both $\beta_{j}^{(\ell)}$ and $\alpha_{j}^{(\ell)}$are scaled by $DR^{(\ell)}/DR^{(1)}$.
%\vspace{-.2in}
\label{tab:Storage-Nodes-Parameters}}}
}{\scriptsize \par}

{\scriptsize{}}%
\begin{tabular}{|c|>{\centering}p{2.53cc}|>{\centering}p{2.53cc}|c|c|c|}
\multicolumn{1}{>{\centering}p{2.53cc}}{{\scriptsize{}Node 1}} & \multicolumn{1}{>{\centering}p{2.53cc}}{{\scriptsize{}Node 2}} & \multicolumn{1}{>{\centering}p{2.53cc}}{{\scriptsize{}Node 3}} & \multicolumn{1}{c}{{\scriptsize{}Node 4}} & \multicolumn{1}{c}{{\scriptsize{}Node 5}} & \multicolumn{1}{c}{{\scriptsize{}Node 6}}\tabularnewline
\hline 
{\scriptsize{}$82.00$} & {\scriptsize{}$76.53$} & {\scriptsize{}$71.06$} & {\scriptsize{}$65.6$} & {\scriptsize{}$60.13$} & {\scriptsize{}$54.66$}\tabularnewline
\hline 
\end{tabular}{\scriptsize \par}

{\scriptsize{}}%
\begin{tabular}{|c|>{\centering}p{2.53cc}|>{\centering}p{2.53cc}|c|c|c|}
 \multicolumn{1}{>{\centering}p{2.53cc}}{{\scriptsize{}Node 7}} & \multicolumn{1}{>{\centering}p{2.53cc}}{{\scriptsize{}Node 8}} & \multicolumn{1}{>{\centering}p{2.53cc}}{{\scriptsize{}Node 9}} & \multicolumn{1}{c}{{\scriptsize{}Node 10}} & \multicolumn{1}{c}{{\scriptsize{}Node 11}} & \multicolumn{1}{c}{{\scriptsize{}Node 12}}\tabularnewline
\hline 
{\scriptsize{}$49.20$} & {\scriptsize{}$44.28$} & {\scriptsize{}$39.36$} & {\scriptsize{}$34.44$} & {\scriptsize{}$29.52$} & {\scriptsize{}$24.60$}\tabularnewline
\hline 
\end{tabular}{\scriptsize \par}
%\vspace{-.2in}
\end{table}
{\scriptsize \par}

%\begin{table}[b]
%\caption{{\small{}Storage Node Parameters Used in our Simulation for $\ell=1$
%(Shift $\beta_{j}^{(1)}=10msec$ and rate $\alpha_{j}^{(1)}$ in 1/s).
%For $1<\ell\leq10$, both $\beta_{j}^{(\ell)}$ and $\alpha_{j}^{(\ell)}$
%are scaled by $DR^{(\ell)}/DR^{(1)}$.}}
%{\small{}}%
%\begin{tabular}{|c|c|c|c|c|c|c|}
%\multicolumn{1}{c}{} & \multicolumn{1}{c}{\textbf{\small{}Node 1}} & \multicolumn{1}{c}{\textbf{\small{}Node 2}} & \multicolumn{1}{c}{\textbf{\small{}Node 3}} & \multicolumn{1}{c}{\textbf{\small{}Node 4}} & \multicolumn{1}{c}{\textbf{\small{}Node 5}} & \multicolumn{1}{c}{\textbf{\small{}Node 6}}\tabularnewline
%\hline 
%{\small{}$\alpha_{j}^{(1)}$} & {\small{}$18.2298$} & {\small{}$24.0552$} & {\small{}$11.8750$} & {\small{}$17.0526$} & {\small{}$26.1912$} & {\small{}$23.9059$}\tabularnewline
%\hline 
%\end{tabular}{\small \par}
%
%{\small{}}%
%\begin{tabular}{|c|c|c|c|c|c|c|}
%\multicolumn{1}{c}{} & \multicolumn{1}{c}{\textbf{\small{}Node 7}} & \multicolumn{1}{c}{\textbf{\small{}Node 8}} & \multicolumn{1}{c}{\textbf{\small{}Node 9}} & \multicolumn{1}{c}{\textbf{\small{}Node 10}} & \multicolumn{1}{c}{\textbf{\small{}Node 11}} & \multicolumn{1}{c}{\textbf{\small{}Node 12}}\tabularnewline
%\hline 
%{\small{}$\alpha_{j}^{(1)}$} & {\small{}$27.006$} & {\small{}$21.3812$} & {\small{}$9.9106$} & {\small{}$24.9589$} & {\small{}$26.5288$} & {\small{}$21.8067$}\tabularnewline
%\hline 
%\end{tabular}\label{tab:Storage-Nodes-Parameters}
%\end{table}

\subsection{Parameter  Setup}

 We simulate our algorithm in a distributed storage cache system of $m=12$ distributed nodes, where some segments, i.e., $L_{j,i}$, of video file $i$ are stored in the storage cache nodes and thus servered from the cache nodes. The non-cached segments are severed from the data-center. Without loss of generality, we assume $e_j=40$, $d_j=20$ (unless otherwise explicitly stated) and $r=1000$ files, whose sizes are generated based on Pareto distribution \cite{arnold2015pareto} (as it is a commonly used distribution for file sizes \cite{Vaphase}) with shape factor of $2$ and scale of $300$, respectively. While we stick in the simulation to these parameters, our analysis and results remain applicable for any setting given that the system maintains stable conditions under the chosen parameters. Since we assume that the video file sizes are not heavy-tailed, the first $1000$ file-sizes that are less than 60 minutes are chosen.  We also assume that the segment service time  follows a shifted exponential distribution whose parameters are depicted in Table \ref{tab:Storage-Nodes-Parameters}, where the different values of server rates $\alpha_{j,\beta_j}^{(d)}$, $\alpha_{j,\beta_j}^{(\overline{d})}$ and  $\alpha_{j,\nu_j}^{(e)}$ are summarized. These values are extracted from our testbed (explained below) where the largest value of $\alpha_{j}$ corresponds to a bandwidth of 110Mbps and the smallest value is corresponding to a bandwidth of 25Mbps. Unless explicitly stated, the arrival rate for the first $500$ files is $0.002s^{-1}$ while for the next $500$ files is set to be $0.003s^{-1}$. Segment size $\tau$ is set to be equal to $4$ seconds and the cache servers are assumed to store only $35\%$ out of the total number of video file segments. When generating video files, the size of each video file is rounded up to the multiple of $4$ seconds.  In order to initialize our algorithm, we assume uniform scheduling,  $\pi_{i,j}=k/n$, $p_{j,\nu_j} = 1/e_j$, $q_{j,\beta_j} = 1/d_j$. Further, we choose  $t_{i}=0.01$, $w_{j, \nu_j}^{(e)}=1/e_j $, $w_{j, \beta_j}^{(\overline{d})}=1/d_j$ and $w_{j, \beta_j}^{(d)}=1/d_j$. However, these choices of the initial parameters may not be feasible. Thus, we modify the parameter initialization to be closest norm feasible solutions.

\subsection{Baselines}
We  compare our proposed approach with five strategies, which are described as follows. 
 \begin{enumerate}[leftmargin=0cm,itemindent=.5cm,labelwidth=\itemindent,labelsep=0cm,align=left]

%2
\item {\em Projected Equal Server-PSs Scheduling, Optimized Auxiliary variables, Cache Placement and Bandwidth Wights (PEA):} Starting with the initial solution mentioned above, the problem in \eqref{eq:optfun} is optimized over the choice of $\boldsymbol{t}$, $\boldsymbol{w}$, and $\boldsymbol{L}$ (using Algorithms \ref{alg:NOVA_Alg1}, \ref{alg:NOVA_Alg3},  and
\ref{alg:NOVA_Alg5}, 
respectively)  using alternating minimization. Thus, the values of $\pi_{i,j}$, $p_{i,j,\nu_j}$,$q_{i,j,\beta_j}$ will be approximately close to $k/n$, $1/e_j$ and $1/d_j$, respectively,  for all $i,j,\nu_j,\beta_j$.

%In this strategy, the initialization is as mentioned above

%we set $q_{i,j}^{(\ell)}=k/n$ $\forall i, \ell, j$ on the placed servers. Then, we modify the initialization of $\boldsymbol{q}$ to be closest norm feasible solution  given the values of $\boldsymbol{p}$, $\boldsymbol{t}$, $\boldsymbol{b}$ and $\boldsymbol{w}$. Finally, an alternating optimization over , $\boldsymbol{t}$, $\boldsymbol{b}$, $\boldsymbol{w}$ and $\boldsymbol{p}$ is performed to the objective using Algorithms \ref{alg:NOVA_Alg1}, \ref{alg:NOVA_Alg3},  \ref{alg:NOVA_Alg4}, and \ref{alg:NOVA_Alg5},  respectively.

%3

\item {\em Projected Equal Bandwidth, Optimized Access Servers and PS scheduling Probabilities, Auxiliary variables and cache placement (PEB):} Starting with the initial solution mentioned above, the problem in \eqref{eq:optfun} is optimized over the choice of $\widetilde{\boldsymbol{\pi}}$
, $\boldsymbol{t}$, and $\boldsymbol{L}$ (using Algorithms \ref{alg:NOVA_Alg1Pi}, \ref{alg:NOVA_Alg1},  and \ref{alg:NOVA_Alg5}, respectively) using alternating minimization. Thus, the bandwidth allocation weights, $w_{j,\nu_j}^{(e)}$, $w_{j,\beta_j}^{(\overline{d})}$, $w_{j,\beta_j}^{(d)}$ will be approximately $1/e_j$, $1/d_j$, and $1/d_j$, respectively.

%% 3
\item {\em Projected Proportional Service-Rate, Optimized Auxiliary variables, Bandwidth Wights, and Cache Placement (PSP):} In the initialization, the access probabilities among the servers, are given as  $\ensuremath{\pi_{i,j}=\frac{\mu_{j}}{\sum_{j}\mu_{j}},\,\forall i,j}$
.
This policy assigns servers proportional to their service rates. The choice of all parameters are then modified to the closest norm feasible solution.  Using this initialization, the problem in \eqref{eq:optfun} is optimized over the choice of $\boldsymbol{t}$, $\boldsymbol{w}$,  and $\boldsymbol{L}$ (using Algorithms \ref{alg:NOVA_Alg1},  \ref{alg:NOVA_Alg3}, and \ref{alg:NOVA_Alg5}, respectively)  using alternating minimization.

%% 4
\item {\em Projected Equal Caching, Optimized Scheduling Probabilities, Auxiliary variables and Bandwidth Allocation Weights (PEC):}
In this strategy, we divide the cache size equally among the video files. Using this initialization, the problem in \eqref{eq:optfun} is optimized over the choice of $\ensuremath{\widetilde{\boldsymbol{\pi}}}$, $\boldsymbol{t}$, and $ \boldsymbol{w}$  (using Algorithms  \ref{alg:NOVA_Alg1Pi}, \ref{alg:NOVA_Alg1}, and \ref{alg:NOVA_Alg3}, respectively)  using alternating minimization.

%% 4
\item {\em Projected Caching-Hottest files, Optimized Scheduling Probabilities, Auxiliary variables and Bandwidth Allocation Weights (CHF):}
In this strategy, we cache the video files that have the highest arrival rates (i.e., hottest files) at the distributed storage caches. Using this initialization, the problem in \eqref{eq:optfun} is optimized over the choice of $\ensuremath{\widetilde{\boldsymbol{\pi}}}$, $\boldsymbol{t}$, and $ \boldsymbol{w}$  (using Algorithms  \ref{alg:NOVA_Alg1Pi}, \ref{alg:NOVA_Alg1}, and \ref{alg:NOVA_Alg3}, respectively)  using alternating minimization.

%% 5
\item {\em Fixed-t Algorithm:}
In this strategy, we optimize all optimization variables except the auxiliary variable ${\bf t}$ where it is assigned a fixed value equals to $0.01$.

\end{enumerate}

\begin{figure*}[htbp]
\begin{minipage}{.32\textwidth}
	\centering
\includegraphics[trim=0in 0in 4in 0.0in, clip, width=\textwidth]{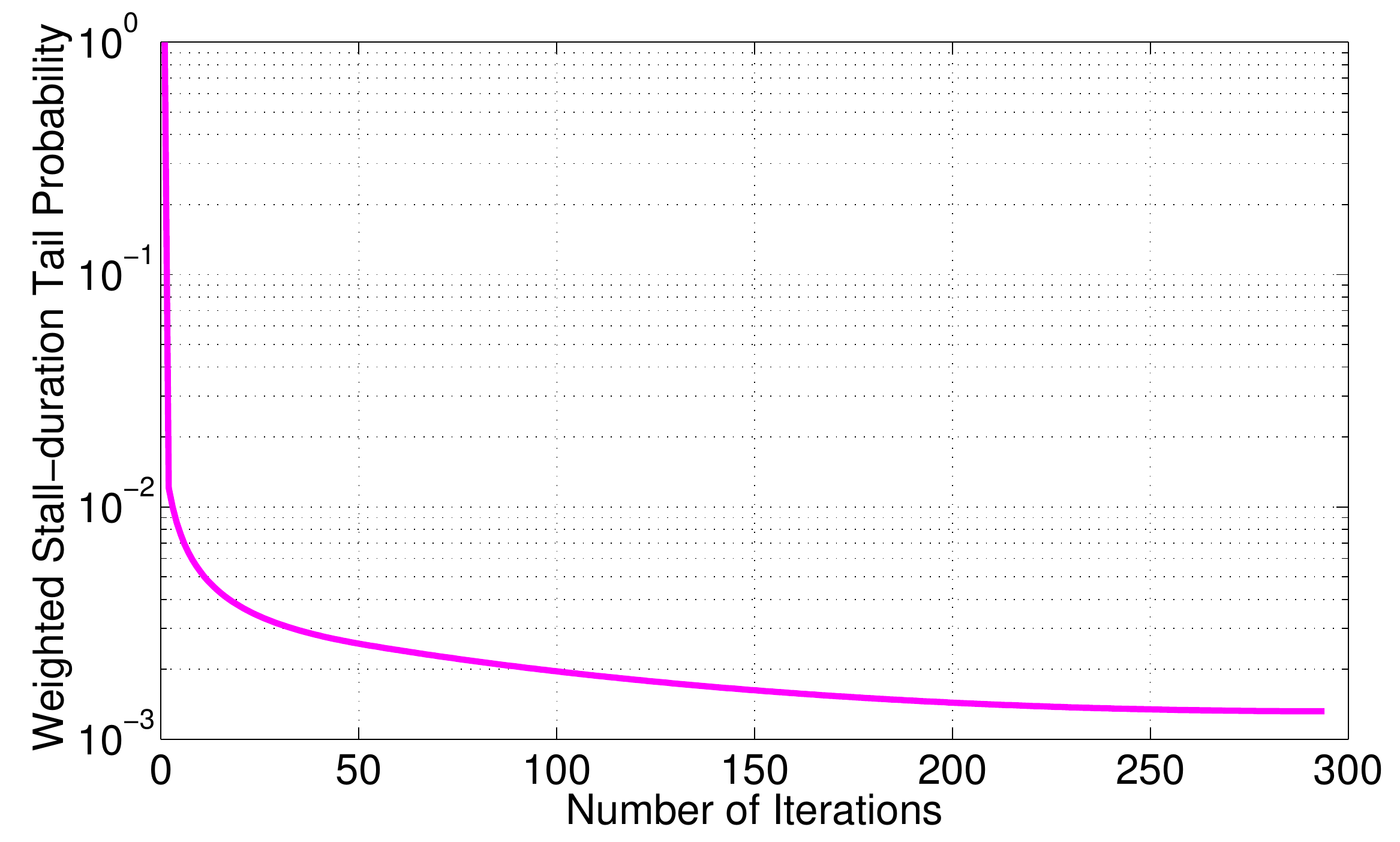}
\vspace{-.16in}
\captionof{figure}{ \small Convergence of weighted stall-duration tail probability.}
\label{fig:ConvgStalltail}
\end{minipage}%
\hspace{.05in}
\begin{minipage}{.32\textwidth}
	\centering
	\includegraphics[trim=0in 0in 5in 0.0in, clip, width=\textwidth]{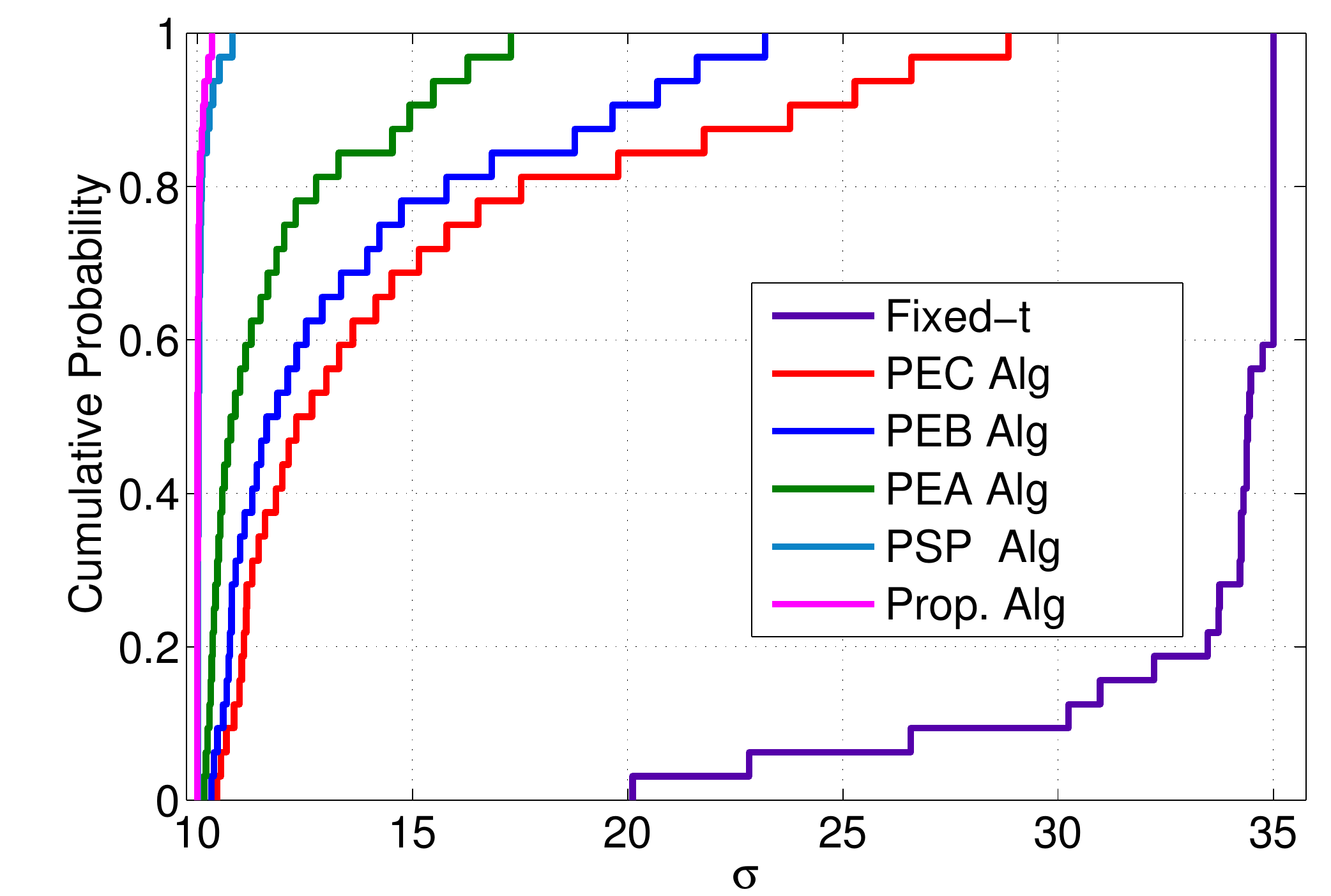}
\vspace{-.2in}
\captionof{figure}{ \small Cumulative Density Function of the weighted stall-duration tail probability.}
\label{fig:Cdf}
\end{minipage}
\hspace{.05in}
\begin{minipage}{.32\textwidth}
	\centering
\includegraphics[trim=0in 0in 4.8in 0.0in, clip, width=\textwidth]{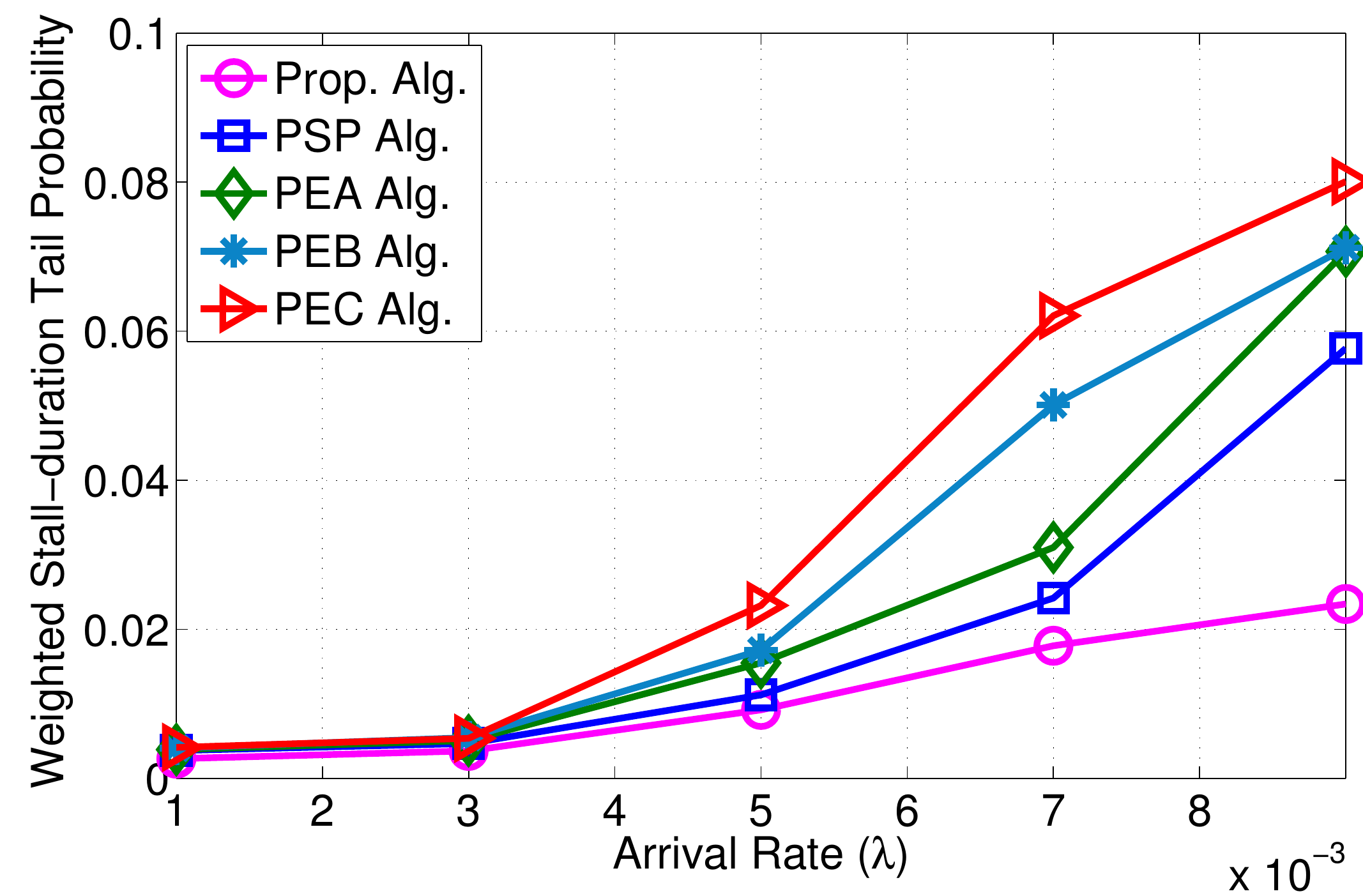}
\vspace{-.2in}
\captionof{figure}{ \small Weighted stall-duration tail probability versus arrival rate of video files. }%Here, we compare our proposed approach with the four considered baselines.}
\label{fig:arrRate}
\end{minipage}
\end{figure*}

%\begin{figure}[t]
%	\centering
%	\includegraphics[trim=0in 0in 0in 0.0in, clip, width=.6\textwidth]{figs_up/conveg_new3}
%	%\vspace{-.16in}
%	\captionof{figure}{Convergence of weighted stall-duration tail probability.}
%	\label{fig:ConvgStalltail}
%	%	\vspace{-.2in}
%\end{figure}
%
%
%
%\begin{figure}[t]
%	\centering
%	\includegraphics[trim=0in 0in 0in 0.0in, clip, width=.7\textwidth]{figs_up/cdf}
%	%\vspace{-.16in}
%	\captionof{figure}{Cumulative Density Function of the weighted stall-duration tail probability.}
%	\label{fig:Cdf}
%	%	\vspace{-.2in}
%\end{figure}
%
%
%
%\begin{figure}[t]
%	\centering
%	\includegraphics[trim=0in 0in 0in 0.0in, clip, width=.7\textwidth]{figs_up/WSDTP_vs_arrRate_allAlg}
%	%\vspace{-.16in}
%	\captionof{figure}{Weighted stall-duration tail probability versus arrival rate of video files. Here, we compare our proposed approach with the four considered baselines.}
%	\label{fig:arrRate}
%	%	\vspace{-.2in}
%\end{figure}
%

\subsection{Numerical Results}

\begin{figure*}[htbp]
	\begin{minipage}{.32\textwidth}
		\centering
		\includegraphics[trim=0in 0in 5in 0.0in, clip, width=\textwidth]{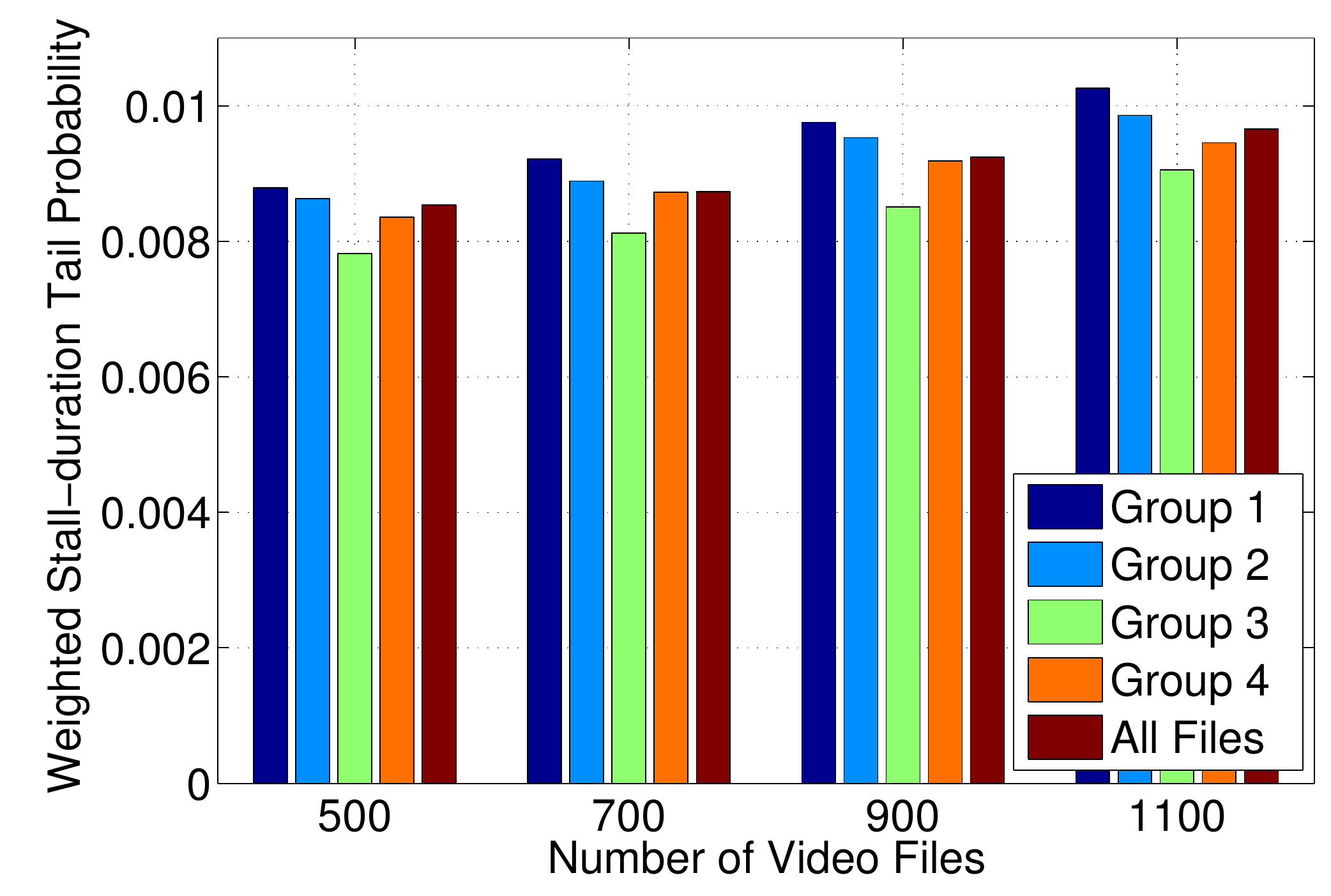}
%		\vspace{-.3in}
		\captionof{figure}{\small Weighted stall-duration tail probability for different number of files.
			We vary the number of files per group from 500 files to 1100 files with an
			increment step of 200 files. For each set, the base file arrival rate $\lambda_i$ is scaled
			by $4, 6, 12, 8$ respectively. }%Hence, the weights of the files in the objective function are proportionally increased, e.g., "Group 1" is weighted the lowest and "Group 3" is weighted the highest.}
		\label{fig:group}
	\end{minipage}%
	\hspace{.05in}
	\begin{minipage}{.32\textwidth}
		\centering
		\includegraphics[trim=0in 0in 5in 0.0in, clip, width=\textwidth]{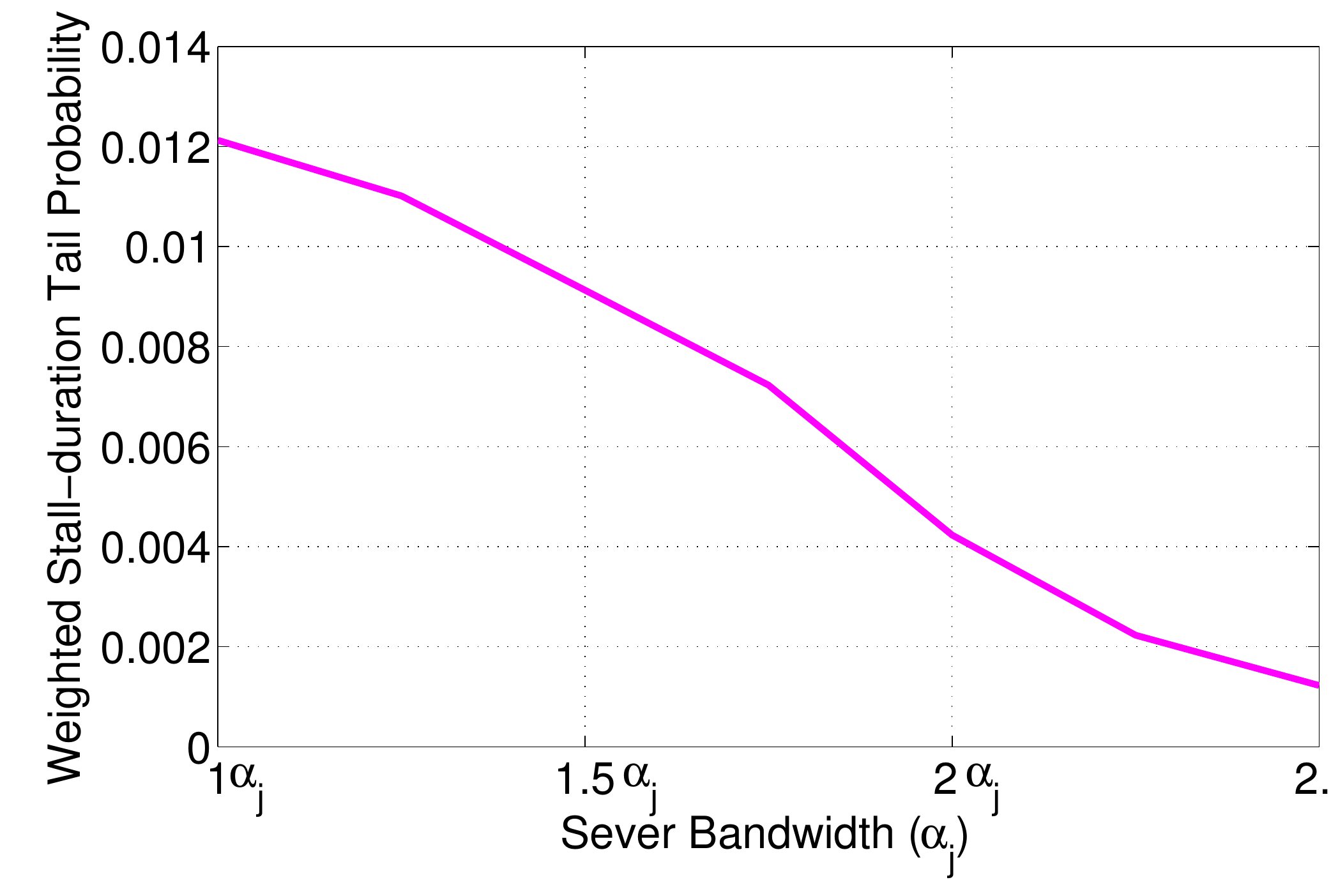}
%		\vspace{-.3in}
		\captionof{figure}{\small Weighted stall-duration tail probability for different scaling for the server bandwidth $\alpha_j$. We scale up $\alpha_j$ by $1, 1.25, 1.50, 1.75,  2.0$}. 
		\label{fig:alphaScaling}
	\end{minipage}
	\hspace{.05in}
	\begin{minipage}{.32\textwidth}
		\centering
		\includegraphics[trim=0in 0in 5.6in 0.0in, clip, width=\textwidth]{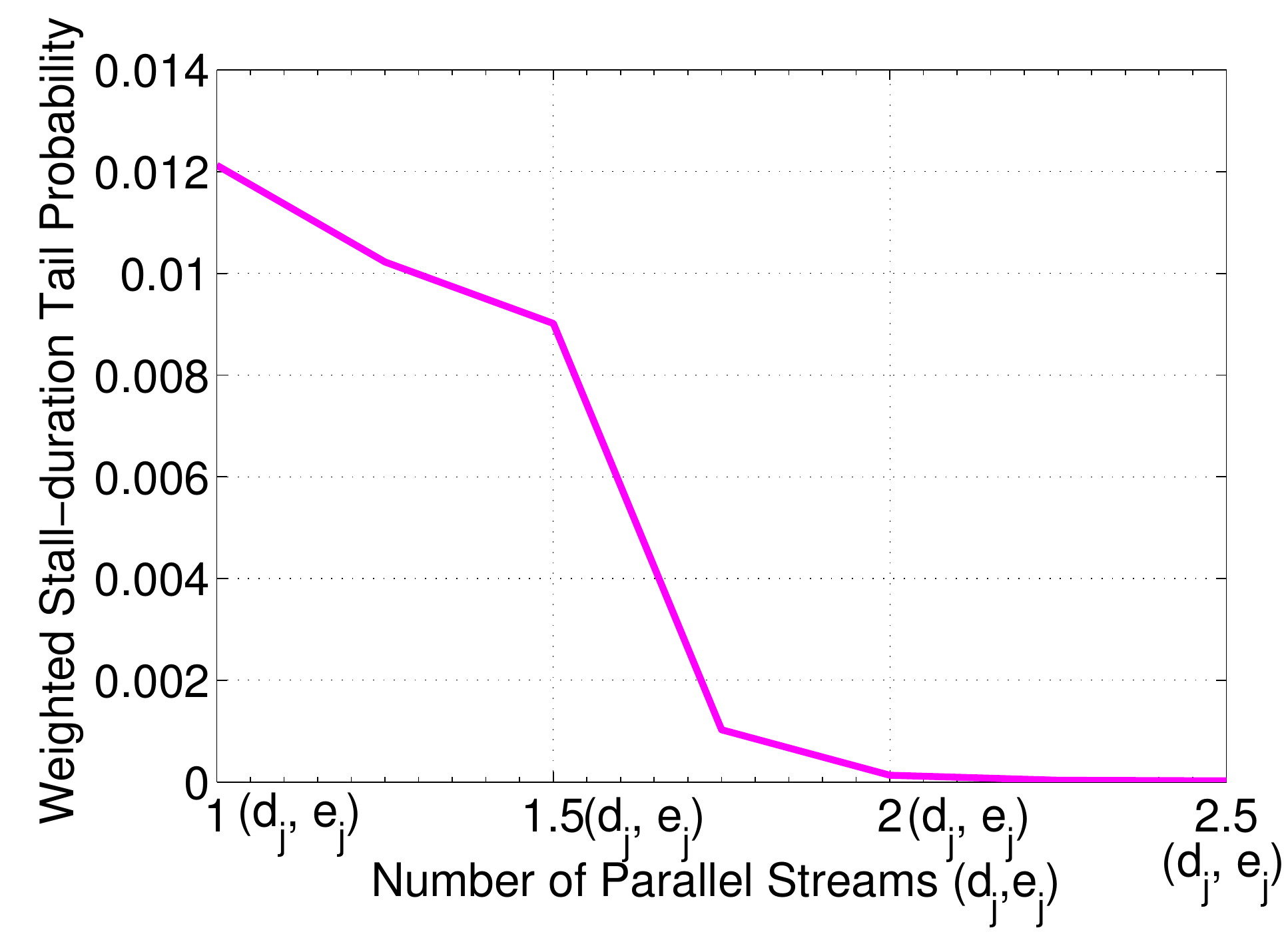}
%		\vspace{-.3in}
		\captionof{figure}{ \small Weighted stall-duration tail probability for different number of parallel streams $d_j$ and $e_j$. We scale up the number of parallel streams by $1, 1.25, 1.50, 1.75, 2.0$}
		\label{fig:PSs}
	\end{minipage}
%	 \vspace{-0.20in}
\end{figure*}

\subsubsection*{Convergence of the Proposed Algorithm} 
Figure \ref{fig:ConvgStalltail} shows the convergence of our proposed algorithm, which alternatively optimizes the weighted stall duration tail probability of all files over scheduling
probabilities $\widetilde{\boldsymbol{\pi}}$, auxiliary variables $\boldsymbol{t}$, bandwidth allocation weights $\boldsymbol{w}$ and cache placement $\boldsymbol{L}$. We
see that for $r = 1000$ video files of size $600$s with
$m = 12$ cache storage nodes, the weighted stall duration tail probability converges to
the optimal value within less than 300 iterations.

\subsubsection*{Weighted  SDTP} 

In Figure \ref{fig:Cdf}, we plot the cumulative distribution function (CDF)  of weighted stall duration tail probability with $\sigma$ (in seconds) for different strategies including  our proposed algorithm,
PSP algorithm, PEA algorithm, PEB algorithm, PEC algorithm, and Fixed t algorithm. We note that our proposed algorithm for jointly optimizing $\ensuremath{\widetilde{\boldsymbol{\pi}}}$,
$\ensuremath{\boldsymbol{w}}$, $\boldsymbol{t}$ and $\boldsymbol{L}$
provides significant improvement over considered strategies as weighted stall duration tail probability reduces by orders of magnitude. For example, our proposed algorithm shows that the weighted stall-duration tail probability will not exceed $11$s, which is much lower when comparing to other considered strategies. Further, uniformly accessing servers, equally allocating bandwidth and cache are unable to optimize the request scheduler based on factors like cache placement, request arrival rates, and different stall weights, thus leading to much higher stall duration tail probability. Since the Fixed t policy performs significantly
worse than the other considered policies, we do not include this policy in the rest of the paper.

\subsubsection*{Effect of Arrival Rates} Figure \ref{fig:arrRate} shows the effect of increasing system workload, obtained by varying the arrival rates of the video files from $0.25\lambda$ to $2\lambda$, where $\lambda$ is the base arrival rate, on the stall duration tail probability for video lengths generated based on Pareto distribution defined above.
We notice a significant improvement in the QoE metric with the proposed strategy as compared to the baselines. For instance, at the
arrival rate of $10\lambda_b$, where $\lambda_b$ is the base arrival rate defined above, the proposed strategy reduces the weighted stall duration tail probability by about $200\%$ as compared to the nearest strategy, i.e., PSP Algorithm.

\subsubsection*{Effect of Video File Weights on the Weighted SDTP}
While weighted stall duration tail probability increases as arrival
rate increases, our algorithm assigns differentiated latency for different
video file groups as captured in Figure \ref{fig:group} to maintain the QoE at a
lower value. Group 3 that has highest weight $w_{3}$ (i.e., most
tail stall sensitive) always receive the minimum stall duration tail
probability even though these files have the highest arrival rate.
Hence, efficiently reducing the stall tail probability of the high
arrival rate files which accordingly reduces the overall weighted
stall duration tail probability. In addition, we note that efficient
server/PSs access probabilities $\ensuremath{\widetilde{\boldsymbol{q}}}$
help in differentiating file latencies as compared to the strategy
where minimum queue-length servers are selected to access the content
obtaining lower weighted tail latency probability.

\subsubsection*{Effect of Scaling up bandwidth of the Cache Servers and Datacenter}

We show the effect of increasing the server bandwidth on the weighted
stall duration tail probability in Figure \ref{fig:alphaScaling}. Intuitively, increasing
the storage node bandwidth will increase the service rate of the
storage nodes thus reducing the weighted stall duration tail probability.

\subsubsection*{Effect of the Parallel Connections $d_j$ and $e_j$}

To study the effect of the number of the parallel connections, we
plot in Figure \ref{fig:PSs} the weighted stall duration tail probability for varying the number
of parallel streams, $d_{j}$ and $e_{j}$ for our proposed algorithm.
We vary the number of PSs from the $d_{j}=20$, $e_{j}=40$ to $d_{j}=50$,
$e_{j}=100$ with increment step of $5$, and $10$, respectively.
We can see that increasing $d_{j}$ and $e_{j}$ improve the performance
since some of the bandwidth splits can be zero thus giving the lower
$d_{j}(e_{j})$ solution as one of the possible feasible solution.
Increasing the number of PSs results in decreasing the stall durations since more video files can be
streamed concurrently. We note that for $d_{j}\geq40$ and $e_{j}\geq80$,
the weighted stall duration tail probability is almost zero. However,
the streaming servers may only be able to handle a limited number
of parallel connections which limit the choice of both $e_{j}$ and
$d_{j}$ in the real systems.

%\begin{figure}[t]
%	\centering
%	\includegraphics[trim=0in 0in 0in 0.0in, clip, width=.7\textwidth]{figs_up/WSDTP_versus_arrRate_ours}
%			%\vspace{-.16in}
%		\captionof{figure}{Weighted stall-duration tail probability for different number of files.
%We vary the number of files per group from 500 files to 1100 files with an
%increment step of 200 files. For each set, the base file arrival rate $\lambda_i$ is scaled
%by $4, 6, 12, 8$ respectively. Hence, the weights of the files in the objective function are proportionally increased, e.g., "Group 1" is weighted the lowest and "Group 3" is weighted the highest.}
%		\label{fig:group}
%	%	\vspace{-.2in}
%\end{figure}
%
%
%\begin{figure}[t]
%	\centering
%	\includegraphics[trim=0in 0in 0in 0.0in, clip, width=.7\textwidth]{figs_up/wsdtp_vs_serverBW}
%			%\vspace{-.16in}
%		\captionof{figure}{Weighted stall-duration tail probability for different scaling for the server bandwidth $\alpha_j$. We scale up $\alpha_j$ by $1, 1.25, 1.50, 1.75,  2.0$}. 
%		\label{fig:alphaScaling}
%	%	\vspace{-.2in}
%\end{figure}
%
%
%\begin{figure}[t]
%	\centering
%	\includegraphics[trim=0in 0in 0in 0.0in, clip, width=.8\textwidth]{figs_up/wsdtp_vs_pllstreams}
%			%\vspace{-.16in}
%		\captionof{figure}{Weighted stall-duration tail probability for different number of parallel streams $d_j$ and $e_j$. We scale up the number of parallel streams by $1, 1.25, 1.50, 1.75, 2.0$}
%		\label{fig:PSs}
%	%	\vspace{-.2in}
%\end{figure}
%
%
\subsection{Testbed Configuration}
\label{testbed}
% Testbed configuration
\begin{table}[t]
	   \caption{\small Testbed Configuration}
  \small
   \centering
   %\topcaption{Table captions are better up top} % requires the topcapt package
  \resizebox{.48\textwidth}{!}{
%\begin{tabular*}{.5\textwidth}{@{} llr @{}}
   \begin{tabular}{@{} llr @{}} % Column formatting, @{} suppresses leading/trailing space
      \hline
      \multicolumn{2}{c}{Cluster Information}         \\
      %\cmidrule(r){1-2} % Partial rule. (r) trims the line a little bit on the right; (l) & (lr) also possible
      \hline
      Control Plane     &       Openstack Kilo    \\ 
      VM flavor                 &     1 VCPU, 2GB RAM, 20G storage (HDD)  \\ 
      \hline
      \hline
      \multicolumn{2}{c}{Software Configuration} \\ 
      \hline
      Operating System       &  Ubuntu Server 16.04 LTS \\ 
      Origin Server(s)          &  Apache Web Server~\cite{apacheweb}: Apache/2.4.18 (Ubuntu)     \\ 
      Cache Server(s)           & Apace Traffic Server~\cite{trafficserv} 6.2.0 (build \# 100621)    \\ 
      Client                & Apache JMeter~\cite{jmeter} with HLS plugin~\cite{hlsplugin} \\
      \hline
   \end{tabular}
}
%   \caption{\small Testbed Configuration}
   \label{tbl:testbed}
%   \vspace{-0.3in}
\end{table}

\begin{figure}[ht]
	\centering
	\includegraphics[viewport=50 110 710 500, scale=0.35, clip=true]{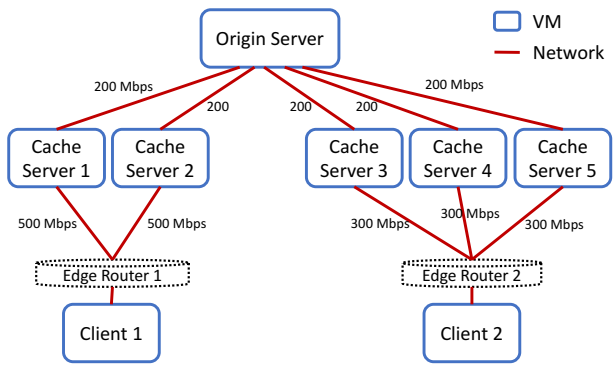}
%	\vspace{-0.18in}
	\captionof{figure}{\small Testbed in the cloud}.
	\label{fig:alphaScaling}
%	\vspace{-0.4in}
\end{figure}

We constructed an experimental environment in a virtualized cloud environment 
managed by Openstack~\cite{openstack}. 

We allocated one VM for an origin server and 5 VMs for cache servers intended
to simulate two locations (i.e., different states). The schematic of our testbed is illustrated in Figure~\ref{fig:alphaScaling} .   
One VM per location is used for generating client workloads. 
Table~\ref{tbl:testbed} summarizes a detailed configuration used for the experiments.  
For client workload, we exploit a popular HTTP-trafic generator, Apache JMeter, with a plug-in 
that can generate traffic using HTTP Streaming protocol. 
We assume the amount of available bandwidth between origin server and each cache server is 200 Mbps, 
500 Mbps between cache server 1/2 and edge router 1, and 300 Mbps between cache server 3/4/5 and edge router 2. 
In this experiments, to allocate bandwidth to the clients, we throttle the client (i.e., JMeter) traffic according to the plan generated by our algorithm.  
We consider $500$ threads (i.e., users) and set $e_j=40$, $d_j=20$. Based on one week trace from our production system, we estimate the aggregate arrival rates at edge router 1 and router 2 to be $\Lambda_1 = 0.01455s^{-1}$, $\Lambda_2 = 0.02155s^{-1}$, respectively. Then, HLS sampler (i.e., request) is sent every $3$s. We assume $40\%$ of the segments are stored in the cache and hence the remaining segments are servers from origin server. The video files are $300$s of length and the segment length is set to be $8$s. For each segment, we used JMeter built-in reports to estimate the downloaded time of each segment and then plug these times into our model to get the SDTP.

\begin{figure}[ht]
	\centering\includegraphics[scale=0.40]{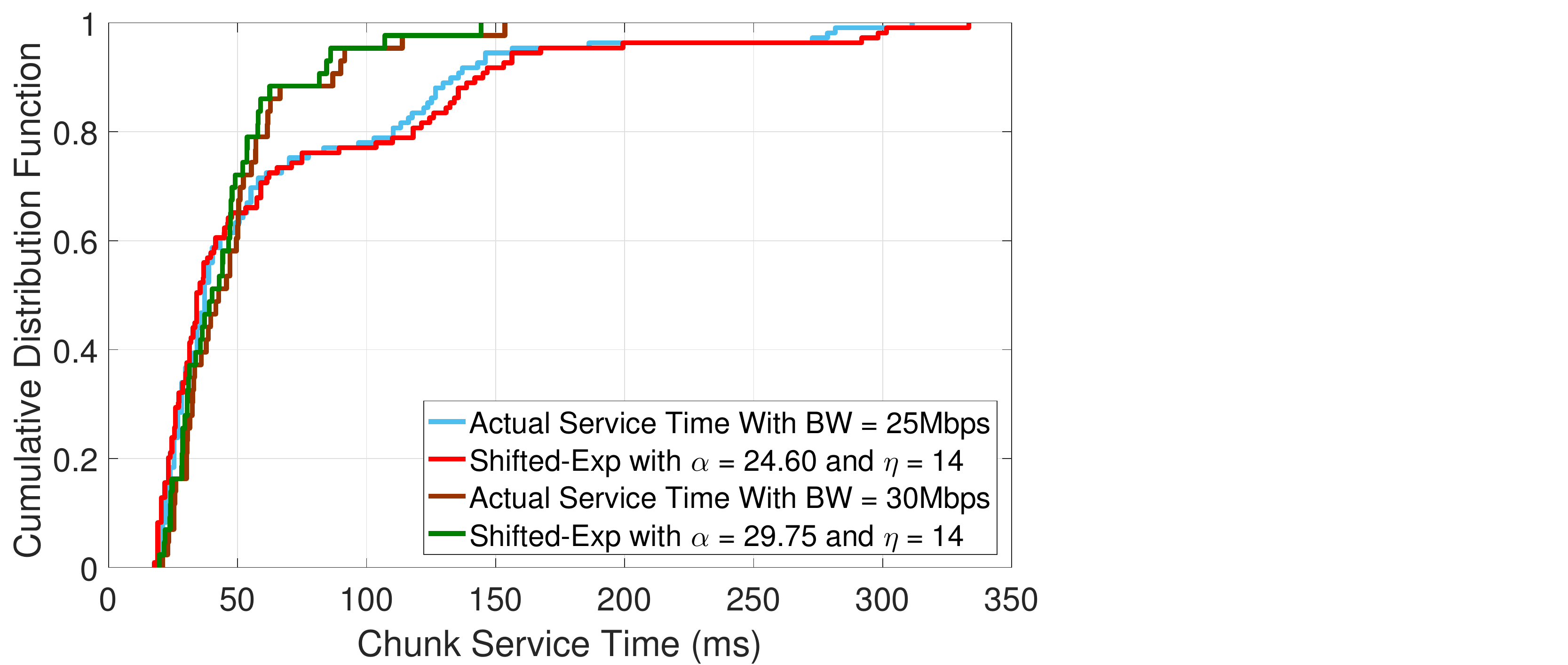}
%	\vspace{-.45in}
	\caption{\small
		Comparison of actual chunk service time distribution and shifted-exponential distribution with the corresponding mean and shift. It verifies that the actual service time of a chunk can be well approximated by a shifted exponential distribution.
		\label{fig:serTimeFig}}
%	\vspace{-.25in}
\end{figure}

\textit{Service Time Distribution:} We first run  experiments to measure the actual service time distribution in our cloud environment. Figure \ref{fig:serTimeFig} depicts the cumulative distribution function (CDF) of the chunk service
time for different bandwidths. Using these results, we show that the service time of the chunk can be well approximated by a shifted-exponential distribution with a rate of 24.60s, 29.75s for a bandwidth of 25 Mbps and 30 Mbps, respectively. These results also verify that actual service time does not follow an exponential distribution. This observation has also been made earlier in \cite{Yu-TON16}. Further, the parameter for the exponential is almost proportional to the bandwidth while the shift is nearly constant, which validates the model.

%This observation is also evident because a typical real service distribution is unlikely to have positive probabilities for even very small service times.

\begin{figure}[ht]
%\centering\includegraphics[scale=0.50]{figs_up/cdf_actual_impl_v3}
\centering\includegraphics[trim=0.01in 0.25in 1.9in 0.0in, clip, width=0.49\textwidth]{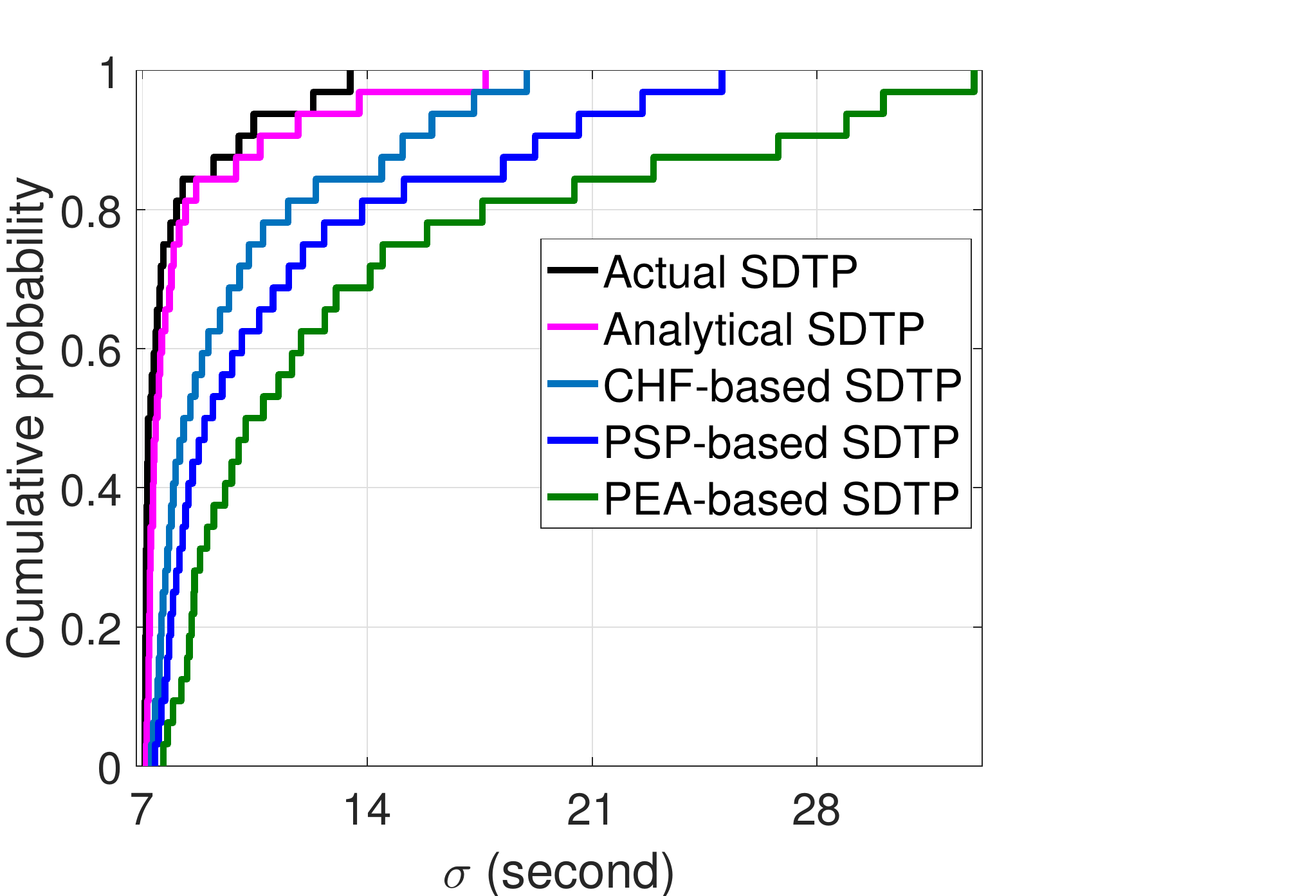}
%\vspace{-.45in}
\caption{\small Comparison of implementation results of our SDTP Algorithm to Analytical SDTP and PEA-based SDTP.
\label{fig:impFig}}
%\vspace{-.25in}
\end{figure}

\textit{SDTP Comparisons:} Figure \ref{fig:impFig} shows four different policies where we compare the actual SDTP, analytical SDTP, PSP, and PEA based SDTP algorithms. We see that the analytical SDTP is very close to the actual SDTP measurement on our testbed. To the best of our knowledge, this is the first work to jointly consider all key design degrees of freedom, including bandwidth allocation among different parallel streams, cache content placement,  the request scheduling, and the modeling variables associated with the SDTP bound.

\begin{figure}[ht]
	\centering\includegraphics[scale=0.29]{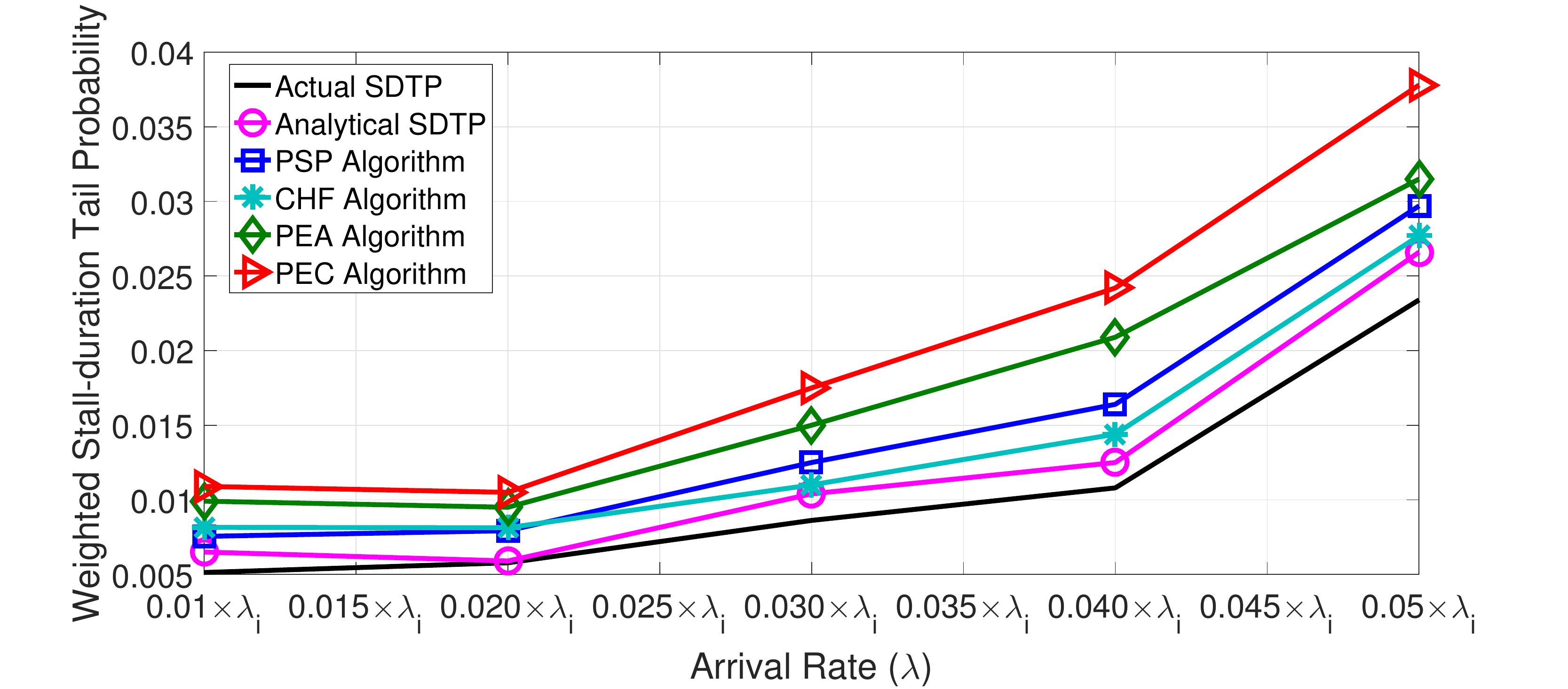}
%	\vspace{-.45in}
	\caption{\small Comparison of implementation results of our SDTP algorithm to analytical SDTP and PEA-based SDTP when the arrival rate is varied from $0.01$ to $0.05$ with an increment step of $0.005$.
		\label{fig:arrRateFig}}
%	\vspace{-.25in}
\end{figure}

\textit{Arrival Rates Comparisons:} Figure \ref{fig:arrRateFig} shows the effect of increasing system workload, obtained by varying the arrival rates of the video files from $0.01s^{-1}$ to $0.05s^{-1}$ with an increment step of $0.005s^{-1}$ on the stall duration tail probability. We notice a significant improvement of the QoE metric with the proposed strategy as compared to the baselines. Further, the gap between the analytical bound and actual SDTP is small which validates the tightness of our proposed SDTP bound. 

\begin{figure}[ht]
	\centering\includegraphics[scale=0.39]{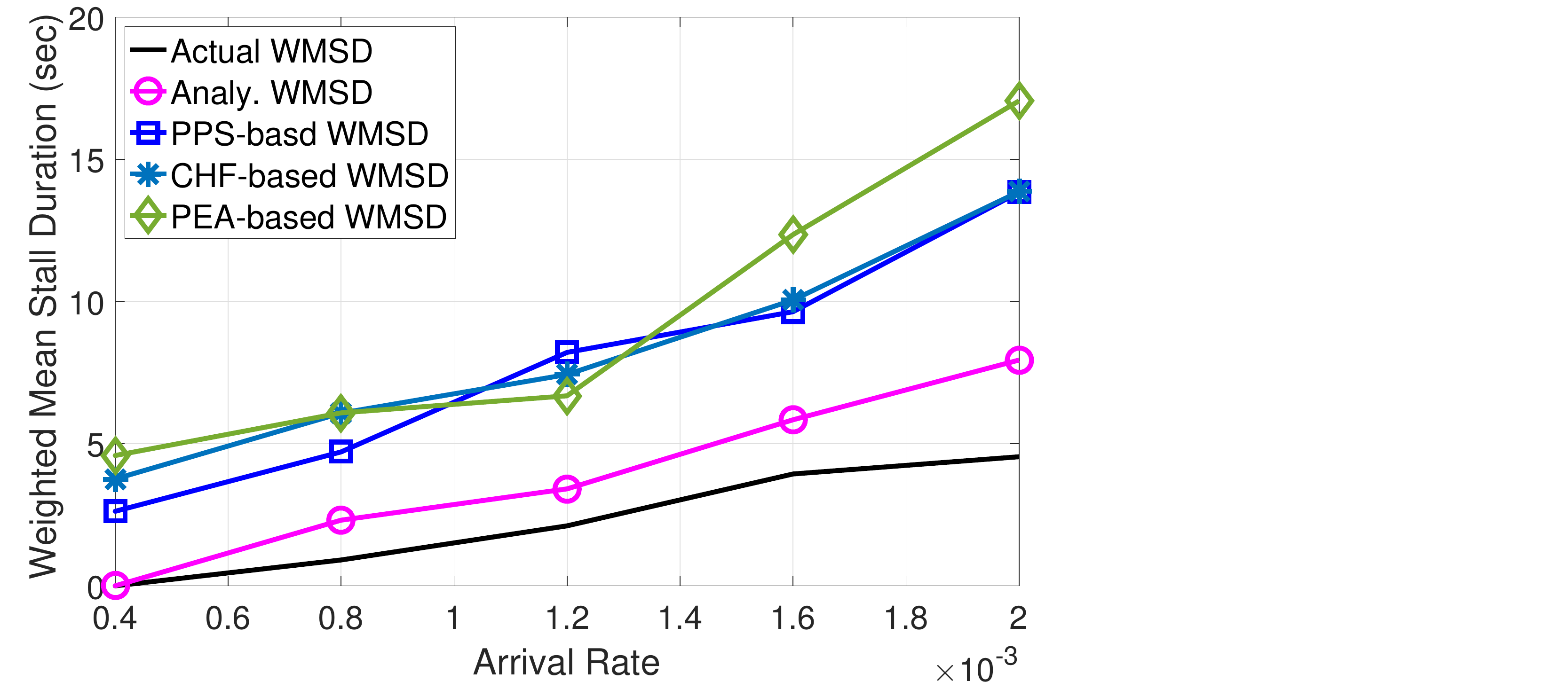}
	%	\vspace{-.45in}
	\caption{\small Weighted mean stall duration versus the arrival rate for different policies.
		\label{fig:WMSD}}
	%	\vspace{-.25in}
\end{figure}

\textit{Mean Stall Duration Comparisons:} We further plots the weighted mean stall duration (WMSD) in Figure \ref{fig:WMSD}. As expected, the proposed approach achieves the lowest stall durations and the gap between the analytical and experimental results is small and thus the proposed bound is tight. Also, caching hottest files does not help much since caching later segments is not necessary as they can be downloaded when the earlier segments are being played. Thus, prioritizing earlier segments over later ones for caching is more helpful in reducing the stalls than caching complete video files.

%{\color{blue} need to talk about how we generate workload in detail, e.g., parse 1 month trace from production systems, extract a certain metrics and apply to etc, etc. then need to discuss our methodology, e.g., what the goal was and how we did exp., comparing the results from non-caching scenario to one for our scenario, in order to show X.}  

%\begin{figure}[ht]
%	\centering\includegraphics[scale=0.30]{newFigs/NoVideos_expm}
%%	\vspace{-.45in}
%	\caption{\small Comparison of implementation results of our SDTP Algorithm to Analytical SDTP and PEA-based SDTP.
%		\label{fig:impFig}}
%%	\vspace{-.25in}
%\end{figure}
\section{Conclusion}\label{sec:conc}

In this paper, a CDN-based Over-the-top video streaming  is studied, where the video content is partially stored on distributed cache servers. We consider optimizing the weighted stall duration tail probability by considering two-stage probabilistic scheduling for the choice of servers and the parallel streams between the server and the edge router. Using the two-stage probabilistic scheduling, upper bound on the stall duration tail probability is characterized.  Further, an optimization problem that minimizes the weighted stall duration tail probability is formulated, over the choice of two-stage probabilistic scheduling, bandwidth allocation, cache placement, and auxiliary variables.  An efficient algorithm is proposed to solve the optimization problem and the experimental results depict the improved performance of the algorithm as compared to the considered baselines.

%\newpage
%\clearpage
%\bibliographystyle{plain}
%\bibliographystyle{ACM-Reference-Format}
\bibliographystyle{IEEEtran}
%\pagebreak{}
\bibliography{vidStallRef,allstorage,Tian,ref_Tian2,ref_Tian3,Vaneet_cloud,Tian_rest}
%\vspace{-.6in}

\begin{IEEEbiography}[{\includegraphics[width=1in,height=1.25in,clip,keepaspectratio]{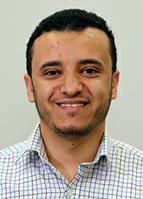}}]{Abubakr O. Al-Abbasi (S'11)} received the B.Sc. and M.Sc. degrees in electronics and electrical communications engineering from Cairo University, Cairo, Egypt, in 2010, and 2014, respectively. He is currently pursuing the Ph.D. degree with Purdue University, USA. From 2011 to 2012, he was a Communications and Networks Engineer with Huawei Company. From 2014 to 2016, he was a Research Assistant with Qatar University, Doha, Qatar. His research interests are in the areas of wireless communications and networking, media streaming, heterogeneous wireless networks, compressive sensing with applications to communications, and signal processing for communications.
\end{IEEEbiography}

%\vspace{-.5in}

\begin{IEEEbiography}[{\includegraphics[width=1in,height=1.25in,clip,keepaspectratio]{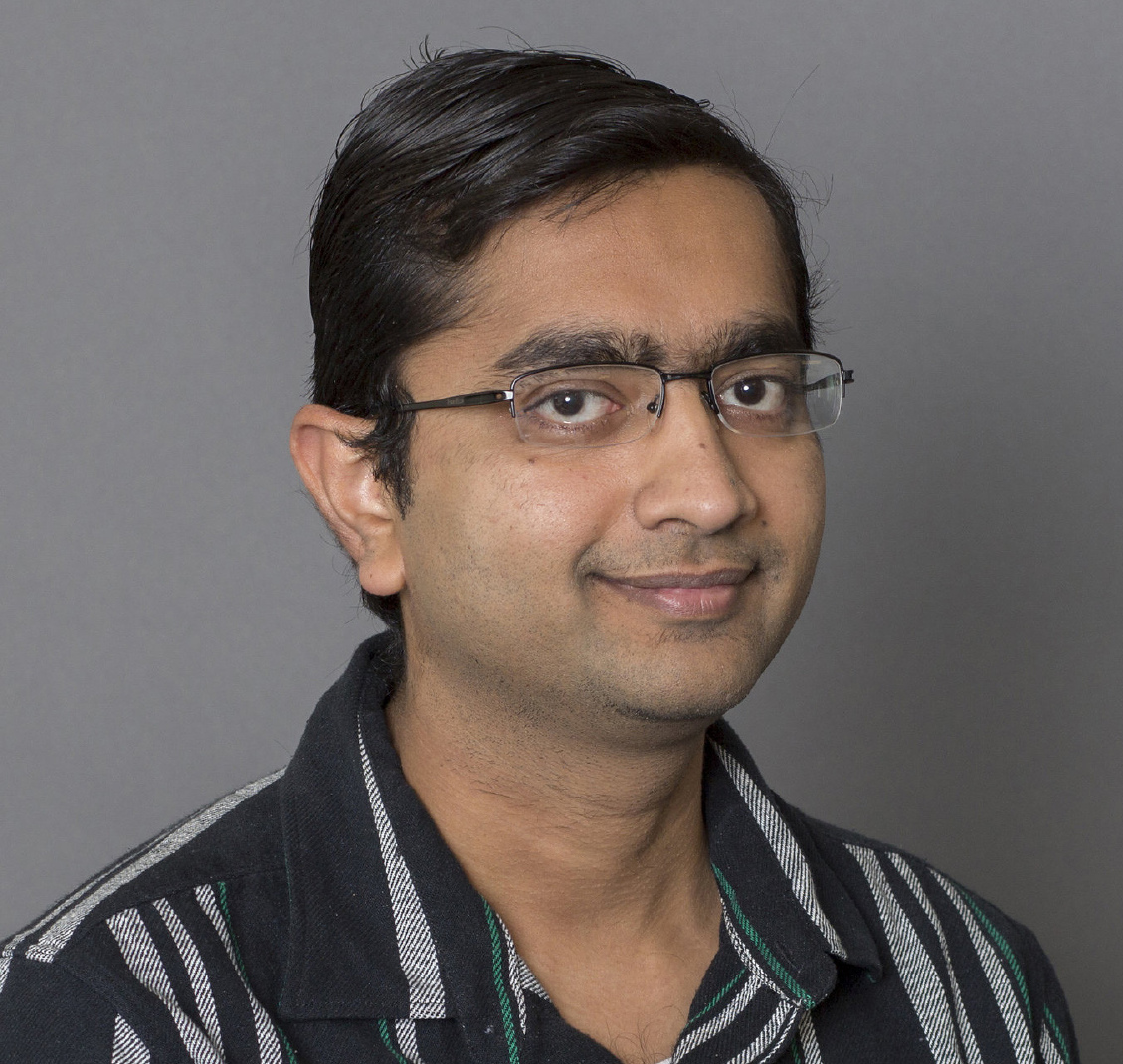}}]{Vaneet Aggarwal (S'08 - M'11 - SM'15)}
	received the B.Tech. degree in 2005 from the Indian Institute of Technology, Kanpur, India, and the M.A. and Ph.D. degrees in 2007 and 2010, respectively from Princeton University, Princeton, NJ, USA, all in Electrical Engineering.
	
	He is currently an Assistant Professor at Purdue University, West Lafayette, IN. Prior to this, he was a Senior Member of Technical Staff Research at AT\&T Labs-Research, NJ, and an Adjunct Assistant Professor at Columbia University, NY. His research interests are in applications of statistical, algebraic, and optimization techniques to distributed storage systems, machine learning,  and wireless systems. Dr. Aggarwal was the recipient of Princeton University's Porter Ogden Jacobus Honorific Fellowship in 2009. He also received the 2017 Jack Neubauer Memorial Award recognizing the Best Systems Paper published in the {\em IEEE Transactions on Vehicular Technology}. He is serving on the editorial board of the {\em IEEE Transactions on Communications} and the {\em IEEE Transactions on Green Communications and Networking}. 
\end{IEEEbiography}
\vspace{-.5in}

\begin{IEEEbiography}[{\includegraphics[width=1in,height=1.25in,clip,keepaspectratio]{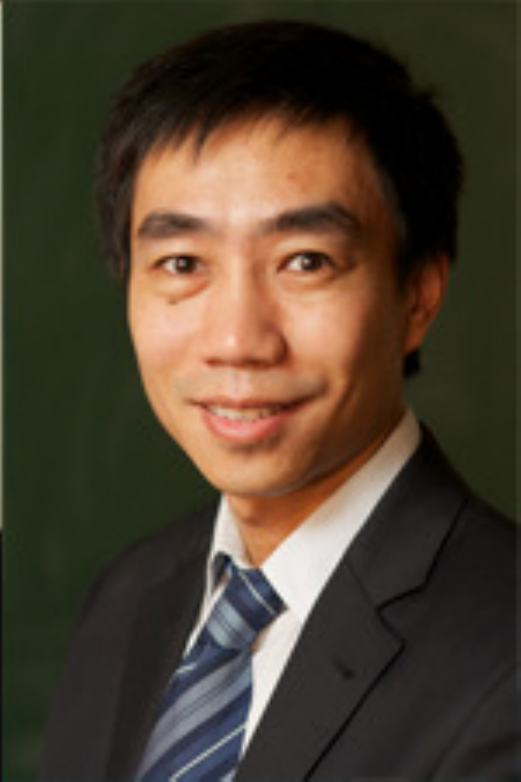}}]{Tian Lan (S'03-M'10)} received the B.A.Sc. degree from the Tsinghua University, China in 2003, the M.A.Sc. degree from the University of Toronto, Canada, in 2005, and the Ph.D. degree from the Princeton University in 2010. Dr. Lan is currently an Associate Professor of Electrical and Computer Engineering at the George Washington University. His research interests include cloud resource optimization, mobile networking, storage systems and cyber security. Dr. Lan received the 2008 IEEE Signal Processing Society Best Paper Award, the 2009 IEEE GLOBECOM Best Paper Award, and the 2012 INFOCOM Best Paper Award.
\end{IEEEbiography}

\vspace{-.5in}

\begin{IEEEbiography}[{\includegraphics[width=1in,height=1.25in,clip,keepaspectratio]{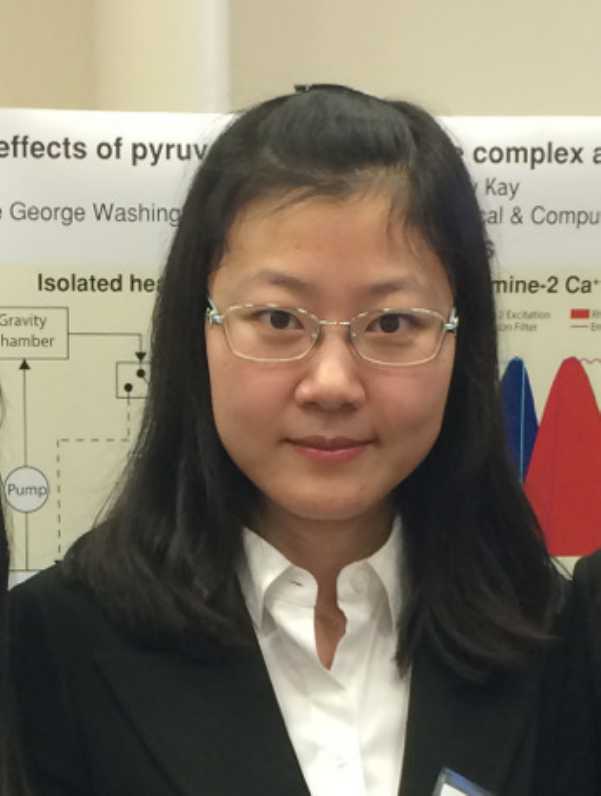}}]{Yu Xiang} received the B.A.Sc. degree from Harbin Institute of Technology in 2010, and the Ph.D. degree from George Washington University in 2015, both in Electrical Engineering. She is now a senior inventive scientist at AT\&T Labs-Research. Her current research interest are in cloud resource optimization, distributed storage systems and cloud storage charge-back.
\end{IEEEbiography}

\vspace{-.5in}

\begin{IEEEbiography}[{\includegraphics[width=1in,height=1.25in,clip,keepaspectratio]{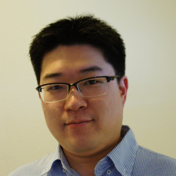}}]{Moo-Ryong Ra}
Moo-Ryong Ra is a principal inventive scientist at AT\&T Labs Research. He is interested in the broad area of systems and networking. In AT\&T, more focus is given to the following areas -- software-defined storage for cloud platform/infrastructure in the context of AT\&T integrated cloud and edge cloud, video storage and delivery, RDMA networking for next generation storage/memory, etc. Before joining AT\&T, he had built several interesting cloud-enabled mobile sensing systems to better understand the interaction between smart mobile devices and the cloud infrastructure. He earned a Ph.D. degree from Computer Science Department at University of Southern California. Contact him at mra@research.att.com. 
\end{IEEEbiography}
\vspace{-.5in}

\begin{IEEEbiography}[{\includegraphics[width=1in,height=1.25in,clip,keepaspectratio]{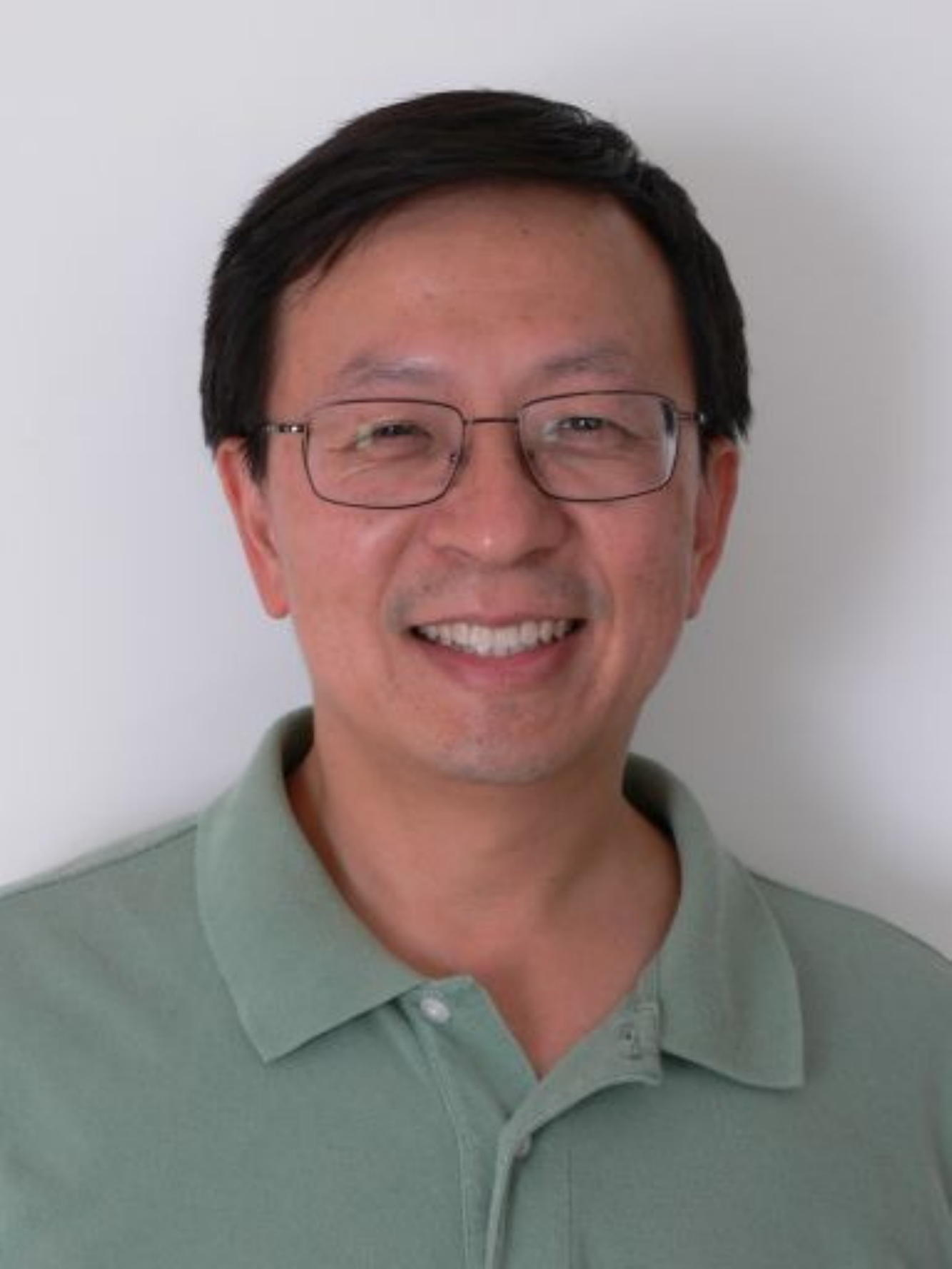}}]{Yih-Farn (Robin) Chen}
	is a Director of Inventive Science, leading the Cloud Platform Software Research department at AT\&T Labs  - Research.  His current research interests include cloud computing, software-defined storage, mobile computing, distributed systems, World Wide Web, and IPTV.  He holds a Ph.D. in Computer Science from University of California at Berkeley, an M.S. in Computer Science from University of Wisconsin, Madison,  and a B.S. in Electrical Engineering from National Taiwan University. Robin is an ACM Distinguished Scientist and a Vice Chair of the International World Wide Web Conferences Steering Committee (IW3C2).  He also serves on the editorial board of IEEE Internet Computing.
\end{IEEEbiography}

\newpage
\clearpage
\clearpage
\setcounter{page}{1}
\appendices
%\section{Appendices}
%\section{Proof of Lemma \label{tailLemma}}

%\section{A schematic of Our TestBed }

\section{Proof of Lemma \ref{PoisArr2ndQ} }
\label{2ndQArr}

We note  that the arrivals at the econd queue in  two $M/M/1$ tandem networks are Poisson is well known in the queuing theory literature \cite{gross011}. The service distribution of the first queue in this paper is a shifted exponential distribution. We note that the deterministic shift also does not change the distribution since the number of arrivals in any time window in the steady state will be the same. Thus, the arrival distribution in the second queue will still be Poisson. 

%
%%Here, we need to show that the arrivals still Poisson when the service time is shifted exponential. 
%
%
%%Let $X(t)$ denote the number of customers in the system at time $t$. 
%
%
%%Also, let $SE$ denote the shifted-exponential distribution. 
%
%%Since the $M/SE/1$ process is a birth and death process, it follows that ${X(t), t \geq 0}$ is time reversible. Going forward in time, the time epochs at which $X(t)$ increases by $1$ constitute a Poisson process since these are just the arrival times of customers of an exponential with shifted mean \cite{gross011}. Hence, by time reversibility the time points at which
%$%X(t)$ increases by $1$ when we go backward in time also constitute a Poisson
%%process. But these latter epochs are exactly the points of time when customers
%%depart. Hence, the departure times constitute a Poisson process with rate $\lambda$.
%

\section{Proof of Lemma \ref{tailLemma} }

\label{lemmaStall}

We have 
%\begin{align}
%\mathbb{E}\left[e^{t_iD_{i,j,\beta_{j},\nu_{j}}^{(v)}}\right] & =\mathbb{E}\left[e^{t_i(\max_{y=L_{j,i}}^v U_{i,j,\beta_{j},g,y})}\right]\nonumber \\
% & =\mathbb{E}\left[\max_{y=L_{j,i}}^v\,\,e^{t_iU_{i,j,\beta_{j},g,y}}\right]\nonumber \\
% & \leq\sum_{y=L_{j,i}}^ve^{t_iU_{i,j,\beta_{j},g,y}}\nonumber \\
% & =\left[e^{t_iU_{i,j,\beta_{j},g,L_{j,i}}}\right]+\sum_{w=L_{j,i}+1}^{v}\frac{(1-\rho_{j,\beta_{j}}^{(d)})t_iB_{j,\beta_{j}}^{(d)}(t_i)}{t_i-\Lambda_{j,\beta_{j}}^{(d)}(B_{j,\beta_{j}}^{(d)}(t_i)-1)}\left(\frac{\alpha_{j,\beta_{j}}^{(d)}}{\alpha_{j,\beta_{j}}^{(d)}-t_i}\right)^{w-L_{j,i}-1}\left(\frac{\alpha_{j,\beta_{j}}^{(c)}}{\alpha_{j,\beta_{j}}^{(c)}-t_i}\right)^{v-w+1}
%\end{align}

\begin{align}
\mathbb{E}\left[e^{t_{i}D_{i,j,\beta_{j},\nu_{j}}^{(v)}}\right] & =\mathbb{E}\left[e^{t_{i}(\max_{y=L_{j,i}}^{v}U_{i,j,\beta_{j},g,y})}\right]\nonumber\\
& =\mathbb{E}\left[\max_{y=L_{j,i}}^{v}\,\,e^{t_{i}U_{i,j,\beta_{j},g,y}}\right]\nonumber\\
& \leq\sum_{y=L_{j,i}}^{v}e^{t_{i}U_{i,j,\beta_{j},g,y}}\nonumber\\
& =\left[e^{t_{i}U_{i,j,\beta_{j},g,L_{j,i}}}\right]+\nonumber\\
& \sum_{w=L_{j,i}+1}^{v}\frac{(1-\rho_{j,\beta_{j}}^{(d)})t_{i}B_{j,\beta_{j}}^{(d)}(t_{i})}{t_{i}-\Lambda_{j,\beta_{j}}^{(d)}(B_{j,\beta_{j}}^{(d)}(t_{i})-1)}\times\nonumber\\
& \left(\frac{\alpha_{j,\beta_{j}}^{(d)}e^{\eta_{j,\beta_{j}}^{(d)}}}{\alpha_{j,\beta_{j}}^{(d)}-t_{i}}\right)^{w-L_{j,i}-1}\left(\frac{\alpha_{j,\beta_{j}}^{(\overline{d})}e^{\eta_{j,\beta_{j}}^{(\overline{d})}}}{\alpha_{j,\beta_{j}}^{(\overline{d})}-t_{i}}\right)^{v-w+1}
 \label{dt_v_apend}
\end{align}

Using some algebraic manipulations, we can further simplify (\ref{dt_v_apend}) to obtain the following. 

\begin{align*}
 & \text{\text{Pr}}\left(\Gamma^{(i)}\geq\sigma\right)\leq\\
& \left.\qquad\qquad\sum_{v=1}^{L_{i}}e^{(\widetilde{\sigma}-\overline{\sigma})t_{i}}\mathbb{E}\left[e^{t_{i}D_{i,j,\beta_{j},\nu_{j}}^{(v)}}\right]\right]\\
& =\sum_{j=1}^{m}\pi_{i,j}\left[e^{-t_{i}\sigma}+\sum_{\nu_{j}=1}^{e_{j}}p_{i,j,\nu_{j}}\sum_{\beta_{j}=1}^{d_{j}}q_{i,j,\beta_{j}}\right.\\
& \left(\sum_{v=1}^{L_{j,i}}e^{(\widetilde{\sigma}-\overline{\sigma})t_{i}}\mathbb{E}\left[e^{t_{i}D_{i,j,\beta_{j},\nu_{j}}^{(v)}}\right]+\right.\\
& \left.\left.\sum_{v=L_{j,i}+1}^{L_{i}}e^{(\widetilde{\sigma}-\overline{\sigma})t_{i}}\mathbb{E}\left[e^{t_{i}D_{i,j,\beta_{j},\nu_{j}}^{(v)}}\right]\right)\right]\\
& \overset{(g)}{\leq}\sum_{j=1}^{m}\pi_{i,j}\left[e^{-t_{i}\sigma}+\sum_{\nu_{j}=1}^{e_{j}}p_{i,j,\nu_{j}}\sum_{\beta_{j}=1}^{d_{j}}q_{i,j,\beta_{j}}\times\right.\\
& e^{(\widetilde{\sigma}-\overline{\sigma})t_{i}}\left(\sum_{v=1}^{L_{j,i}}\frac{(1-\rho_{j,\nu_{j}}^{(e)})t_{i}B_{j,\nu_{j}}^{(e)}(t_{i})}{t_{i}-\Lambda_{j,\nu_{j}}^{(e)}(B_{j,\nu_{j}}^{(e)}(t_{i})-1)}\left(\frac{\alpha_{j,\nu_{j}}^{(e)}e^{\eta_{j,\nu_{j}}^{(e)}}}{\alpha_{j,\nu_{j}}^{(e)}-t_{i}}\right)^{v}+\right.\\
& \frac{(1-\rho_{j,\beta_{j}}^{(\overline{d})})t_{i}B_{j,\beta_{j}}^{(\overline{d})}(t_{i})}{t_{i}-\Lambda_{j,\beta_{j}}^{(\overline{d})}(B_{j,\beta_{j}}^{(\overline{d})}(t_{i})-1)}\left(\frac{\alpha_{j,\beta_{j}}^{(\overline{d})}e^{\eta_{j,\beta_{j}}^{(\overline{d})}}}{\alpha_{j,\beta_{j}}^{(\overline{d})}-t_{i}}\right)^{L_{i}-L_{j,i}}\!\!\!\!\!\!\!\!\!\!\!\!\!\!\!\!\!\!\!(L_{i}-L_{j,i}){\bf 1}_{L_{i}>L_{j,i}}\\
 & +\sum_{v=L_{j,i}+1}^{L_{i}}\sum_{w=L_{j,i}+1}^{v}\frac{(1-\rho_{j,\beta_{j}}^{(d)})t_{i}B_{j,\beta_{j}}^{(d)}(t_{i})}{t_{i}-\Lambda_{j,\beta_{j}}^{(d)}(B_{j,\beta_{j}}^{(d)}(t_{i})-1)}\times\\
& \left.\left.\left(\frac{\alpha_{j,\beta_{j}}^{(d)}e^{\eta_{j,\beta_j}^{(d)}}}{\alpha_{j,\beta_{j}}^{(d)}-t_{i}}\right)^{w-L_{j,i}-1}\left(\frac{\alpha_{j,\beta_{j}}^{(\overline{d})}e^{\eta_{j,\beta_j}^{(\overline{d})}}}{\alpha_{j,\beta_{j}}^{(\overline{d})}-t_{i}}\right)^{L_{i}-w+1}\right)\right]\\
 & =\sum_{j=1}^{m}\pi_{i,j}\left[e^{-t_{i}\sigma}+\sum_{\nu_{j}=1}^{e_{j}}p_{i,j,\nu_{j}}\sum_{\beta_{j}=1}^{d_{j}}q_{i,j,\beta_{j}}e^{(\widetilde{\sigma}-\overline{\sigma})t_{i}}\times\right.\\
& \left(\frac{M_{j,\nu_{j}}^{(e)}(t_{i})(1-\rho_{j,\nu_{j}}^{(e)})t_{i}B_{j,\nu_{j}}^{(e)}(t_{i})}{t_{i}-\Lambda_{j,\nu_{j}}^{(e)}(B_{j,\nu_{j}}^{(e)}(t_{i})-1)}\frac{\left((M_{j,\nu_{j}}^{(e)}(t_{i}))^{L_{j,i}}-1\right)}{M_{j,\nu_{j}}^{(e)}(t_{i})-1}{\bf 1}_{L_{j,i}>0}+\right.\\
&  \frac{(1-\rho_{j,\beta_{j}}^{(\overline{d})})t_{i}B_{j,\beta_{j}}^{(\overline{d})}(t_{i})}{t_{i}-\Lambda_{j,\beta_{j}}^{(\overline{d})}(B_{j,\beta_{j}}^{(\overline{d})}(t_{i})-1)}\left(\frac{\alpha_{j,\beta_{j}}^{(\overline{d})}e^{\eta_{j,\beta_{j}}^{(\overline{d})}}}{\alpha_{j,\beta_{j}}^{(\overline{d})}-t_{i}}\right)^{L_{i}-L_{j,i}}\!\!\!\!\!\!\!\!\!(L_{i}-L_{j,i}){\bf 1}_{L_{i}>L_{j,i}}+\\
 & \frac{(1-\rho_{j,\beta_{j}}^{(d)})t_{i}B_{j,\beta_{j}}^{(d)}(t_{i})\left(\widetilde{M}_{j,\beta_{j}}^{(d,c)}(t_{i})\right)^{L_{j,i}+1}}{t_{i}-\Lambda_{j,\beta_{j}}^{(d)}(B_{j,\beta_{j}}^{(d)}(t_{i})-1)}\times\\
&  \left.\left.\left(\frac{\alpha_{j,\beta_{j}}^{(d)}-t_{i}}{\alpha_{j,\beta_{j}}^{(d)}e^{\eta_{j,\beta_{j}}^{(d)}}}\right)^{L_{j,i}+1}\left(\frac{\alpha_{j,\beta_{j}}^{(\overline{d})}e^{\eta_{j,\beta_{j}}^{(\overline{d})}}}{\alpha_{j,\beta_{j}}^{(\overline{d})}-t_{i}}\right)^{L_{i}+1}\times\right.\right.\\
\end{align*}
\begin{align*}
%\end{align*}
%
%\begin{align*}
 & \left(\frac{(1-\rho_{j,\beta_{j}}^{(d)})t_{i}B_{j,\beta_{j}}^{(d)}(t_{i})\left(\widetilde{M}_{j,\beta_{j}}^{(d,\overline{d})}(t_{i})\right)^{L_{j,i}+1}}{t_{i}-\Lambda_{j,\beta_{j}}^{(d)}(B_{j,\beta_{j}}^{(d)}(t_{i})-1)}+\right.\\
 & \left.\left.\left.\frac{\widetilde{M}_{j,\beta_{j}}^{(d,\overline{d})}(t_{i})\left(\left(\widetilde{M}_{j,\beta_{j}}^{(d,\overline{d})}(t_{i})\right)^{L_{i}-L_{j,i}-1}-1\right)}{\left(\widetilde{M}_{j,\beta_{j}}^{(d,\overline{d})}(t_{i})-1\right)^{2}}\right){\bf 1}_{L_{i}>L_{j,i}}\right)\right]
\end{align*}
where $\widetilde{M}_{j,\beta_{j}}^{(d,\overline{d})}(t)=\frac{\alpha_{j,\beta_{j}}^{(d)}(\alpha_{j,\beta_{j}}^{(\overline{d})}-t)e^{\eta_{j,\beta_j}^{(d)}t}}{\alpha_{j,\beta_{j}}^{(\overline{d})}(\alpha_{j,\beta_{j}}^{(d)}-t)e^{\eta_{j,\beta_j}^{(\overline{d})}t}},\,\,\,\forall j,\beta_{j}$ and\\
 $M_{j,\nu_{j}}^{(e)}(t_i)= \frac{\alpha_{j,\nu_{j}}^{(e)}e^{\eta_{j,\beta_j}^{(e)}}}{\alpha_{j,\nu_{j}}^{(e)}-t_i}$. The step (g) follows by substitution of the moment generating functions, and the rest of the steps use the sum of geometric and Arithmetico-geometric sequences. This proves the statement of the Lemma.

\section{Proof of Results in Section  \ref{sec:algo}} \label{apd_ww}

\subsection{Proof of Lemma \ref{cnv_t}}
The constraints \eqref{eq:mgf_beta_d}--\eqref{eq:mgf_beta_e} are separable for each $t_i$ and due to symmetry of the three constraints it is enough to prove convexity of $E(t)=\sum_{f=1}^{r}\pi_{f,j}q_{f,j,\beta_{j}}\lambda_{f}\left(\frac{\alpha}{\alpha-t_{i}}\right)^{L_{f}-L_{j,f}}-\left(\Lambda_{j,\beta_{j}}+t_{i}\right)$. Thus, it is enough to prove that $E''(t)\ge 0$. 
We further note that it is enough to prove that $D''(t)\ge 0$, where $D(t) = \left(\frac{\alpha}{\alpha-t_{i}}\right)^{L_{f}-L_{j,f}
}$. 
This follows since 
\begin{align}
D^{'}(t) & =(L_{f}-L_{j,f})(1-\frac{t}{\alpha})^{L_{j,f}-L_{f}-1}\times(1/\alpha)\ge0 \\
D^{''}(t) & =(L_{f}^{2}-L_{j,f}^{2}+L_{f}-L_{j,f})(1-\frac{t}{\alpha})^{L_{j,f}-L_{f}-2} \nonumber \\
& \,\,\,\,\,\,\,\,\,\,\,\,\,\,\,\,\,\,\,\,\, \times(1/\alpha^{2})\ge0
\end{align}

%The first derivative of $D(t)$ is given as

%\begin{equation}
%	D'(t) =\frac{\pi_{ij}\lambda_{i}L_{i}\alpha_{j,\nu_j}^{L_i}e^{\beta_{j,\nu_j}tL_{i}}}{\left(\alpha_{j,\nu_j}-t\right)^{L_i+1}}\left[\beta_{j,\nu_j}\left(\alpha_{j,\nu_j}-t\right)+1\right]
%\end{equation}

%Note that $E'(t)\ge0$ since $\alpha_{j,\nu_j}\get$.  Differentiating it again to get the second derivative, we get the second derivative as follows. 
%\begin{equation}
%	E''(t)=\left(1+\frac{1}{\alpha_{j,\nu_j}-t}\right)\left((\alpha_{j,\nu_j}-t)\,L_{i}+L_{i}+1\right)-1 
%	\label{eq:E_22}
%\end{equation}

%Since $\alpha_{j,\nu_j}\get$, $E''(t)$ given in  \eqref{eq:E_22} is non-negative, which proves the Lemma. 
\subsection{Proof of Lemma \ref{cnv_ww}} 

The constraint \eqref{eq:t_leq_alpha_beta_d}--\eqref{eq:mgf_beta_e}  are separable for each  $\alpha_{j,\beta_j}^{(d)}$, $\alpha_{j,\beta_j}^{(c)}$ and $\alpha_{j,\nu_j}^{(e)}$, respectively. Thus, it is enough to prove convexity of the following three equations
%$E_{1}(\alpha_{j,\beta_{j}}^{(d)})=\sum_{f=1}^{r}\pi_{f,j}\lambda_{f}q_{f,j,\beta_{j}}\left(\frac{\alpha_{j,\beta_{j}}^{(d)}}{\alpha_{j,\beta_{j}}^{(d)}-t}\right)^{L_{f}-L_{j,f}}-\left(\Lambda_{j,\beta_{j}}^{(d)}+t\right)$,

\begin{align*}
\ensuremath{} & E_{1}(\alpha_{j,\beta_{j}}^{(d)})=\\
& \ensuremath{\sum_{f=1}^{r}\pi_{f,j}\lambda_{f}q_{f,j,\beta_{j}}\left(\frac{\alpha_{j,\beta_{j}}^{(d)}}{\alpha_{j,\beta_{j}}^{(d)}-t}\right)^{L_{f}-L_{j,f}}-\left(\Lambda_{j,\beta_{j}}^{(d)}+t\right)}
\end{align*}

%$E_{2}(\alpha_{j,\beta_{j}}^{(c)})=\sum_{f=1}^{r}\pi_{f,j}\lambda_{f}q_{f,j,\beta_{j}}\left(\frac{\alpha_{j,\beta_{j}}^{(c)}}{\alpha_{j,\beta_{j}}^{(c)}-t}\right)^{L_{f}-L_{j,f}}-\left(\Lambda_{j,\beta_{j}}^{(c)}+t\right)$

\begin{align*}
\ensuremath{} & E_{2}(\alpha_{j,\beta_{j}}^{(\overline{d})})\\
& =\sum_{f=1}^{r}\pi_{f,j}\lambda_{f}q_{f,j,\beta_{j}}\left(\frac{\alpha_{j,\beta_{j}}^{(\overline{d})}}{\alpha_{j,\beta_{j}}^{(\overline{d})}-t}\right)^{L_{f}-L_{j,f}}-\left(\Lambda_{j,\beta_{j}}^{(\overline{d})}+t\right)\\
& \ensuremath{E_{3}(\alpha_{j,\nu_{j}}^{(e)})=}\\
& \sum_{f=1}^{r}\pi_{f,j}\lambda_{f}p_{f,j,\nu_{j}}\left(\frac{\alpha_{j,\nu_{j}}^{(e)}}{\alpha_{j,\nu_{j}}^{(e)}-t}\right)^{L_{f}-L_{j,f}}-\left(\Lambda_{j,\nu_{j}}^{(e)}+t\right)
\end{align*}
%and $E_{3}(\alpha_{j,\nu_{j}}^{(e)})=\sum_{f=1}^{r}\pi_{f,j}\lambda_{f}p_{f,j,\nu_{j}}\left(\frac{\alpha_{j,\nu_{j}}^{(e)}}{\alpha_{j,\nu_{j}}^{(e)}-t}\right)^{L_{f}-L_{j,f}}-\left(\Lambda_{j,\nu_{j}}^{(e)}+t\right)$ 
for $t<\alpha_{j,\beta_j}^{(d)}$, $t<\alpha_{j,\beta_j}^{(c)}$, and $t<\alpha_{j,\nu_j}^{(e)}$, respectively. Since there is only a single index $j$, $\beta_j$, and $\nu_j$, here, we ignore the subscripts and superscripts for the
rest of this proof and prove for only one case due to the symmetry. Thus, it is enough to prove that $E_1''(\alpha)\ge 0$  for $t<\alpha$. We further note that it is enough to prove that $D_1''(\alpha)\ge 0$, where $\ensuremath{D_{1}(\alpha)=\left(1-\frac{t}{\alpha}\right)^{L_{j,i}-L_{i}}}$. This holds since,

\begin{align}
D_{1}^{'}(\alpha) & =(L_{j,i}-L_{i})(1-\frac{t}{\alpha})^{L_{j,i}-L_{i}-1}\times(t/\alpha^{2})\\
D_{1}^{''}(\alpha) & =((L_{j,i}-L_{i})^2-L_{j,i}+L_{i})(1-\frac{t}{\alpha})^{L_{j,i}-L_{i}-2} \times \nonumber \\
& ,\,\,\,\,\,\,\,\,\,\,\,\,\,\,\,\,\,\,\,\,\,\,\,\ (t/\alpha^{3})\ge0
\end{align}

\section{Algorithm Pseudo-codes for the Sub-problems}\label{apdx_alg}

\begin{algorithm}[ht]
\caption{NOVA Algorithm to solve Server Access and PSs selection Optimization sub-problem\label{alg:NOVA_Alg1Pi}}

\begin{enumerate}
\item \textbf{Initialize} $\nu=0$, $k=0$,$\gamma^{\nu}\in\left(0,1\right]$,
$\epsilon>0$,$\widetilde{\boldsymbol{\pi}}^{0}$ such that $\widetilde{\boldsymbol{\pi}}^{0}$
is feasible ,
\item \textbf{while} $\mbox{obj}\left(k\right)-\mbox{obj}\left(k-1\right)\geq\epsilon$
\item $\quad$//\textit{\small{}Solve for $\widetilde{\boldsymbol{\pi}}^{\nu+1}$ with
given $\widetilde{\boldsymbol{\pi}}^{\nu}$}{\small \par}
\item $\quad$\textbf{Step 1}: Compute $\boldsymbol{\widehat{\pi}}\left(\widetilde{\boldsymbol{\pi}}^{\nu}\right),$
the solution of $\boldsymbol{\widehat{\pi}}\left(\widetilde{\boldsymbol{\pi}}^{\nu}\right)=$$\underset{\widetilde{\boldsymbol{\pi}}}{\text{argmin}}$
$\boldsymbol{\widetilde{U}}\left(\widetilde{\boldsymbol{\pi}},\widetilde{\boldsymbol{\pi}}^{\nu}\right)\,\, $ s.t.
$\left(\eqref{eq:rho_beta_d}--\eqref{eq:pi_q_p_gz}\right)$, $\left(\eqref{eq:mgf_beta_d}--
\eqref{eq:mgf_beta_e}\right)$, solved using  projected gradient descent

\item $\quad$\textbf{Step 2}: $\ensuremath{\widetilde{\boldsymbol{\pi}}^{\nu+1}=\widetilde{\boldsymbol{\pi}}^{\nu}+\gamma^{\nu}\left(\widehat{\boldsymbol{\pi}}\left(\widetilde{\boldsymbol{\pi}}^{\nu}\right)-\widetilde{\boldsymbol{\pi}}^{\nu}\right)}$.
\item $\quad$//\textit{\small{}update index}{\small \par}
\item \textbf{Set} $\ensuremath{\nu\leftarrow\nu+1}$
\item \textbf{end while}
\item \textbf{output: }$\ensuremath{\widehat{\boldsymbol{\pi}}\left(\widetilde{\boldsymbol{\pi}}^{\nu}\right)}$
\end{enumerate}
\end{algorithm}

\begin{algorithm}[ht]
	\caption{NOVA Algorithm to solve Auxiliary Variables  Optimization sub-problem\label{alg:NOVA_Alg1}}
	
	\begin{enumerate}
		\item \textbf{Initialize} $\nu=0$, $\gamma^{\nu}\in\left(0,1\right]$, $\epsilon>0$, $\boldsymbol{t}^{0}$ 
		such that $\boldsymbol{t}^{0}$  is feasible,
		\item \textbf{while} $\mbox{obj}\left(\nu\right)-\mbox{obj}\left(\nu-1\right)\geq\epsilon$
		\item $\quad$//\textit{\small{}Solve for  $\boldsymbol{t}^{\nu+1}$ with
			given $\boldsymbol{t}^{\nu}$}{\small \par}
		\item $\quad$\textbf{Step 1}: Compute $\boldsymbol{\widehat{t}}\left(\boldsymbol{t}^{\nu}\right),$
		the solution of  $\boldsymbol{\widehat{t}}\left(\boldsymbol{t}^{\nu}\right)=$$\underset{\boldsymbol{t}}{\text{argmin}}$
		$\boldsymbol{\overline{U}}\left(\boldsymbol{t},\boldsymbol{t}^{\nu}\right)$, s.t. \eqref{eq:t_leq_alpha_beta_d}--\eqref{eq:mgf_beta_e}, using projected gradient descent
\item $\quad$\textbf{Step 2}: $\ensuremath{\boldsymbol{t}^{\nu+1}=\boldsymbol{t}^{\nu}+\gamma^{\nu}\left(\widehat{\boldsymbol{t}}\left(\boldsymbol{t}^{\nu}\right)-\boldsymbol{t}^{\nu}\right)}$.
		\item $\quad$//\textit{\small{}update index}{\small \par}
		\item \textbf{Set} $\ensuremath{\nu\leftarrow\nu+1}$
		\item \textbf{end while}
		\item \textbf{output: }$\ensuremath{\widehat{\boldsymbol{t}}\left(\boldsymbol{t}^{\nu}\right)}$
	\end{enumerate}
\end{algorithm}

\begin{algorithm}[ht]
	\caption{NOVA Algorithm to solve Bandwidth Allocation Optimization sub-problem\label{alg:NOVA_Alg3}}	
	\begin{enumerate}
		\item \textbf{Initialize} $\nu=0$, $\gamma^{\nu}\in\left(0,1\right]$, $\epsilon>0$, $\boldsymbol{w}^{0}$ 
		such that $\boldsymbol{w}^{0}$  is feasible,
		\item \textbf{while} $\mbox{obj}\left(\nu\right)-\mbox{obj}\left(\nu-1\right)\geq\epsilon$
		\item $\quad$//\textit{\small{}Solve for  $\boldsymbol{w}^{\nu+1}$ with
			given $\boldsymbol{w}^{\nu}$}{\small \par}
		\item $\quad$\textbf{Step 1}: Compute $\boldsymbol{\widehat{w}}\left(\boldsymbol{w}^{\nu}\right),$
		the solution of  $\boldsymbol{\widehat{w}}\left(\boldsymbol{w}^{\nu}\right)=$$\underset{\boldsymbol{b}}{\text{argmin}}$
		$\boldsymbol{\overline{U}}\left(\boldsymbol{w},\boldsymbol{w}^{\nu}\right)$, s.t. \eqref{eq:w_cde_gz}--
\eqref{eq:sum_w_de},\,\,\eqref{eq:t_leq_alpha_beta_d}--\eqref{eq:mgf_beta_e}, using projected gradient descent
		\item $\quad$\textbf{Step 2}: $\ensuremath{\boldsymbol{w}^{\nu+1}=\boldsymbol{w}^{\nu}+\gamma^{\nu}\left(\widehat{\boldsymbol{w}}\left(\boldsymbol{w}^{\nu}\right)-\boldsymbol{w}^{\nu}\right)}$.
		\item $\quad$//\textit{\small{}update index}{\small \par}
		\item \textbf{Set} $\ensuremath{\nu\leftarrow\nu+1}$
		\item \textbf{end while}
		\item \textbf{output: }$\ensuremath{\widehat{\boldsymbol{w}}\left(\boldsymbol{w}^{\nu}\right)}$
	\end{enumerate}
\end{algorithm}

\begin{algorithm}[ht]
	\caption{NOVA Algorithm to solve Cache Placement Optimization sub-problem\label{alg:NOVA_Alg5}}
	
	\begin{enumerate}
		\item \textbf{Initialize} $\nu=0$, $\gamma^{\nu}\in\left(0,1\right]$, $\epsilon>0$, $\boldsymbol{L}^{0}$ 
		such that $\boldsymbol{L}^{0}$  is feasible,
		\item \textbf{while} $\mbox{obj}\left(\nu\right)-\mbox{obj}\left(\nu-1\right)\geq\epsilon$
		\item $\quad$//\textit{\small{}Solve for  $\boldsymbol{L}^{\nu+1}$ with
			given $\boldsymbol{L}^{\nu}$}{\small \par}
		\item $\quad$\textbf{Step 1}: Compute $\boldsymbol{\widehat{L}}\left(\boldsymbol{L}^{\nu}\right),$
		the solution of  $\boldsymbol{\widehat{L}}\left(\boldsymbol{L}^{\nu}\right)=$$\underset{\boldsymbol{L}}{\text{argmin}}$
		$\boldsymbol{\overline{U}}\left(\boldsymbol{L},\boldsymbol{L}^{\nu}\right)$, s.t. \eqref{eq:rho_beta_d}--
\eqref{eq:rho_nu_e},
\eqref{eq:capConst},
\eqref{eq:mgf_beta_d}-- \eqref{eq:mgf_beta_e}\\
		using projected gradient descent
		\item $\quad$\textbf{Step 2}: $\ensuremath{\boldsymbol{L}^{\nu+1}=\boldsymbol{L}^{\nu}+\gamma^{\nu}\left(\widehat{\boldsymbol{L}}\left(\boldsymbol{L}^{\nu}\right)-\boldsymbol{L}^{\nu}\right)}$.
		\item $\quad$//\textit{\small{}update index}{\small \par}
		\item \textbf{Set} $\ensuremath{\nu\leftarrow\nu+1}$
		\item \textbf{end while}
		\item \textbf{output: }$\ensuremath{\widehat{\boldsymbol{L}}\left(\boldsymbol{L}^{\nu}\right)}$
	\end{enumerate}
\end{algorithm}

\end{document}